\documentclass[aps,pre,twocolumn,superscriptaddress,showpacs,showkeys ]{revtex4-1}

\usepackage{graphicx}                   % for insert file eps
\usepackage{psfig}
\usepackage{hyperref}                   % for hyper refrences
\usepackage{amsmath,amssymb}            % for insert extra math
\usepackage{epstopdf}
\usepackage{placeins}

\usepackage{color}
\definecolor{orange}{rgb}{1,0.5,0}
\definecolor{brown}{rgb}{0.65, 0.16, 0.16}
\definecolor{phlox}{rgb}{0.87, 0.0, 1.0}

\graphicspath{{figs/}}                  % the path of image folder

\begin{document}

    % \title{Persistent Homology of  Weighted Visibility Graph from Fractional Gaussian Noise}
    \title{Persistent Homology of Fractional Gaussian Noise}

    \author{H. Masoomy}
    %\email{hoseingmasoomy@gmail.com}
    \affiliation{Department of Physics, Shahid Beheshti University,  1983969411, Tehran, Iran}
    %\affiliation{Ibn-Sina Multidisciplinary laboratory, Department of Physics, Shahid Beheshti University, Velenjak, Tehran 19839, Iran}

    \author{B. Askari}
    %\email{b.askari@gmail.com}
    \affiliation{Department of Physics, Shahid Beheshti University,  1983969411, Tehran, Iran}

    \author{M. N. Najafi}
    \email{nattagh.najafi@uma.ac.ir}
    %\email{morteza.nattagh@gmail.com}
    \affiliation{Department of Physics, University of Mohaghegh Ardabili, P.O. Box 179, Ardabil, Iran}

    \author{S. M. S. Movahed}
    \email{m.s.movahed@ipm.ir}
    \affiliation{Department of Physics, Shahid Beheshti University,  1983969411, Tehran, Iran}
    % \affiliation{Ibn-Sina Multidisciplinary laboratory, Department of Physics, Shahid Beheshti University, Velenjak, Tehran 19839, Iran}

    \begin{abstract}

        In this paper, we employ the persistent homology (PH) technique to examine the topological properties of fractional Gaussian noise (fGn). We develop the weighted natural visibility graph algorithm, and the associated simplicial complexes through the filtration process are quantified by PH. The evolution of the homology group dimension represented by Betti numbers demonstrates a strong dependency on the Hurst exponent ($H$). The coefficients of the birth and death curve of the $k$-dimensional topological holes ($k$-holes) at a given threshold depend on $H$ which is almost not affected by finite sample size. We show that the distribution function of a lifetime for $k$-holes decays exponentially and the corresponding slope is an increasing function versus $H$, and more interestingly, the sample size effect completely disappears in this quantity. The persistence entropy logarithmically grows with the size of the visibility graph of a system with almost $H$-dependent prefactors. On the contrary, the local statistical features are not able to determine the corresponding Hurst exponent of fGn data, while the moments of eigenvalue distribution ($M_{n}$) for $n\ge1$ reveal a dependency on $H$, containing the sample size effect. Finally, the PH shows the correlated behavior of electroencephalography for both healthy and schizophrenic samples.
    \end{abstract}

    \keywords{Topological Data Analysis, Persistent Homology, Fractional Gaussian Noise, Weighted Natural Visibility Graph, Topological Persistence, Persistence Entropy}

    \maketitle
    \section{Introduction}

    A powerful approach to study different types of data sets ranging from point cloud data and scalar fields to a complex network (graph), particularly high-dimensional data, is called topological data analysis (TDA)~\cite{edelsbrunner2000topological,zomorodian2005computing,zomorodian2005topology,carlsson2009topology,zomorodian2012topological,wasserman2018topological}. TDA as an application of algebraic topology \cite{nakahara2003geometry, munkres2018elements, hatcher2005algebraic} and a branch of computational topology \cite{Edels2010Comp}; it analyzes the shape of high-dimensional complex data in terms of global features (number of connected components, loops, voids, etc.) of topological space underlying the data set. In the persistent homology (PH) technique, as the main part of TDA, the topological approximation of phase space of any type of data set which is called the simplicial complex is assigned to the underlying data, and then topological invariants are computed.

    The PH aims to capture topological evolution of data set by varying scale (parameter), and extracts topological invariants of the data set in each scale summarizing them in different representations, e.g., persistence barcode (PB)~\cite{ghrist2008barcodes,carlsson2005persistence}, persistence diagram (PD)~\cite{patel2018generalized}, persistence landscape (PL)~\cite{bubenik2017persistence}, persistence image (PI)~\cite{adams2017persistence}, persistence surface (PS), and $\beta$-curve, which reveal topological information of the data set. Being robust to noise, PH can clarify the essential features of the systems with high internal degrees of freedom and is capable of classifying data sets~\cite{chazal2017introduction,munch2017user}. The PH technique has attracted much attention due to its vast applications in analyzing complex networks~\cite{serrano2020simplicial,saggar2018towards,horak2009persistent}. Also it has been used in various systems (see, e.g., \cite{topaz2015topological,sizemore2019importance,pranav2017topology,jaquette2020fractal,speidel2018topological,salnikov2018simplicial,bobrowski2020homological,tran2021topological,olsthoorn2020finding} and references therein).

    A mathematical model containing correlation tuned by one parameter (the Hurst exponent~\cite{hurst1951long}) is the fractional Brownian motion (fBm) and its increment is known as fractional Gaussian noise (fGn) \cite{mandelbrot1968fractional}. It is also used to model self-similar phenomena of various types ranging from meteorology, engineering, econophysics, and astronomy to biology (see, e.g., \cite{eghdami2018multifractal,jiang2019multifractal} and references therein). To quantify the properties of a given self-similar data set whose power spectrum behaves as a power-law in the frequency (wavelength) domain, many methods have been proposed. A well-studied method is
    multi-fractal detrended fluctuation analysis (MFDFA)~\cite{peng1994mosaic,Peng95,Kantelhardt}. Taking into account the higher-order detrended covariance, the multi-fractal detrended cross-correlation analysis (MFDXA) has been introduced~\cite{mf-dxa}. Despite many advantages brought by the mentioned methods, the impact of more complicated trends and finite-size effects have not been diminished completely in many previous approaches~\cite{kunhu,trend2,physa,trend3}.
    On the other hand, the finite size of time-series affects the accurate estimation of the Hurst exponent by some of previous methods~\cite{xu2005quantifying}.

    Although different methods can be found in the literature to determine the scaling exponent of fBm or fGn series, very little attention has been paid to deal with the topological properties and the probable capability of TDA for estimation of the Hurst exponent. In this regard, knowing a given time series belongs to an fGn class, some relevant questions can be raised: (i) Does the topological aspect of time series depend on the Hurst exponent? (ii) What are the effects of sample size, trends, and irregularity of fGn signal on the PH of topological motifs generated from the data set? (iii) How is the multifractality expressed by PH? Motivated by the mentioned questions, in this paper, we concentrate on the topological properties (homology group) of fGn.

    A way to construct the higher dimensional manifold from a typical time-series in order to evaluate the evolution of $k$-dimensional holes through the filtration process is assigning a weighted graph to the underlying time-series. There are various methods to assign a network to a typical time-series, e.g., the proximity network, cycle network, visibility graph, correlation network, recurrence network, and transition network (see, e.g., \cite{zou2019complex} and references therein).  The idea of the visibility graph (VG) is a complex network constructed by considering the \textit{visibility} algorithm, proposed by Lacasa \textit{et al} as a novel way to analyze time-series in terms of complex network language~\cite{lacasa2008time} proposed as an alternative method to estimate $H$ for fBm and fGn series ~\cite{lacasa2009visibility}.

    In this paper, we rely on statistical and topological properties (homology group) of natural VGs (NVGs) associated with such signals. First, we propose a weighted version of natural VGs (WNVGs) by defining a well-defined weight function to quantify the quality of visibility between data points of the signal for extracting nonsensitive (robust) features in the presence of possible noises in the signal. Then, we apply the filtration process on the weighted clique simplicial complex corresponding to WNVG by considering the weight (visibility value) of links as a threshold parameter and higher-order connections ($k$-cliques ; $k>2$) as building blocks of such higher-order structure. Accordingly, we figure out the evolution of robust topological features explaining the global structure of WNVG. Considering other techniques for making networks from time-series and applying PH could provide interesting results, but is time-consuming from computational points of view; the possibility of distinguishing the non-stationary and stationary series, the sensitivity of corresponding results to the value of the Hurst exponent, and the robustness of results with sample size are some of our criteria for taking VG in its weighted version.

    Our research presented in this paper has the following advantages.

    (1) Inspired by network science and a self-similar process characterized by a scaling exponent called the "Hurst exponent", we implement the weighted natural visibility graph (WNVG) to make a network from fractional Gaussian noise (fGn). In this approach, the topological motifs survive; consequently, their evolution concerning self-similar exponents can be examined. Our results approve that such evolution is a robust feature for determining the Hurst exponent.

    (2) We will demonstrate that the statistical analysis of WNVGs such as probability distribution functions of eigenvector, betweenness, and closeness centralities are almost $H$-independent, and therefore, they can not recognize the type of correlation of fGn series.

    (3) We rely on TDA and implement the PH, which is a robust method to examine the evolution of global properties during the filtration process. Almost all relevant results are sensitive enough to the Hurst exponent of the underlying data set and it is also possible to check whether the data are stationary or not. More precisely, irrespective of either stationarity or non-stationarity of underlying times series, to capture the homology generators, the constructed network should be weighted. It is worth noting that the non-stationary behavior may be produced due to the various types of trends superimposed on the intrinsic fluctuation; therefore, in the presence of any prior information regarding trends the pre-processing procedure including at least one of the following algorithms must be employed to produce clean series, e.g., the singular value decomposition algorithm \citep{nagarajan2005minimizing}, the adaptive detrending algorithm \citep{hu2009multifractal}, and empirical mode decomposition \citep{wu2007trend}.

    (4) Finally, we emphasize that the behavior of the local (statistical) observables depend weakly on $H$, whereas the coefficients of global (topological) observables are almost strongly $H$-dependent and even for a part of measures, the size effect is almost diminished. We notify that TDA provides a new type of measure to quantify the Hurst exponent, and therefore, the scaling exponents of the correlation function, power spectrum, and fractal dimension can be specified with reliable approaches.

    The rest of paper is organized as follows: In the next section, our methodology is introduced briefly. The construction of VGs for a time-series and the key idea of topological network analysis are clarified in Sec. \ref{Sec:network_analysis}. The numerical results of synthetic fGn time series, which are covered in two subsections, local statistical properties and topological properties, are presented in Sec.~\ref{Sec:results}. The implementation of realistic data is described in Sec. \ref{Sec:eeg}. We give some ideas for applying the proposed method to various problems in Sec.~\ref{sec:summery}, and close the paper with a conclusion. More details on synthetic data generation, statistical network analysis, algebraic topology, and persistent homology are explained in appendices.

    \section{Methodology}\label{Sec:network_analysis}
    %%%%%%%%%%%%%%%%%%%%%%%%%%%%%%%%%%%%%%%%%%%%%%%%%%%%%%%%%%%%%%%%%%%%%

    \begin{figure*}
        \begin{center}
            \includegraphics[width=0.48\textwidth]{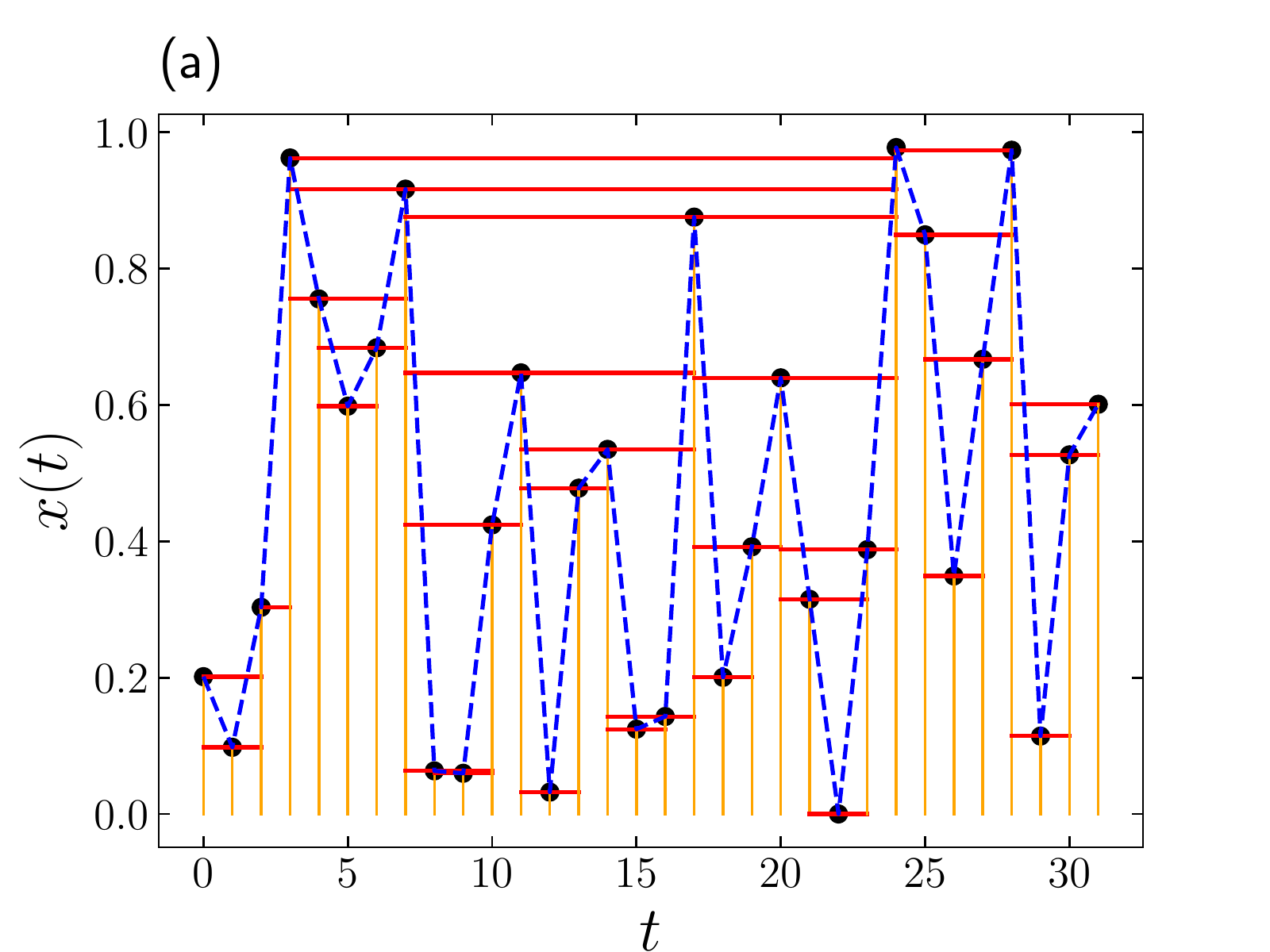}
            \includegraphics[width=0.48\textwidth]{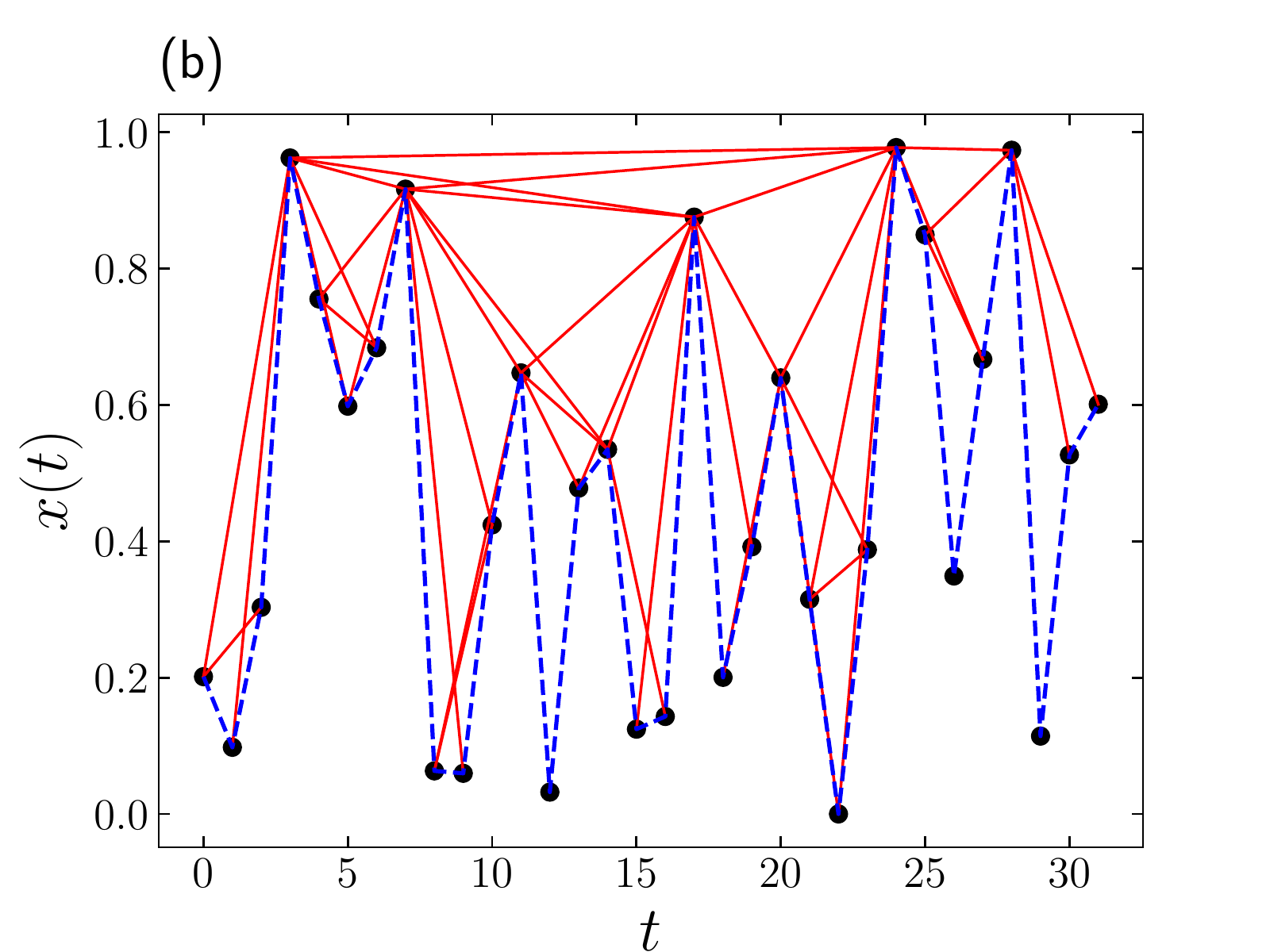}
            \caption{The schematic representation of making a network for a typical data set using VG algorithm. (a) The HVG and (b) NVG of a synthetic fGn with $H=0.5$ (white-noise). To make more sense, we took the sample size equal to $N=32$.}\label{fig:HVG_NVG}
        \end{center}
    \end{figure*}

    In this research, we aim to analyze the complex network of the visibility graph constructed from a fractional Gaussian noise (fGn) (see Appendix A for more details of generating a synthetic fGn series), with an emphasis on the topological aspects. Besides this, we also compute some conventional statistical properties of the mentioned signal.

    \subsection{Visibility Graph Method}\label{Sec:VG}
    Among growing applicability of complex networks in many fields and interdisciplinary branches in science~\cite{costa2011analyzing}, a technique has been suggested which converts a time-series to a network, the so-called visibility graph method~\cite{lacasa2008time}. Generally, suppose $\{x\}:\{x(t_i), i=1,...,N\}$ represents a real-valued time-series. One can construct a network, the so-called visibility graph, denoted by $G=(V,E,w)$, $V \equiv \{v_i\}_{i=1}^{N}$ is the node (vertex) set, and $E$ is link (edge) set and $w$ is a map from $E$ to real numbers.
    The VG is defined by using the bijection as follows,
    \begin{equation}
    f : ~ V\equiv\Bigr\{v_i\Bigr\}_{i=1}^{N} \leftrightarrow \quad T\equiv \Bigr(t_i\Bigr)_{i=1}^{N} \quad ; \quad f(v_i)=t_i
    \end{equation}
    and the connections are constructed according to the \textit{visibility condition} between the nodes, i.e., the nodes $v_i$ and $v_j$ are connected if the node $v_j$ is visible from the node $v_i$ and vice versa, and therefore the resulting graph is undirected (for more details on the properties of VGs, see~\cite{lacasa2008time}). In general, there are two ways to construct a network (graph) from a time-series: the horizontal visibility graph (HVG) \cite{gonccalves2016time,gao2016multiscale,xie2011horizontal} and the natural visibility graph (NVG) \cite{zheng2020visibility,yang2009visibility,ahmadlou2010new}; the former is more sparse than the latter case and in this work we focus on the NVG. In Fig.~\ref{fig:HVG_NVG}, we show how an HVG [panel (a)] and an NVG [panel (b)] for a synthetic fGn series can be constructed. In a binary setup, the corresponding visibility graph chooses the range of the weights from a binary set, $w^{(B)}: E \rightarrow \mathbb{Z}_2 \equiv \{ 0,1 \}$; e.g., for a binary NVG (BNVG) the weight function can be written according to following relation,
    \begin{widetext}
        \begin{align}
        w_{ij}^{(BN)} \equiv\left\{
        \begin{array}{l l}
        1 \qquad \qquad\qquad \qquad ; \qquad \Bigr{|}f(v_{i})-f(v_{j}) \Bigr{|} = 1\\
        \displaystyle \prod_{k=i+1}^{j-1} \Theta \Bigr{(} s_{ij} - s_{ik} \Bigr{)} ~ ; \qquad \Bigr{|}f(v_{i})-f(v_{j}) \Bigr{|} > 1
        \end{array} \right.
        \label{Eq:BWeights}
        \end{align}
    \end{widetext}
    where $\Theta$ is the step function, and $s_{ij}\equiv \frac{x(f(v_{j})) - x(f(v_{i}))}{f(v_{j}) - f(v_{i})}$. The argument of the $\Theta$ function being positive guarantees that the node $v_j$ is visible from the node $v_i$ and vice versa. Since the edge in a BNVG has the weight $0$ or $1$, it is unsuitable for continuous filtering. To take into account the quality of visibility between nodes, we suggest the weighted version of the natural visibility graph (WNVG), by considering the weight function as follows,
    \begin{widetext}
        \begin{equation}
        w_{ij}^{(WN)} \equiv \left\{
        \begin{array}{l l}
        \Bigr{|}s_{ij} \Bigr{|} \qquad\qquad\qquad\qquad \qquad\qquad\qquad\qquad \quad ; \qquad \Bigr{|}f(v_{i})-f(v_{j}) \Bigr{|} = 1 \\
        \Bigr( \displaystyle \prod_{k=i+1}^{j-1} \Theta \Bigr{(} s_{ij} - s_{ik} \Bigr{)}
        \Bigr[ s_{ij} - s_{ik} \Bigr] \Bigr)^{1/(j-i-1)} \quad ; \qquad \Bigr{|}f(v_{i})-f(v_{j}) \Bigr{|} > 1\\
        \end{array} \right.
        \label{Eq:weights}
        \end{equation}
    \end{widetext}
    There are two factors inside the product. In the second branch of Eq. (\ref{Eq:weights}), one is the step function just like the binary graph, and the other is the weight which is proportional to "how visible is the site $j$ from $i$ and vice versa"; i.e., the more distinguishable the data points are in the original time-series, the higher the corresponding weight is in the constructed network. The term $\frac{1}{j-i-1}$ is necessary to make the weights reasonable numbers for comparison reasons. In the absence of this exponent, the more the distance between the nodes is, the higher the corresponding weights are. For both statistical and topological analysis, we use this weight function which admits continuous filtering.
    %%%%%%%%%%%%%%%%%%%%%%%%%%%%%%%%%%%%%%%%%%%%%%%%%%%%%%%%%%%%%%%%%%%%%%%%%%
    \begin{figure*}
        \begin{center}
            \includegraphics[width=0.48\textwidth]{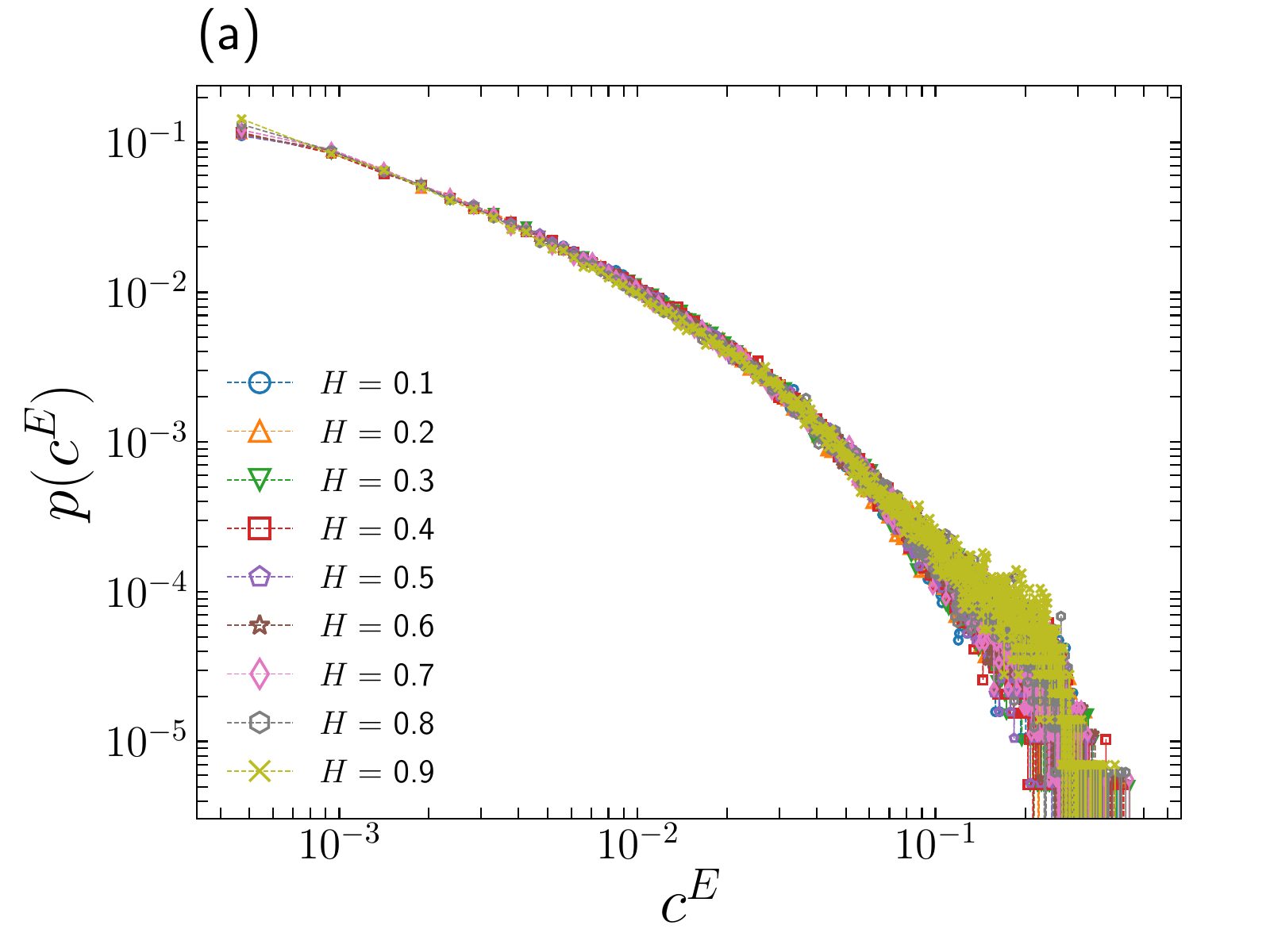}
            \includegraphics[width=0.48\textwidth]{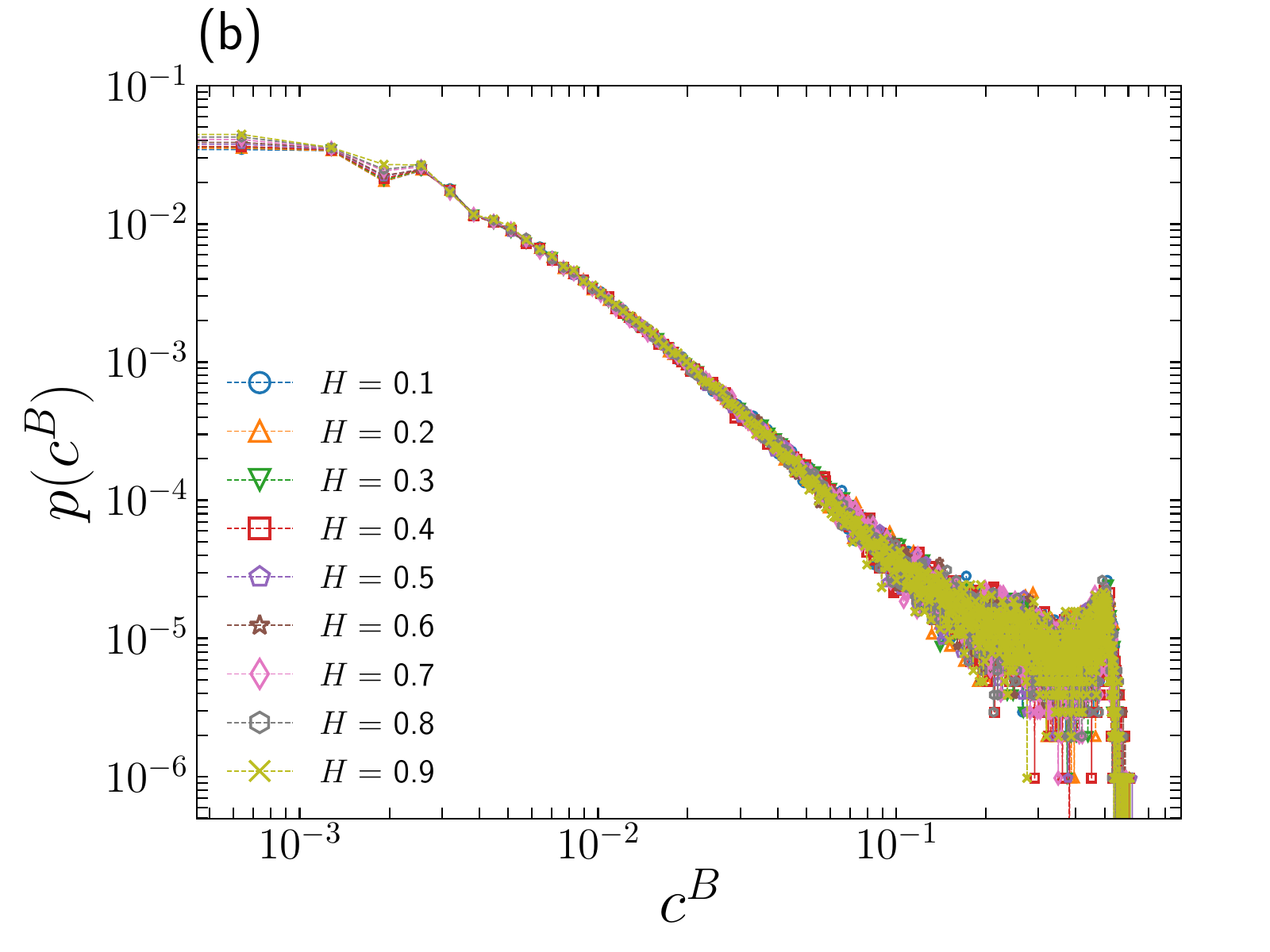}
            \includegraphics[width=0.48\textwidth]{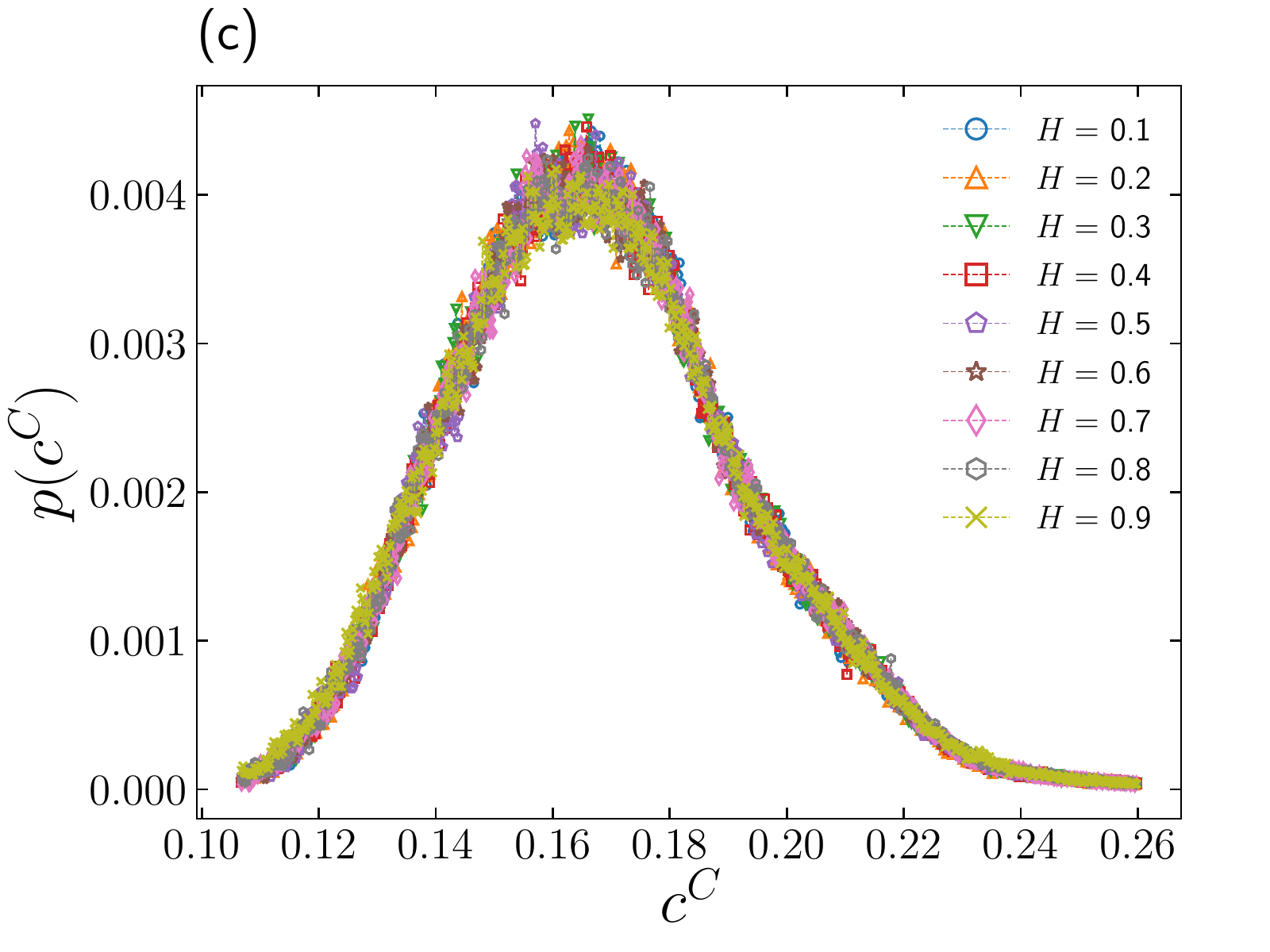}
            \includegraphics[width=0.48\textwidth]{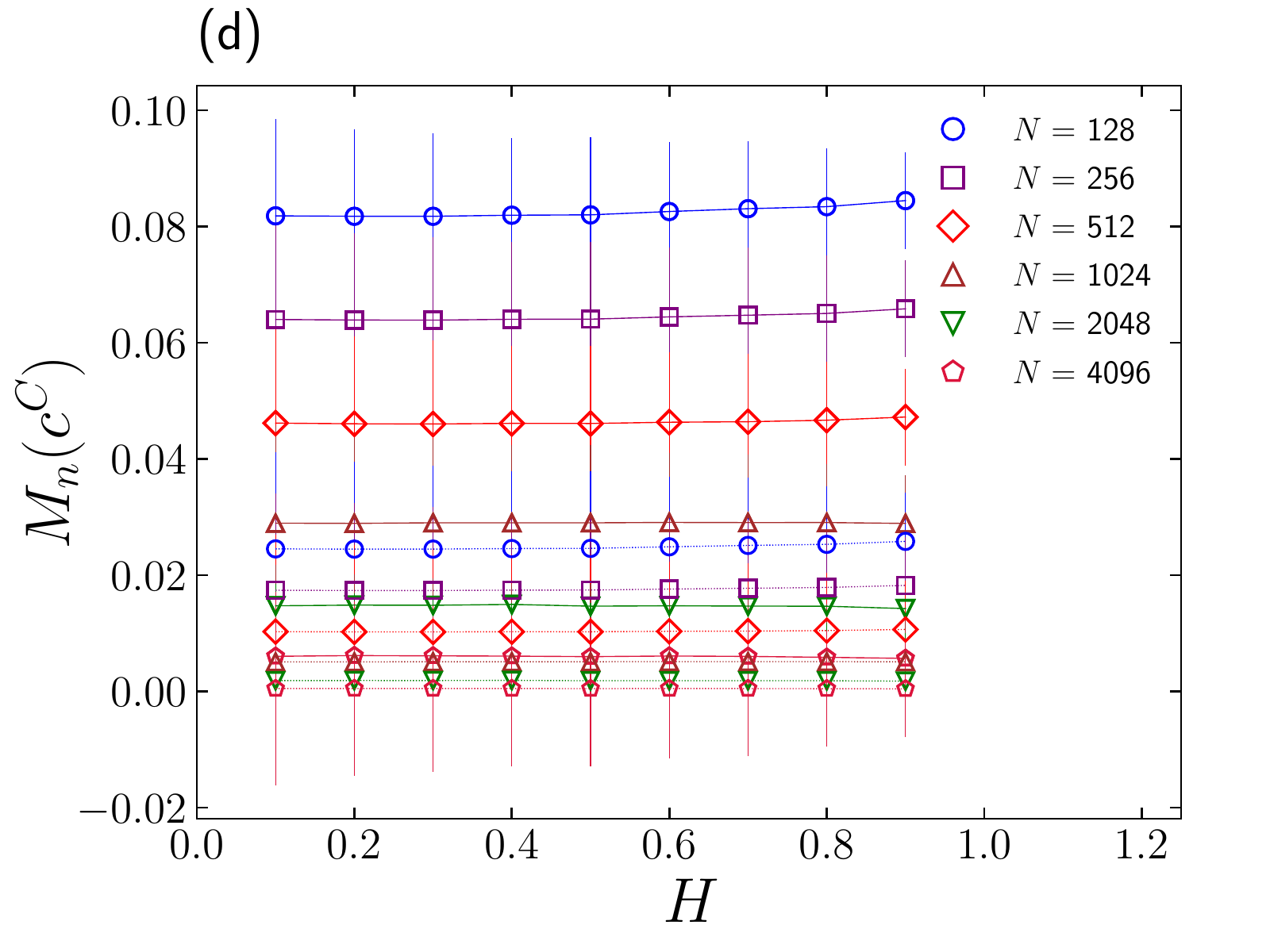}
            \caption{Probability distribution function of local features of  WNVGs constructed from fGns for  various Hurst exponents. Panel (a) is the $p(c^E)$ while panel (b) shows the probability distribution of betweenness centrality for $N=2^{10}$ in log-log scale. Panels (c) and (d) respectively illustrate the $p(c^{C})$ and corresponding moments, $M_n$, for $n=2$ (solid line) and $n=3$ (dashed line) versus $H$. Different symbols are taken for various sizes.  }
            \label{fig:localstat1}
        \end{center}
    \end{figure*}
    %%%%%%%%%%%%%%%%%%%%%%%%%%%%%%%%%%%%%%%%%%%%%%%%%%%%%%%%%%%%%%%%%%%%%%%%%%

    \subsection{Topological Network Analysis} \label{Sec:TDA}
    A conventional approach to quantify the properties of a typical network is determining degree, the number of nodes straightly connected with the underlying node by non-zero weight, and centrality measures such as betweenness centrality, the number of shortest paths between any pair of nodes going through a distinct node, closeness centrality, eigenvector centrality, \textit{etc.} (see Appendix B for more details). Besides the mentioned approach, other formalisms have been proposed in which not only the pairwise connections (links) are examined but also a systematic pipeline is considered to incorporate higher-order connections ($k$-cliques), including the simplicial complex and hypergraph~\cite{bianconi2016network,kovalenko2021growing,young2020hypergraph}.

    To analyze such higher-order structures, one can generalize the well-known statistical quantities based on higher-order connections. On the other hand, for global analysis of these structures newly born topological tools are proposed. By the term "global", we mean the essential features of the object which are not affected by geometric transformations. The well-known topological quantities describing global properties of the space (or any object mapped to a topological space) are Betti numbers. The $k$th Betti number of an object ($\beta_k$) is a topological invariant. The dimension of the $k$-homology group of the topological space corresponding to the object counts the number of $k$-dimensional topological holes ($k$-holes) of the object. Intuitively, $\beta_0$ indicates the number of connected components, $\beta_{1}$ is the number of topological (unshrinkable) loops, and $\beta_{2}$ counts the number of topological (unshrinkable) voids (see the Appendix C for more details). %For weighted complex networks,  it is reasonable to analyze it by considering the weights of links and to assess how these pairwise information change the structure of the network.
    According to the mentioned properties of the topological measures and weight-based analysis of weighted complex networks, the applied algebraic topology tool, known as the persistent homology (PH) technique, is useful to study global structures of the weighted complex network in higher-dimensions~\cite{Edels2010Comp,rote2006computational,zomorodian2009computational}.

    In the PH-based analysis of a weighted complex network, the weighted network is mapped to a weighted clique simplicial complex as a higher-order structure containing any $(k+1)$-clique of the network as the $k$-simplex. Therefore, PH creates a nested sequence of the clique simplicial complex, so-called filtration, by considering the weight as the parameter, such that any link (1-simplex) presents in the complex at a distinct weight if the weight of the link is less than the distinct weight. Accordingly, the $k$-holes of the complex appear and disappear when the weight (threshold) varies. To summarize the evolution of these $k$-dimensional topological features of the system, PH assigns an ordered tuple, the so-called persistence pair (PP), to this global descriptor to visualize the topological variation of the system in the birth-death space. This kind of visualization for the evolution of $k$-holes is called the persistence diagram (PD) for the $k$-homology group. PD for the $k$-homology group includes hidden topological information of the $k$-holes in the weighted complex. For example, one can compute the population of birth, death, and lifetime of $k$-holes using the distribution of PPs in PD for the $k$-homology group. Persistence entropy (PE) for the $k$-homology group is another quantity defined by the Shannon entropy of topological persistence (lifetime) of PPs of the $k$-homology group (see Appendix D for more details).

    \section{Results}\label{Sec:results}

    To evaluate the statistical and topological features for the constructed VG from a series, at first, we focus on the synthetic fractional Gaussian noise (fGn) as a self-similar series. There are some exact and approximate algorithms for generating a self-similar time-series. The Hosking \cite{hosking1984modeling}, Cholesky \cite{dieker2003spectral}, and Davies-Harte methods \cite{davies1987tests} are the examples of exact approach to generate fBm and fGn series. Throughout this paper, we implement the Davies-Harte method to generate fGn series with different self-similar exponents and sizes. To reduce any bias in the nominal Hurst exponent as much as possible, all relevant results are determined by doing an ensemble average over $10^4$ realizations generated by Davies-Harte method for each $H$. To ensure the correctness of the Hurst exponent value associated with each simulated time-series, we have computed again the value of $H$ by our code based on the detrended fluctuation analysis (DFA) method \cite{Kantelhardt} and compare the expected and computed Hurst exponents. Our results show that both Hurst exponents are in agreement with each other.

    Now  we turn to the statistical and topological properties of the VGs constructed from the fGn time-series and investigate their behavior concerning the Hurst exponent. The networks of sizes $N=2^7,2^8,2^9,2^{10},2^{11}$, and $2^{12}$ (for which a desktop with $128$ GB memory is capable of performing matrix operations) are considered. The Python toolbox "NetworkX"{\footnote{\texttt{https://networkx.org/}}} is employed for the matrix operations on the graphs. In the topological analysis, we especially focus on the Betti-$0$ (represented by the $\beta_0$ defined as the number of connected components of the network) and Betti-$1$ (represented by $\beta_1$ defined as the number of loops) features, which are extracted by using the "Dionysus" Python package \cite{dmitriydionysus}. The persistence statistics, containing the lifetime (the interval between birth and death) of the topological features, and its Shannon entropy are also analyzed.

    Each exponent has been estimated by Bayesian statistics accordingly; the
    $\{{\mathcal {D}}\}$ and $\{\Upsilon\}$ reveal the data and model free parameters, respectively. The posterior function is defined by
    \begin{equation}\label{posterior}
    {\mathcal P}(\Upsilon |{\mathcal {D}})=\frac{{\mathcal{L}}({\mathcal {D}}|\Upsilon){\mathcal P}(\Upsilon)}{\int
        {\mathcal{L}}({\mathcal {D}}|\Upsilon){\mathcal P}(\Upsilon)d\Upsilon}
    \end{equation}
    where $\mathcal{L}$ is the likelihood and $\mathcal{P}(\Upsilon)$ is the prior probability function containing all information concerning model parameters. Here we adopt the top-hat function for the prior function whose window's size depends on the expected range of the corresponding  exponent. Taking into account the central limit theorem, the functional form of likelihood becomes multivariate Gaussian, i.e., ${\mathcal{L}}({\mathcal {D}}|\Upsilon)\sim \exp(-\chi^2/2)$.
    The $\chi^2$ for determining the best-fit value for the scaling exponent reads as
    \begin{equation}
    \chi^2(\Upsilon)\equiv \Delta^{\dag}.{C}^{-1}.\Delta
    \end{equation}
    where $\Delta$ is a column vector whose elements are determined by difference between computed value and theoretical form for each measure, and ${C}$ is the corresponding covariance matrix.
    Finally, the best fit value of the considered exponent is computed by maximizing the likelihood probability distribution and the associated errorbar is given by
    \begin{equation}
    68.3\%=\int_{-\sigma_{\Upsilon}}^{+\sigma_{\Upsilon}}{\mathcal{L}}(\mathcal{D}|\Upsilon)d\Upsilon
    \end{equation}
    Subsequently, we report the best value of the scaling exponent at a $1\sigma$
    confidence interval as $\Upsilon_{-\sigma_{\Upsilon}}^{+\sigma_{\Upsilon}}$. \\

    \subsection{Local Statistical Properties}
    By local properties, we mean the properties which are node-dependent and are not necessarily globally defined.
    It has been confirmed that the distribution function of the node degree of VGs of the fBms and fGns is power-law $p(k) \propto k ^ {-\gamma}$ with the exponent $\gamma(H)=3-2H$ and $\gamma(H)=5-2H$ , respectively \cite{lacasa2008time,lacasa2009visibility}. In this subsection, we perform our computation for the WNVG, introduced in this paper.\\

    \begin{figure}
        \begin{center}
            \includegraphics[width=0.48\textwidth]{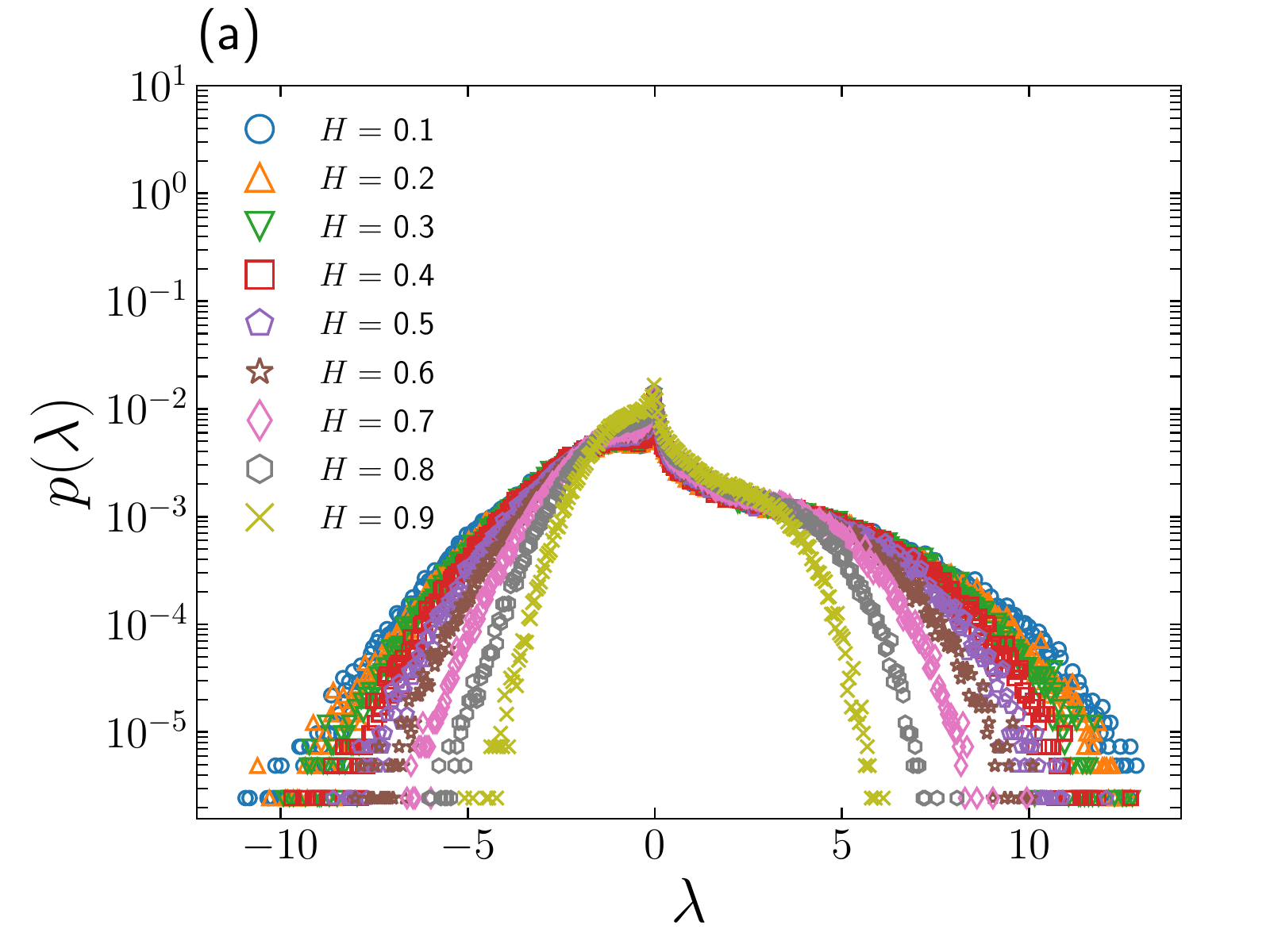}
            \includegraphics[width=0.48\textwidth]{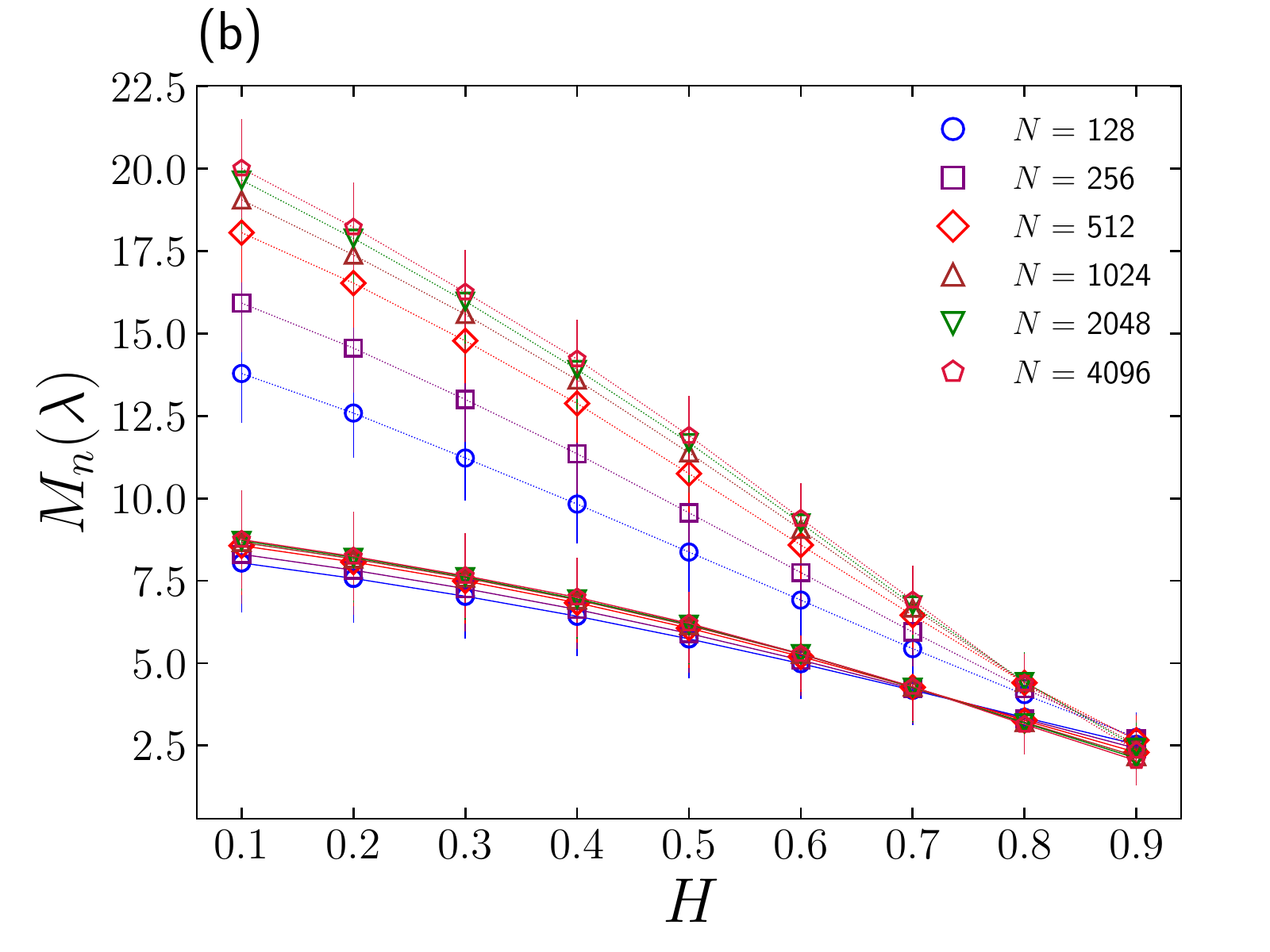}
            \caption{(a) The probability distribution function of eigenvalue for $N=2^{10}$ when the vertical axis is plotted in log-scale. (b) The corresponding moments, $M_n$, for $n=2$ (solid line) and $n=3$ (dashed line) versus $H$. Different symbols are taken for various sizes.}
            \label{fig:localstat2}
        \end{center}
    \end{figure}

    %%%%%%%%%%%%%%%%%%%%%%%%%%%%%%%%%%%%%%%%%%%%%%%%%%%%%%%%%%%%%%%%%%%%%%%%%%%%%%%%
    The probability distribution function of eigenvector ($c^{E}$) and betweenness ($c^{B}$) centralities are indicated in panels (a) and (b) of Fig.~\ref{fig:localstat1} in log-log scale, in terms of $H$ for WNVGs. The scaling behavior of mentioned distributions has been checked by the Kolmogorov-Smirnov test (KS test) and the results confirm that the power-law behavior of the computed $p(c^{E})$ and $p(c^{B})$ is highly $H$- and size-dependent [see panels (a) and (b) of Fig. \ref{fig:localstat1}]. The probability distribution of closeness centrality and corresponding moments $M_{n}$ for $n=2$ and $n=3$ are illustrated in panels (c) and (d) of Fig. \ref{fig:localstat1}, respectively. This part depicts that not only the Hurst dependency is not confirmed but also the overall shape of probability distribution functions of statistical measures depends on the size of the underlying fGn series. Subsequently, such distributions can not discriminate between different fGn series.

    The full spectrum of the eigenvalues [Eq.~(\ref{Eq:eigenvalueproblem})] is illustrated in panel (a) of Fig.~\ref{fig:localstat2}.
    We see that the impact of $H$ is changing the range of the spectrum, and by increasing the Hurst exponent, the range of the spectrum for WNVGs becomes tight. This phenomenon can be understood by recalling that correlations (obtained by increasing $H$) smooth the underlying time-series, causing the corresponding network to have more links at low weight. For more smoothed time-series, the typical slopes for the associated WNVG become down, leading to lower weights according to Eq.~(\ref{Eq:weights}), and equivalently making a shorter range for the distribution of $\lambda$s. Panel (b) of Fig.~\ref{fig:localstat2} indicates different moments for $p(\lambda)$. The solid and dashed lines correspond to $M_2$ and $M_3$, respectively. The $n$th moment of this distribution behaves as an $H$-dependent quantity. The higher the order of moments, the stronger the dependency on sample size is.

    \subsection{Topological Properties}
    %%%%%%%%%%%%%%%%%%%%%%%%%%%%%%%%%%%%%%%%%%%%%%%%
    \begin{figure*}
        \begin{center}
            %\centerline{\includegraphics[width=0.5\textwidth]{Betti0-curve.pdf}}
            %  \centerline{\includegraphics[width=0.5\textwidth]{pdf_PB0.pdf}}
            \includegraphics[width=0.48\textwidth]{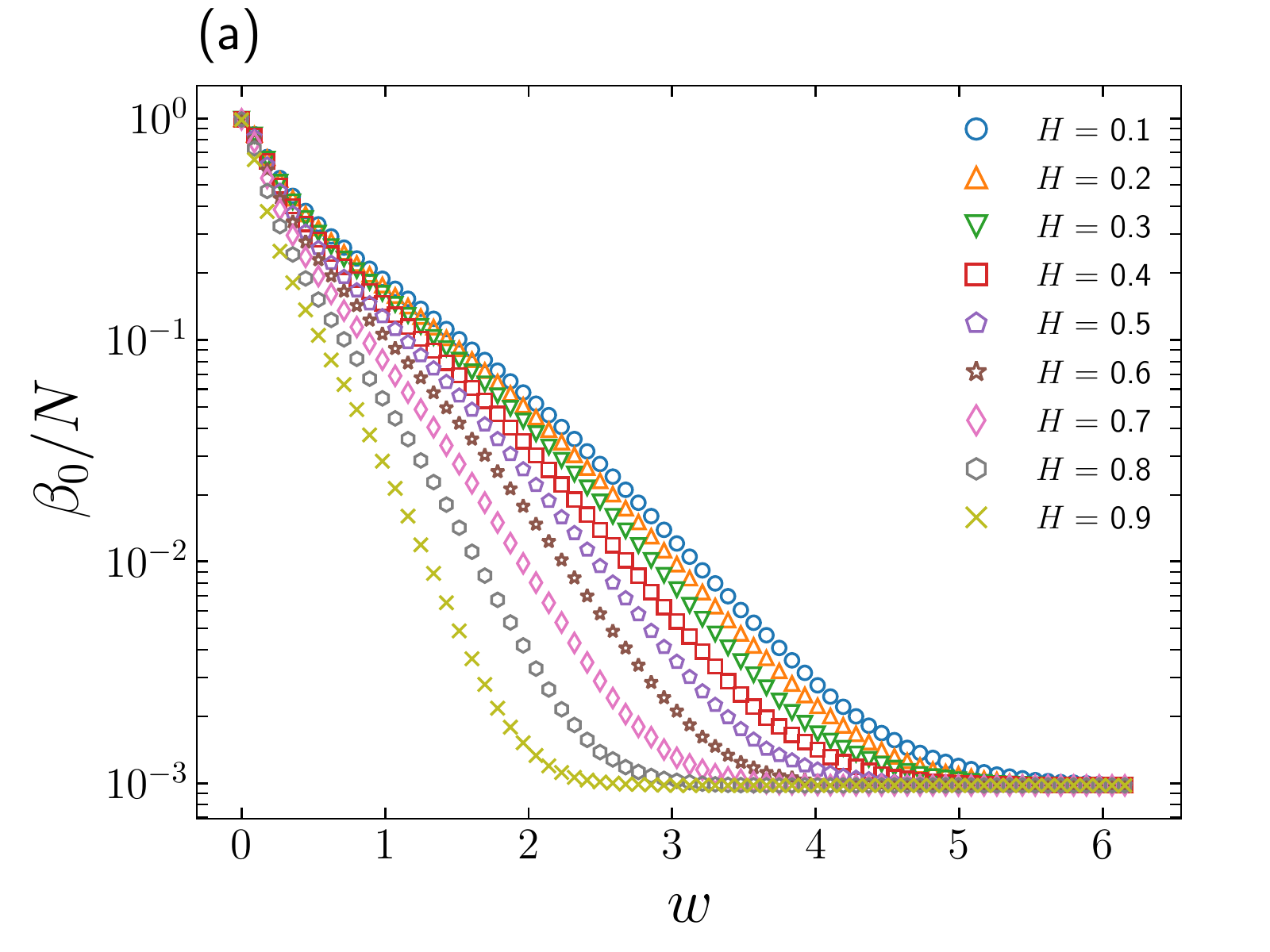}
            \includegraphics[width=0.48\textwidth]{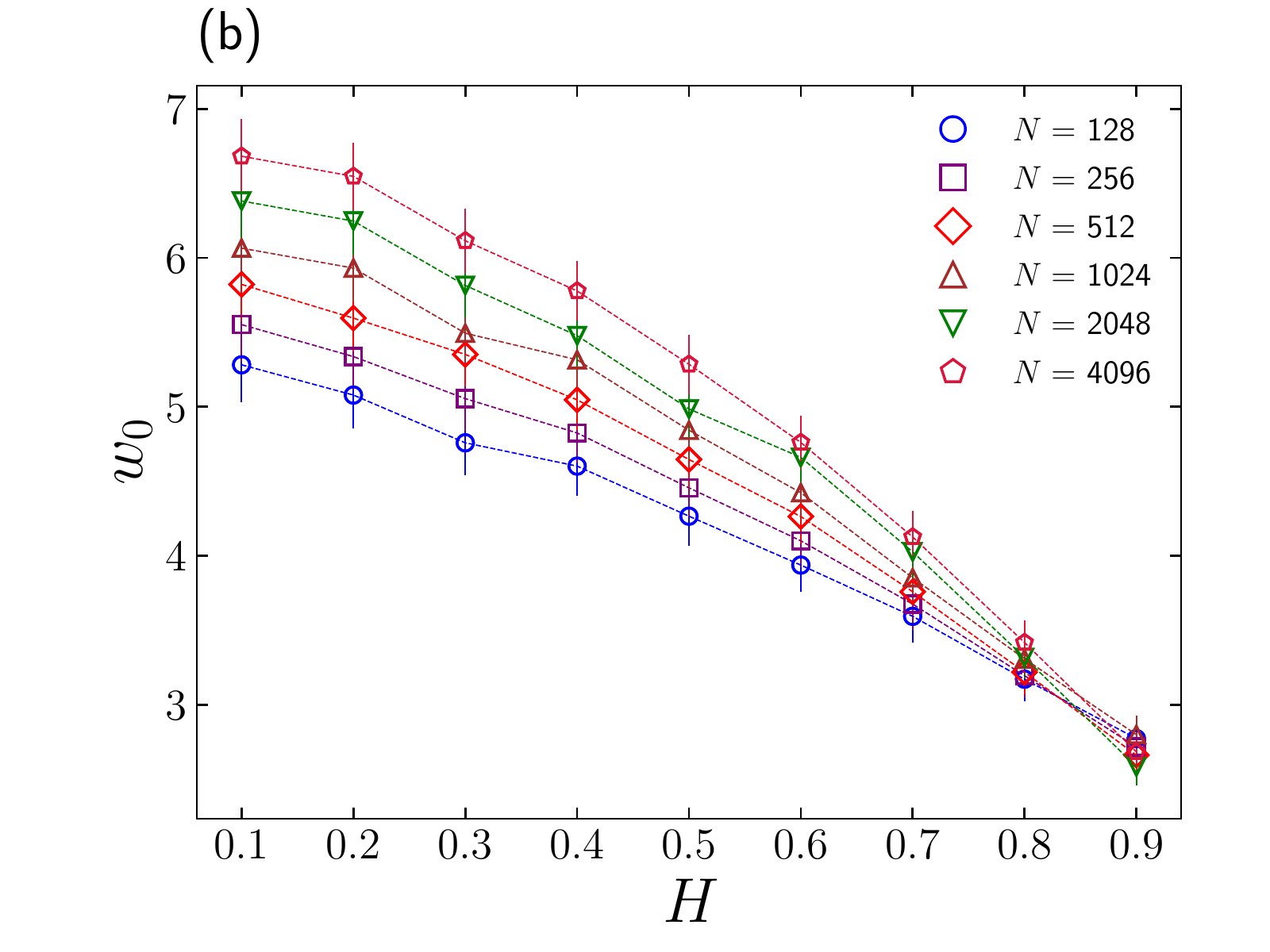}
            \includegraphics[width=0.48\textwidth]{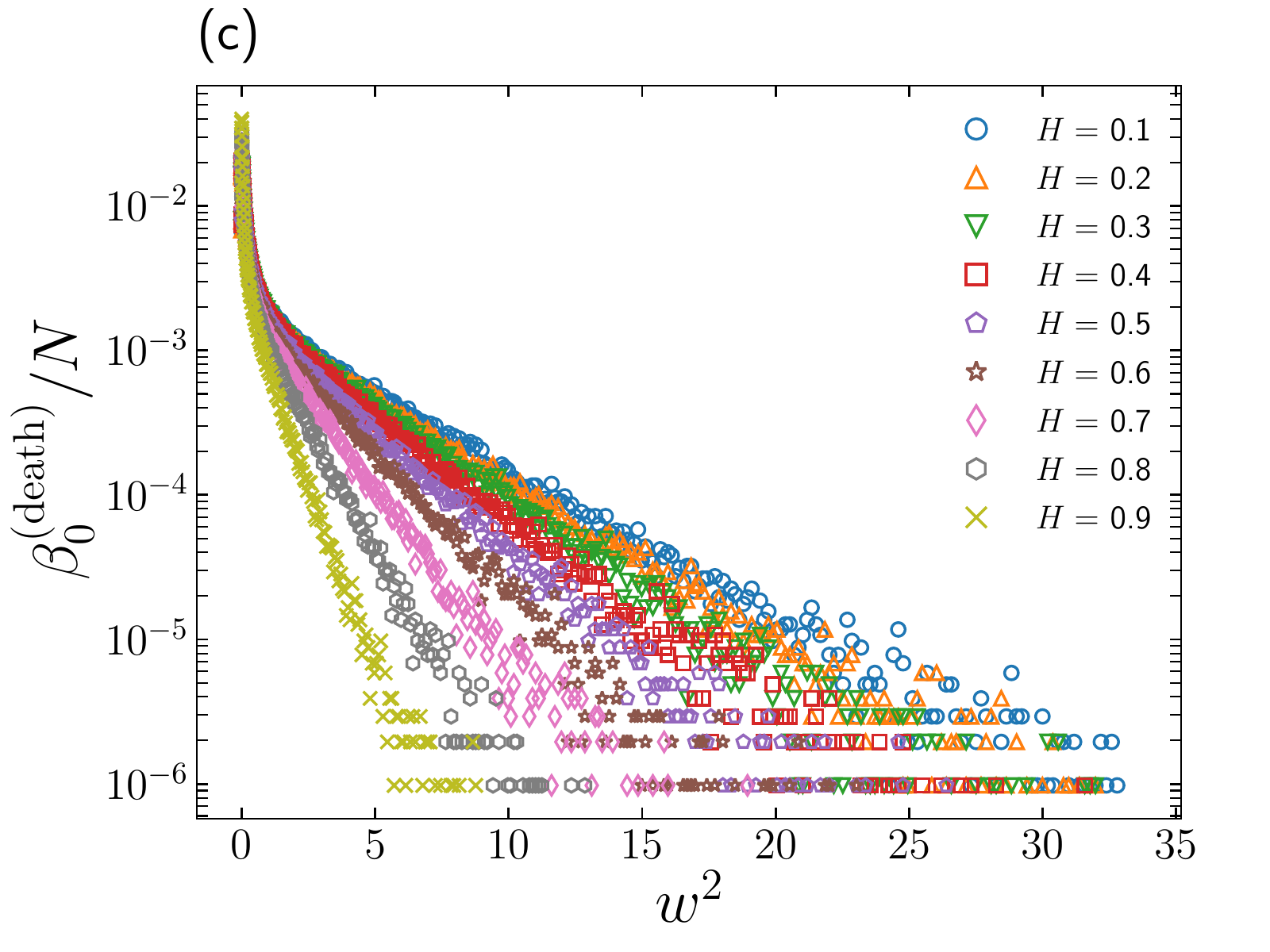}
            \includegraphics[width=0.48\textwidth]{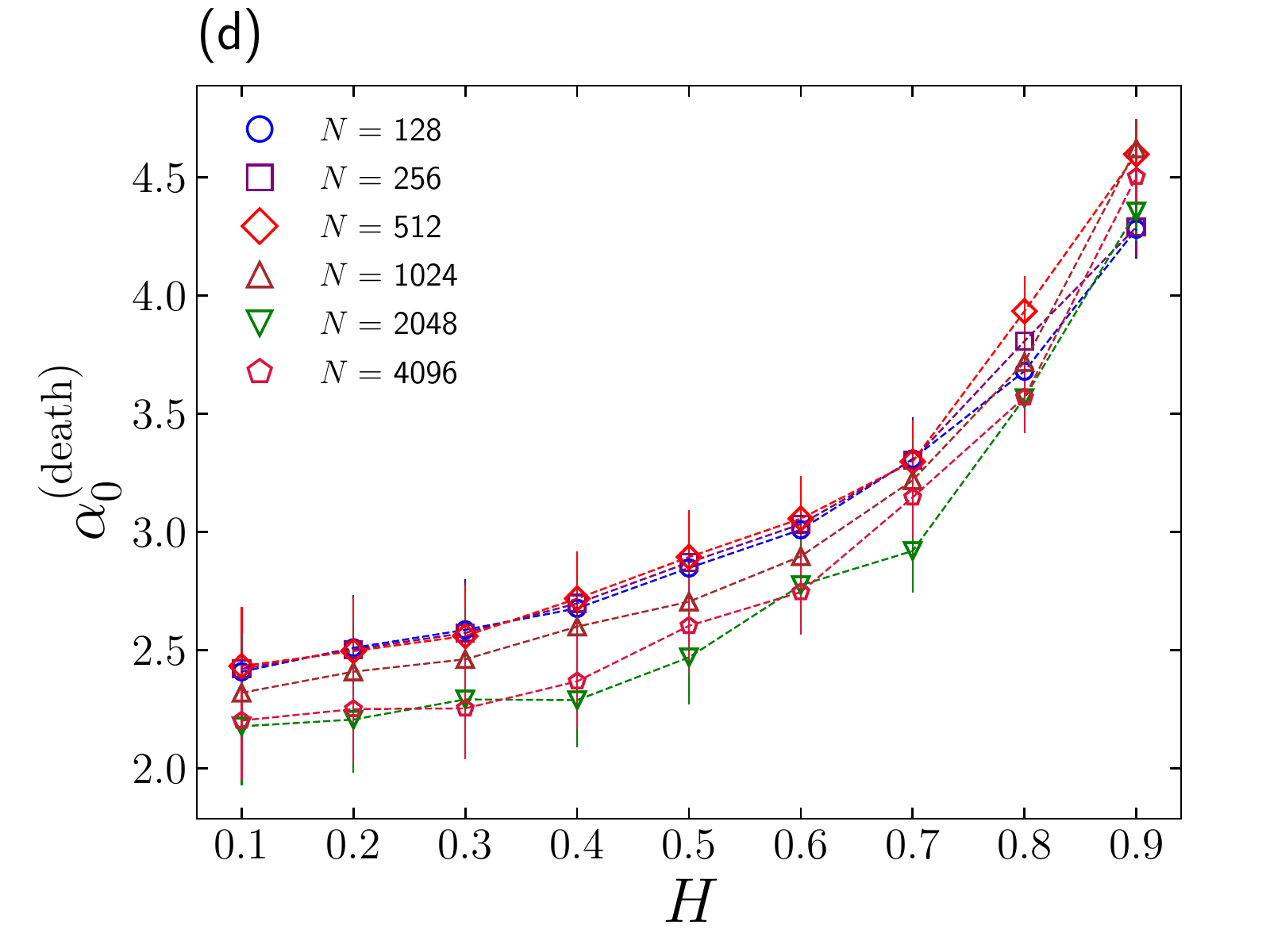}
            \caption{(a) The normalized $\beta_{0}$-curve in log-scale for clique complexes of WNVGs associated with fGns of various Hurst exponent versus threshold. (b) The $w_0$ as a function of $H$. (c) The normalized $\beta_{0}^{(\rm death)}$ in log-scale  as a function of $w^2$ (see the text) for various Hurst exponents as a function of threshold. (d) The value of $\alpha_0^{\rm (death)}$ as a function of Hurst exponent.}
            \label{fig:beta0}
        \end{center}
    \end{figure*}
    %%%%%%%%%%%%%%%%%%%%%%%%%%%%%%%%%%%%%%%%%%%%%%%%%%%%%%

    \begin{figure*}
        \includegraphics[width=0.48\textwidth]{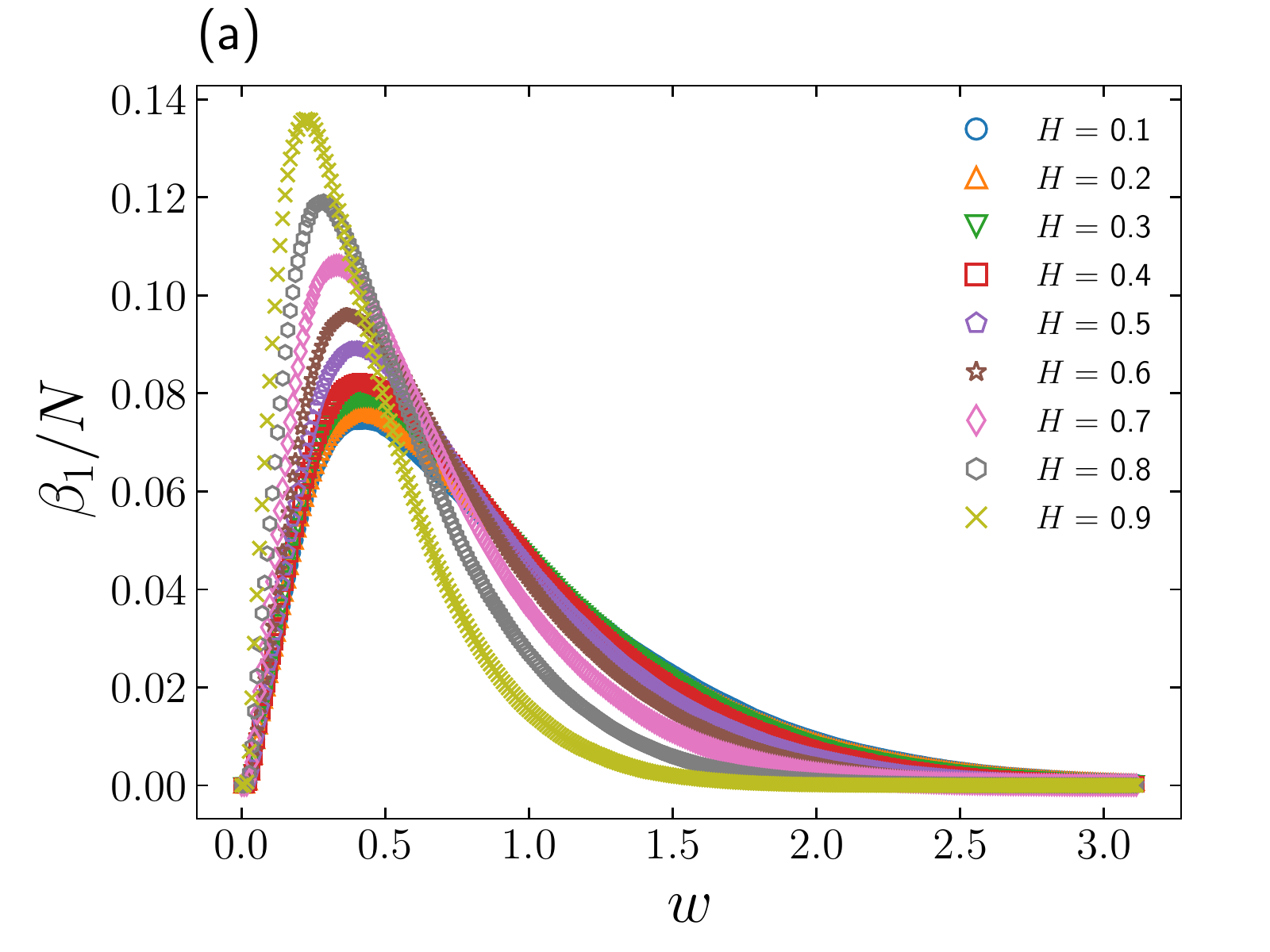}
        \includegraphics[width=0.48\textwidth]{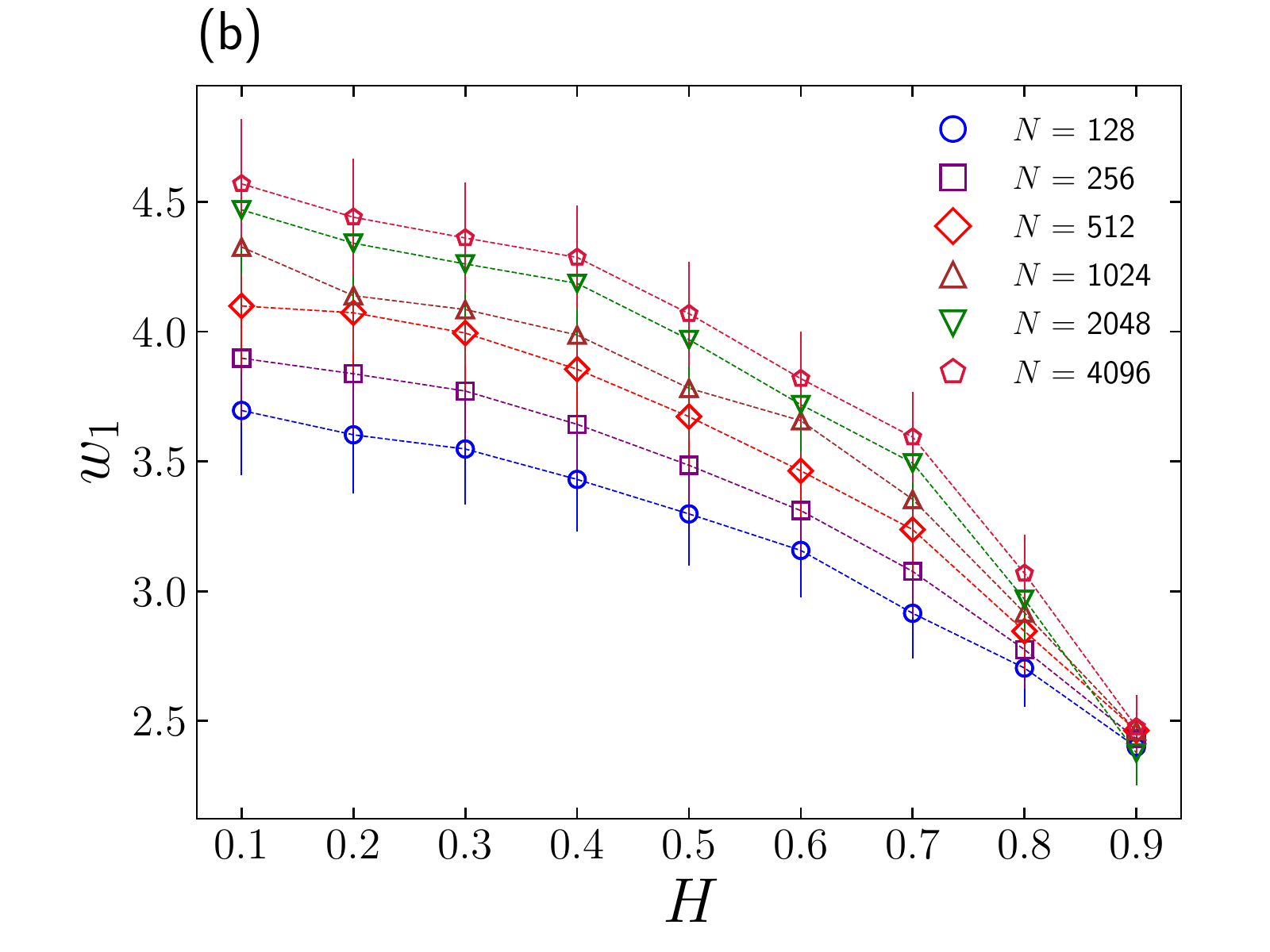}
        \includegraphics[width=0.48\textwidth]{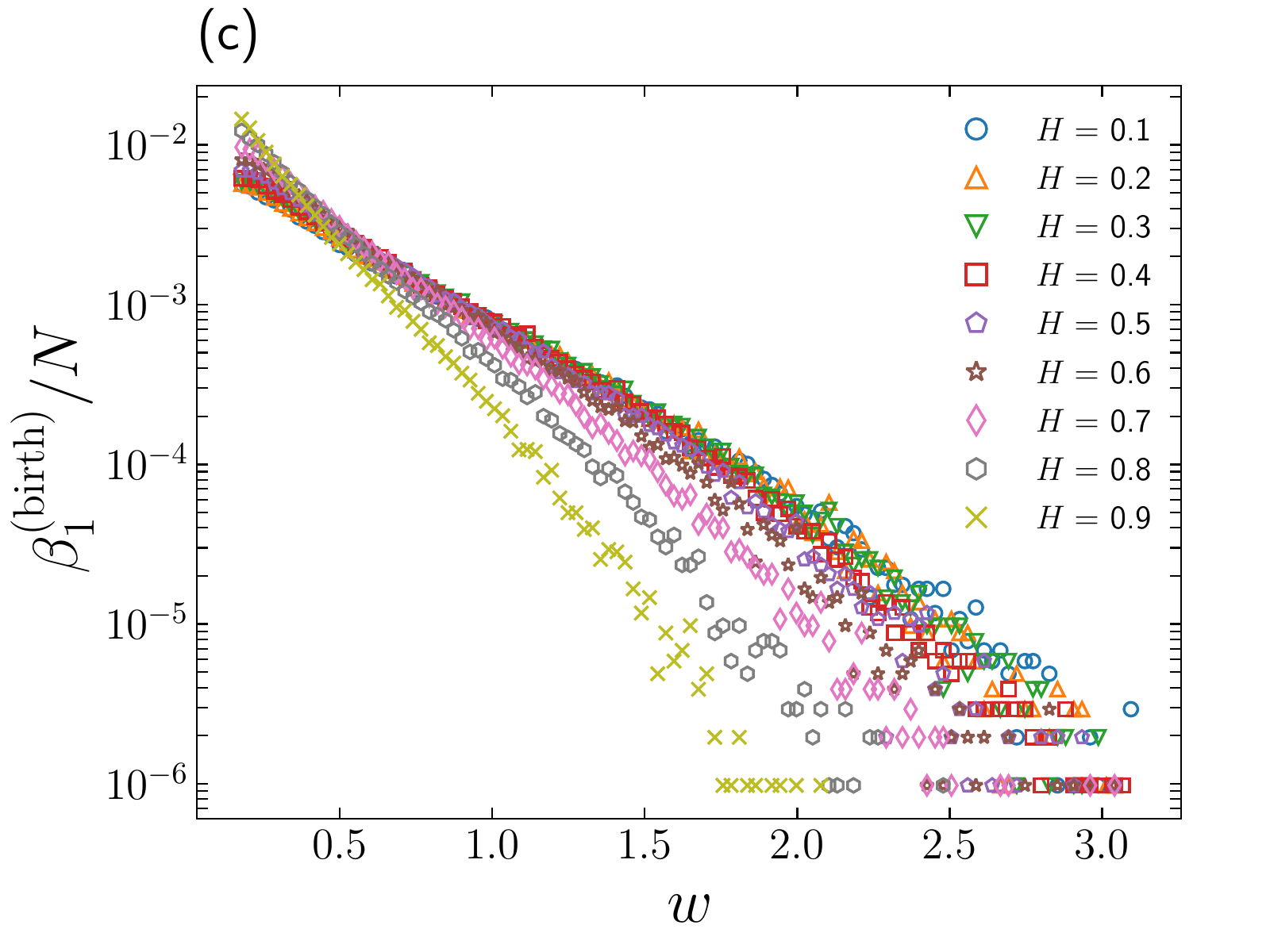}
        \includegraphics[width=0.48\textwidth]{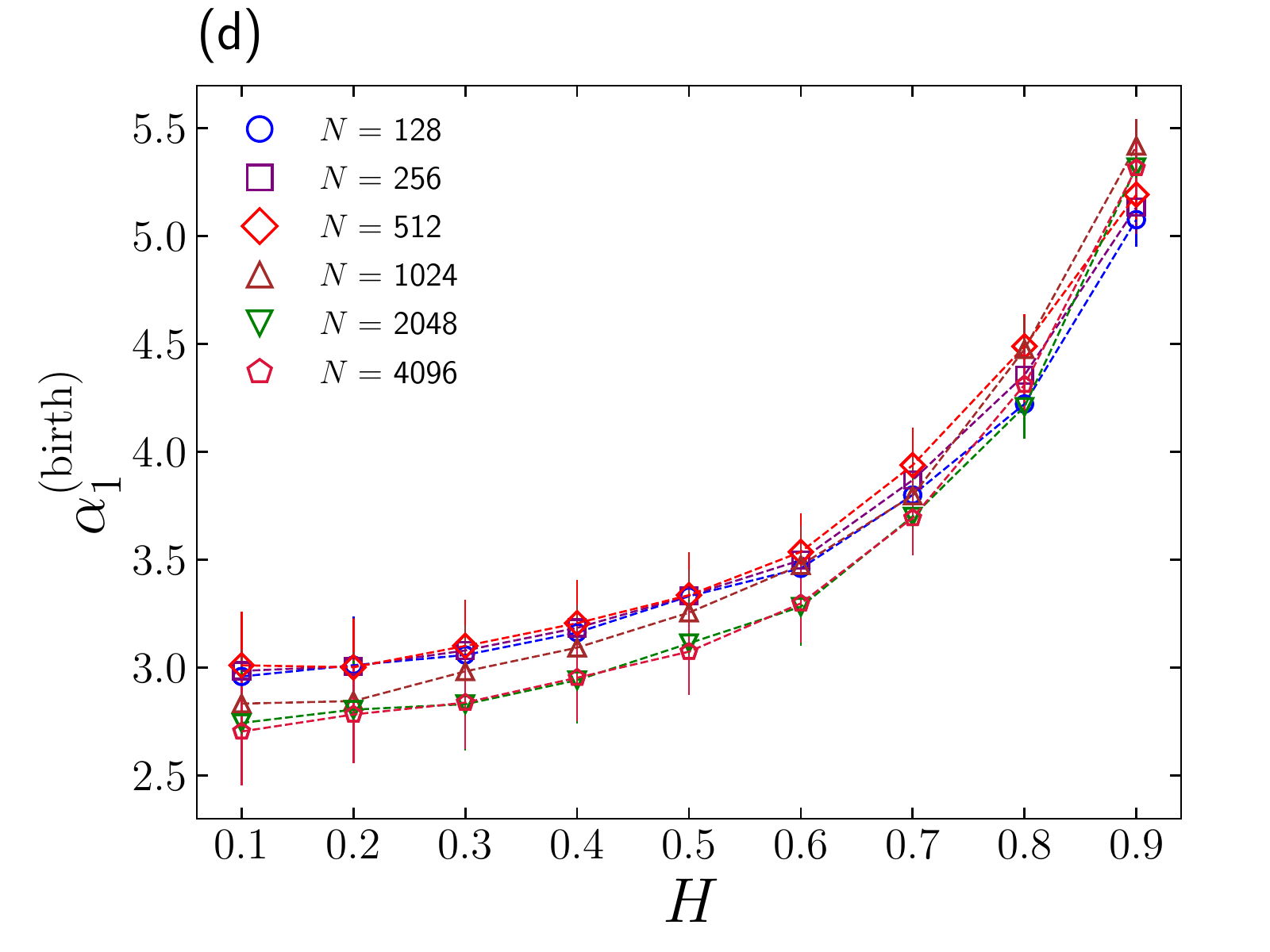}
        \includegraphics[width=0.48\textwidth]{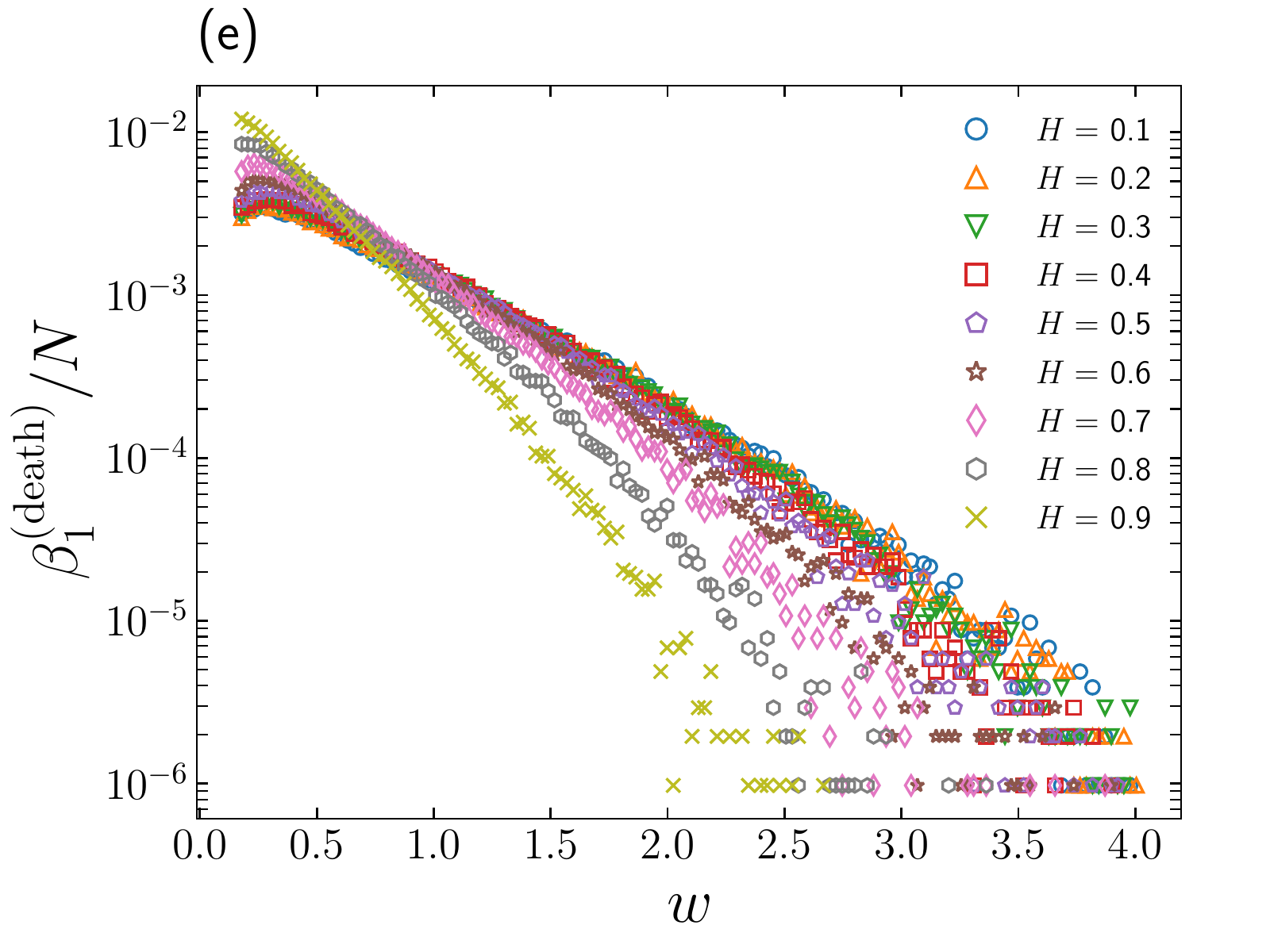}
        \includegraphics[width=0.48\textwidth]{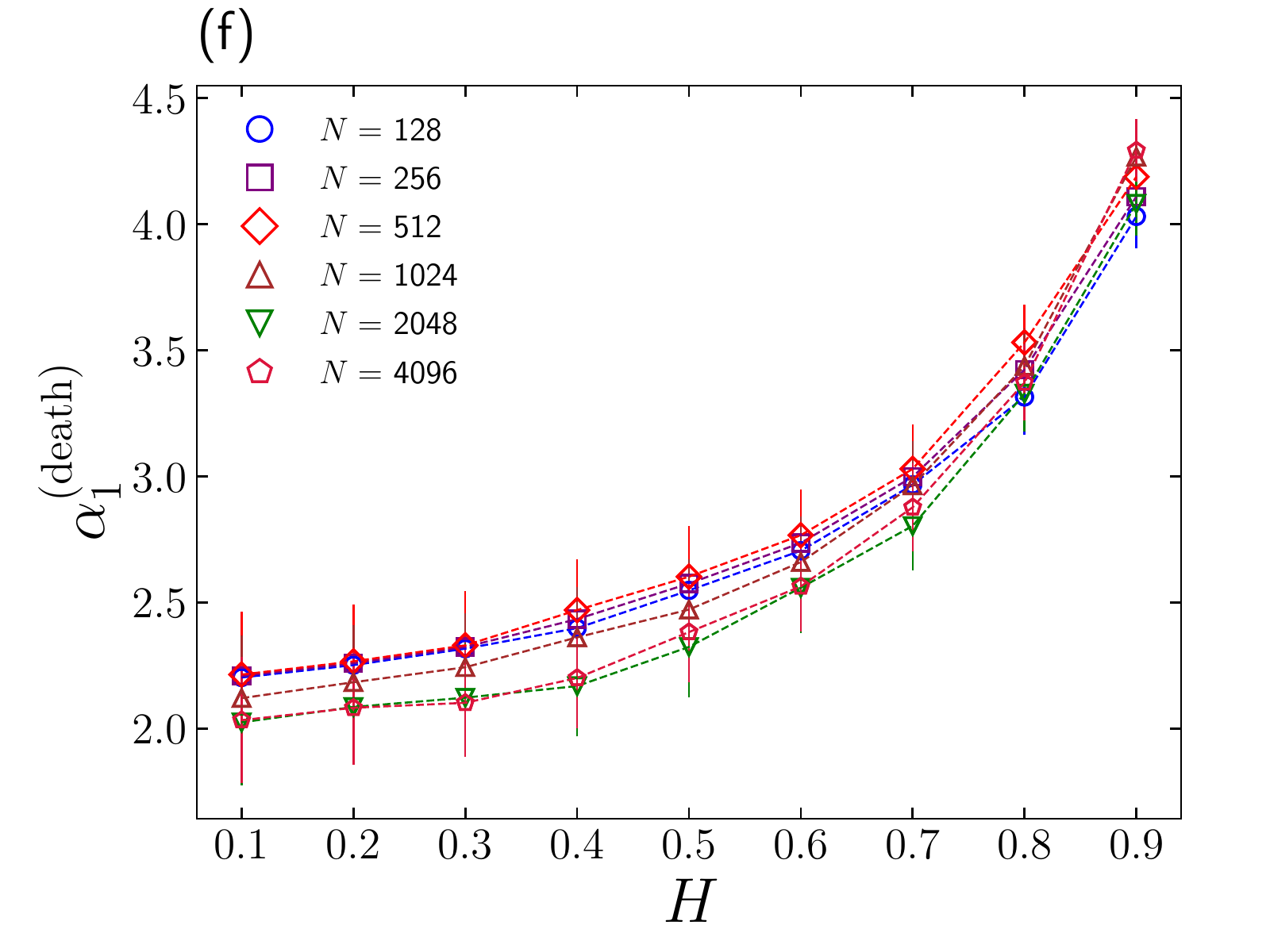}
        \caption{(a) The normalized $\beta_{1}$-curve for clique complexes of WNVGs associated with fGns of various Hurst exponents versus threshold. (b) The $w_1$ as a function of $H$ for different lengths of data set. (c) The distribution of the persistence diagram in log-scale versus the birth-axis as a function of threshold. (d) The corresponding coefficient of $\beta_1^{\rm (birth)}$ in terms of threshold known as $\alpha_1^{\rm (birth)}$ as a function of Hurst exponent. Panels (e) and (f) are the same as panels (c) and (d) respectively just for dying 1-hole statistics. In this plot for computing $\beta_1$, we took $N=2^{10}$.}
        \label{fig:Betti-curve}
    \end{figure*}
    %%%%%%%%%%%%%%%%%%%%%%%%%%%%%%%%%%%%%%
    \begin{figure*}
        \begin{center}
            \includegraphics[width=0.48\textwidth]{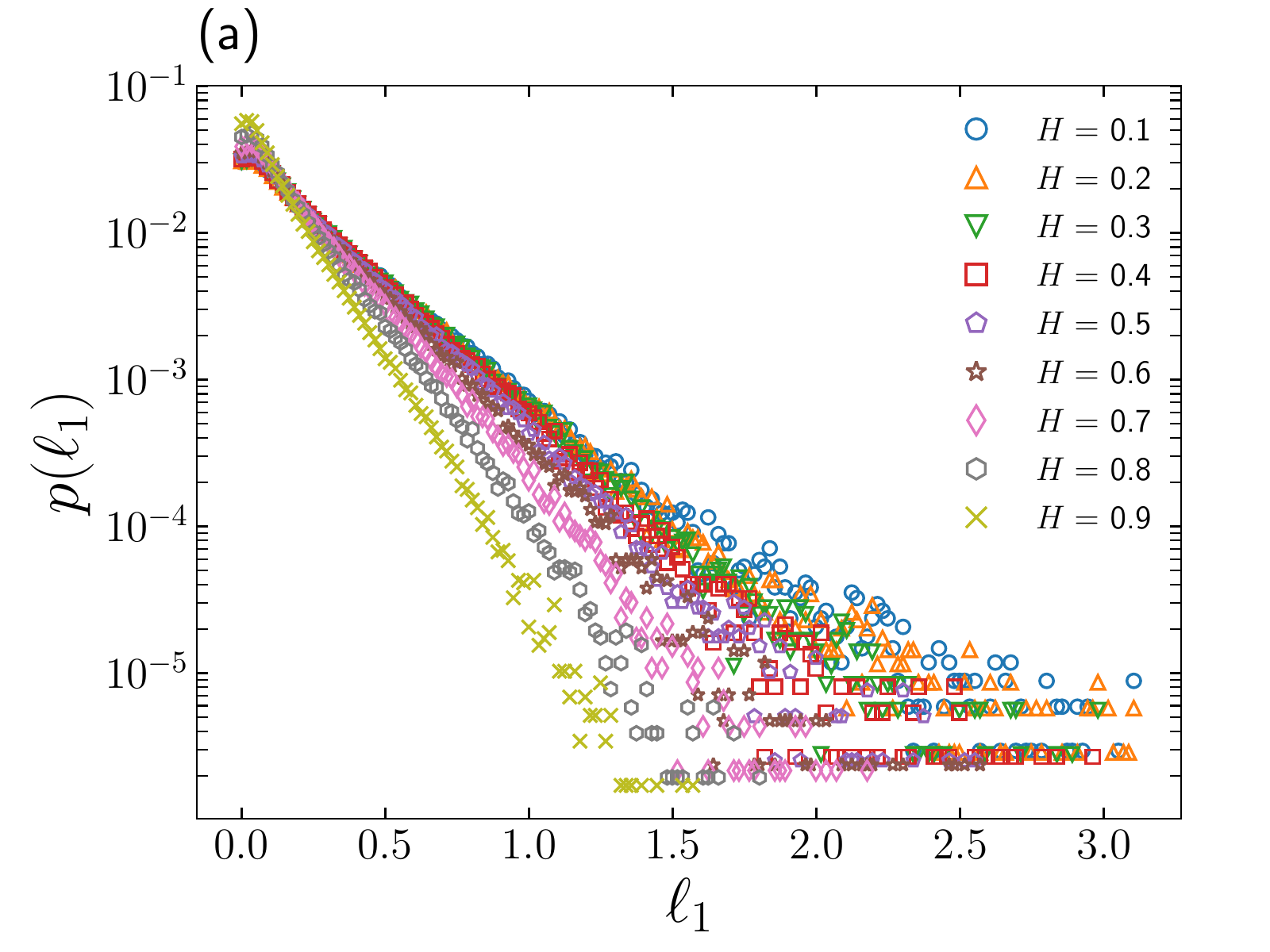}
            \includegraphics[width=0.48\textwidth]{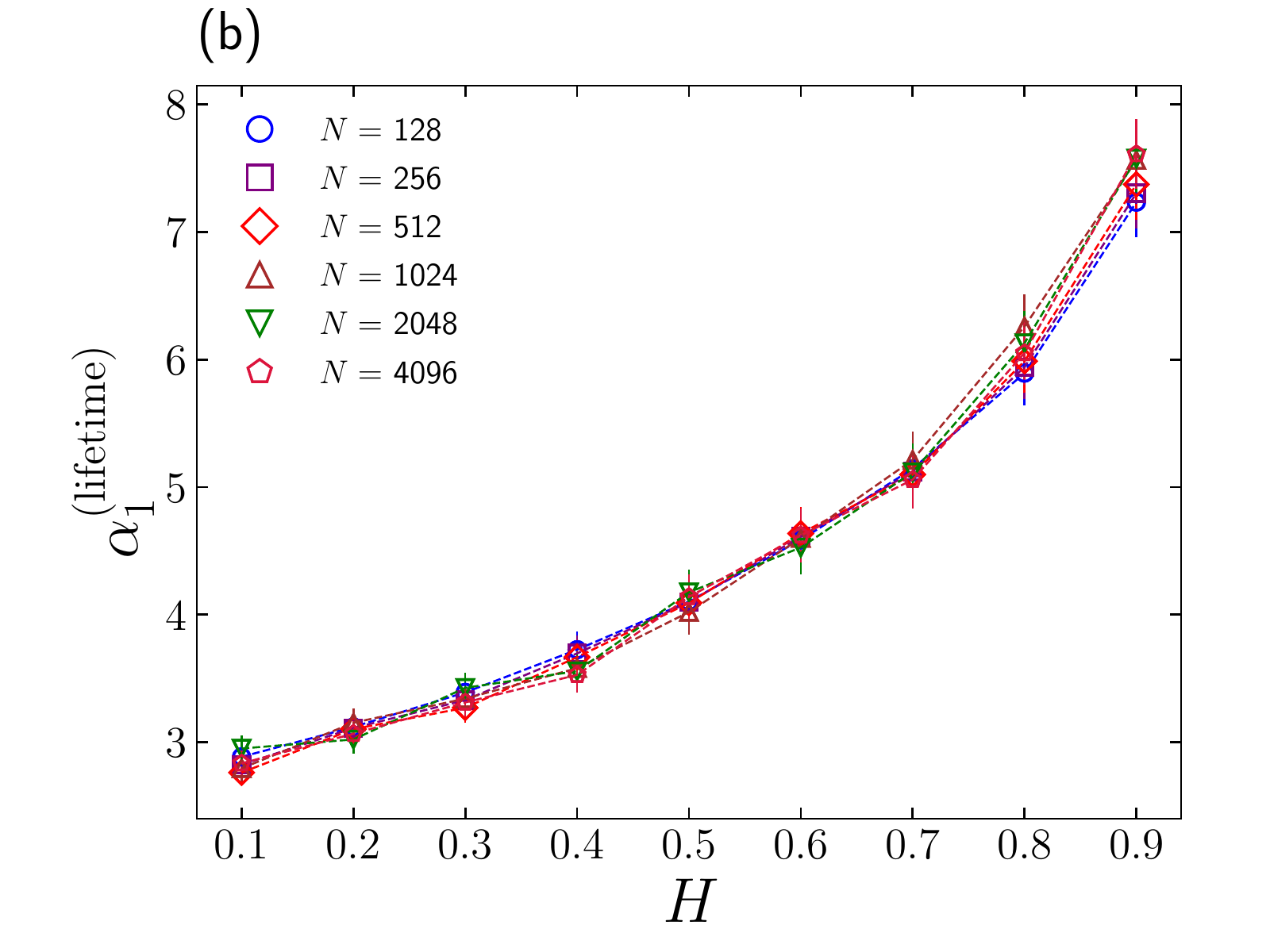}
            \caption{ (a) The probability distribution of lifetime for 1-dimensional holes. Different symbols correspond to various $H$. This diagram has been obtained by doing an ensemble average. We set the vertical axis in log-scale. (b) The $\alpha_1^{({\rm lifetime})}$ coefficient for different sizes represented by different symbols.}
            \label{fig:betti1}
        \end{center}
    \end{figure*}

    %In this subsection, we rely on the properties of topological measures to examine the structures embedded in the networks constructed by visibility graph mapping.
    For any given value of Hurst exponent, the associated BNVG of fGn is a \textit{topological tree}; therefore, the BNVGs are topologically equivalent to each other. In other words, various BNVGs are homeomorphic for different $H$ according to the homeomorphism theorem. The looplessness of BNVGs indicates that these networks do not contain some classes of network motifs corresponding to topological loops, so-called topological motifs \cite{xie2011horizontal}.
    To make more sense, suppose 4 successive data points as $\{x_j\}_{j=i}^{i+3} , (j=1, ... , N-4)$ from a given time-series $\{x_i\}_{i=1}^{N}$ to make a minimal topological loop. From a network perspective, the necessary (not sufficient) condition to construct a topological loop is that $w_{i,i+1}=w_{i,i+3}=w_{i+1,i+2}=w_{i+2,i+3}=1$ and the sufficient condition is $w_{i,i+2}=w_{i+1,i+3}=0$. The weights here are determined by Eq. (\ref{Eq:BWeights}). Geometrically, the necessary condition is equivalent to the case where data points have upward concavity. The visibility condition does not satisfy the sufficient condition, such that either $w_{i,i+2}=1$ (creating two 2-simplices $[v_{i},v_{i+1},v_{i+2}]$ and $[v_{i},v_{i+2},v_{i+3}]$) or $w_{i+1,i+3}=1$ (making two 2-simplices $[v_{i},v_{i+1},v_{i+3}]$ and $[v_{i},v_{i+2},v_{i+3}]$) or even $w_{i,i+2}=w_{i+1,i+3}=1$ (constructing a 3-simplex $[v_{i},v_{i+1},v_{i+2},v_{i+3}]$) killing the topological loop and leads the binary network to be topologically tree. Therefore, the BNVG does not contain a certain class of network motifs equivalent to topological holes, whereas, in the WNVG case these local patterns can be captured as topological loops. Also mentioned loops are extracted through the filtration of the corresponding weighted simplicial complex.
    According to the PH language, the sufficient condition can be treated as $w_{i,i+1},w_{i+1,i+2},w_{i+2,i+3},w_{i,i+3} < w_{i,i+2},w_{i+1,i+3}$; therefore, the associated topological motifs now remain. Subsequently, we only focus on the topological aspects of WNVGs.  Using this terminology, we can analyze local behaviors of a generic fGn using the statistics of corresponding patterns (topological motifs) in WNVGs which are presented as PPs in PD.

    Now, we are going to evaluate topological measures for WNVG constructed from a generic time series. Since in this paper, we are interested in examining the capability of the PH technique in determining the Hurst exponent of a typical self-similar process, therefore we take fGn which is an increment of fBm. Classifying the stationary and non-stationary series is in principle possible when the weighted network is constructed from time series in our approach. Intuitively, by increasing the value of the Hurst exponent, the underlying series becomes more smooth and the statistics of weights grow for lower values of weight. For an fBm data set which is the profile set of fGn \cite{taqqu1995estimators,Kantelhardt}, the maximum value of distribution function of weights occurs at lower weight values. Subsequently, the PH of WNVG which is represented by topological curves can recognize the stationary and non-stationary series. The constructed networks according to the visibility graph possess the footprint of Hurst exponent empirically demonstrated in \cite{lacasa2009visibility}. The higher value of $H$ causes making the denser network. Also, the number of $k$-simplices at lower value of threshold increase in the weighted version of the visibility graph; on the other hand, such simplices influence the evolution of $k$-holes in associated simplicial complexes. Therefore, the coefficients of topological curves depend on Hurst exponent. Hereafter, we consider fGn as our case study.
    Panel (a) of Fig.~\ref{fig:beta0} indicates the $\beta_0$ as a function of filtration parameter (threshold) for clique complexes of WNVG produced for various time-series with different values of Hurst exponent. To reduce the effect of sample size, we divide the Betti numbers by the sample size and call these {\it normalized Betti numbers}. By increasing the value of the Hurst exponent, the normalized $\beta_0$ versus threshold decreases. Namely, the number of connected components in the network decreases with increasing $H$ (the graphs become steeper).
    Interestingly, the slope of normalized $\beta_{0}$ as a function of $w$ is different for different $H$; consequently, the WNVG of fGn sets reaches to the connected regime (path-connected), $\beta_{0}=1$, [in panel (a) of Fig.~\ref{fig:beta0}, we take $N=2^{10}$] by different rates and at different thresholds, $w_0$. Panel (b) of Fig. \ref{fig:beta0} represents the value of $w_0$ as a function of Hurst exponent for different network sizes. As depicted in the mentioned panel, the value of $w_0$ behaves as a sample size-dependent quantity and by increasing $H$, such dependency becomes negligible.

    We define the $\beta_k^{\text{(birth)}}(w)$ and $\beta_k^{\text{(death)}}(w)$ as the number of $k$-dimensional holes that are born and die at a given threshold, $w$, respectively. It turns out that $\beta_0^{\text{(birth)}}=0$ for $w>0$, since at $w=0$ the underlying network has $N$ connected components; therefore, all connected components are born at $w=0$. Panel (c) of Fig.~\ref{fig:beta0} illustrates the $\beta_0^{\text{(death)}}/N$ as a function of $w^2$. As shown in this plot, one of the proper fitting functions to describe the normalized $\beta_0^{\text{(death)}}$ is given by $\beta_0^{\text{(death)}}(w)\sim \exp\left[ -\alpha_0^{\text{(death)}} w^2\right]$ for $2\lesssim w^2\lesssim w^2_{\rm max}$, where the value of $w_{\rm max}$ depends on the $H$ value. The $\alpha_0^{\rm (death)}$ depends on the Hurst exponent as increasing function and it behaves as an almost independent function on size of series [panel (d) of Fig.~\ref{fig:beta0}].
    %%%%%%%%%%%%%%%%%%%%%%%%%%%%%%%%%%%%%%%%%%%%%%%%%%%%%%%%%%%%%%%%%%

    Panel (a) of Fig. \ref{fig:Betti-curve} is devoted to $\beta_1/N$ for various Hurst exponents. As we expect, for a trivial threshold, $w=0$, we have $N$ connected components ($\beta_0=N$) and therefore the number of loops is identically zero ($\beta_1=0$). By increasing the threshold, the higher value of the Hurst exponent leads to a more rapidly increasing rate of $\beta_1$. On the other hand, for a high enough value of the threshold, again the underlying data set behaves like a topological tree without any topological loops. Therefore, the normalized $\beta_1$ goes asymptotically to zero, and such descending is more rapid for higher $H$. We also determine the lowest non-trivial threshold for which there is no loop in the underlying network denoted by $w_1$, and depict this threshold versus $H$ for different samples size in panel (b) of Fig. \ref{fig:Betti-curve}. The $w_1$ is also a size-dependent quantity. Comparing the $\beta_0/N$ and $\beta_1/N$ demonstrate that by increasing threshold value, the WNVGs of the fGn series reach the loopless regime ($\beta_{1}=0$) before appearing in the connected regime (for which $\beta_{0}=1$), irrespective of the Hurst exponent, i.e., $w_{0}(H) > w_{1}(H)$. The quantities $\beta_1^{\text{(birth)}}\propto \exp\left[-\alpha_1^{\text{(birth)}}w \right] $, $\beta_1^{\text{(death)}}\propto \exp\left[-\alpha_1^{\text{(death)}}w \right]$ versus thresholds and corresponding coefficients are illustrated in panels (c)-(f) of Fig. \ref{fig:Betti-curve}, respectively. The $\alpha_{1}^{\rm (birth)}$ and $\alpha_{1}^{\rm (death)}$ are almost size-independent and they grow by increasing $H$.  The local upward concavity behavior of time series satisfying the loop condition and associated topological motifs (1-holes) appears and disappears earlier as the Hurst exponent of the time-series increases.

    Another interesting property to assess is the probability distribution of lifetime for 1-holes which is the difference between the death and birth thresholds of a typical measure in a 1-homology class. Panel (a) of Fig. \ref{fig:betti1} shows the probability distribution of topological 1-dimensional hole lifetime, $\ell_1$, for various synthetic data sets with different values for the Hurst exponent when the vertical axis is plotted in log-scale. Our results confirm that $p(\ell_1)\propto \exp\left[-\alpha_1^{\text{(lifetime)}}\ell_1 \right] $. The $H$-dependency of $\alpha_1^{\rm (lifetime)}$ for various systems sizes is depicted in panel (b) of Fig.~\ref{fig:betti1}. This result confirms that $\alpha_1^{\rm (lifetime)}$ can be considered as a robust measure for determining the Hurst exponent of the fGn signal which is not affected by sample size even compared to $\alpha_0^{\rm (death)}$, $\alpha_1^{\rm (birth)}$, and $\alpha_1^{\rm (death)}$.  The increasing behavior of the lifetime coefficient $\alpha_1^{\rm (lifetime)}$ versus the Hurst exponent also indicates that the anti-correlated signals ($H<0.5$) contain more persistent topological motifs (1-holes). This behavior also demonstrates a considerable difference between the amount of visibility between the inner and the outer nodes which are corresponding to the local upward concavity behaving data points in these signals.

    \begin{figure*}
        \begin{center}
            \includegraphics[width=0.3\textwidth]{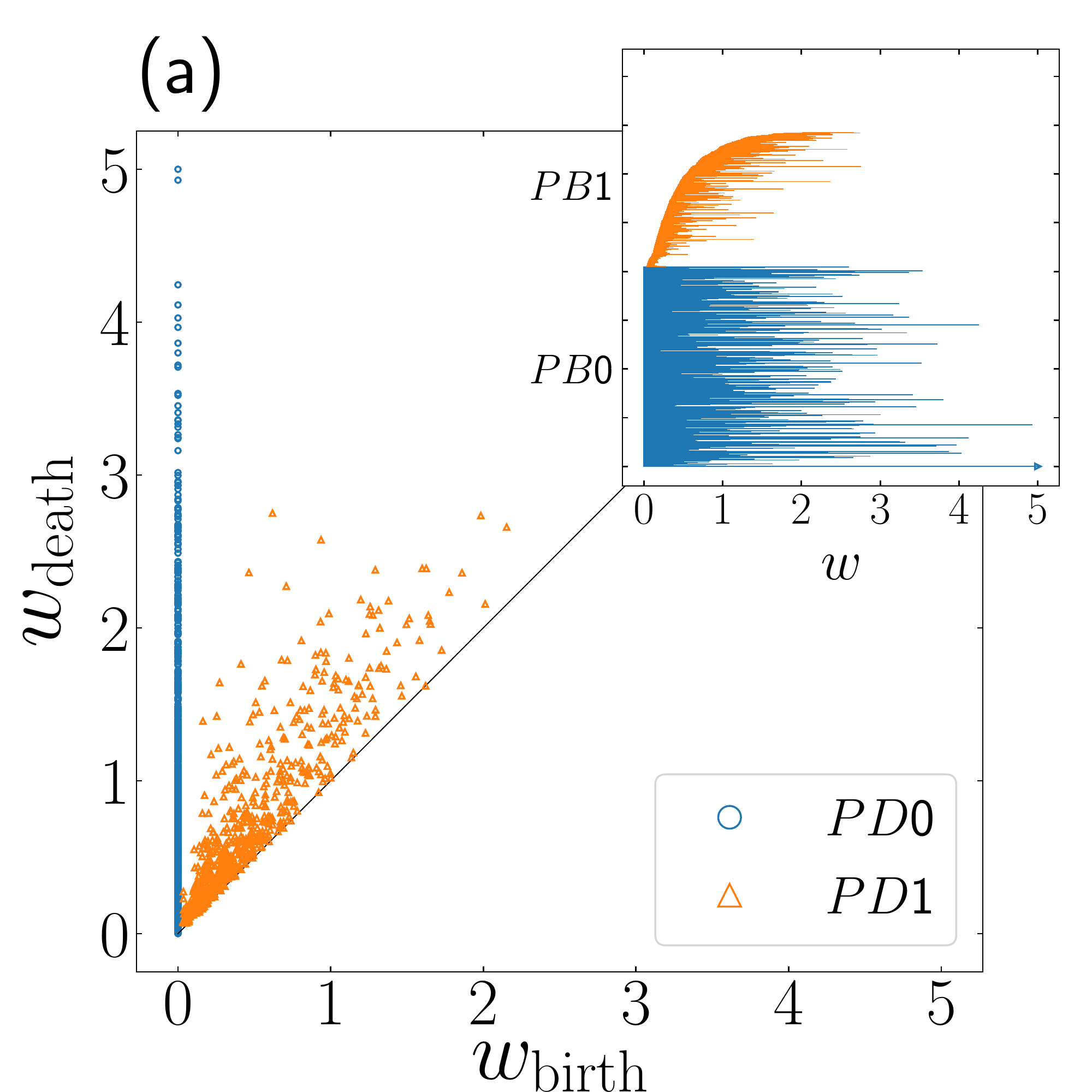}
            \includegraphics[width=0.3\textwidth]{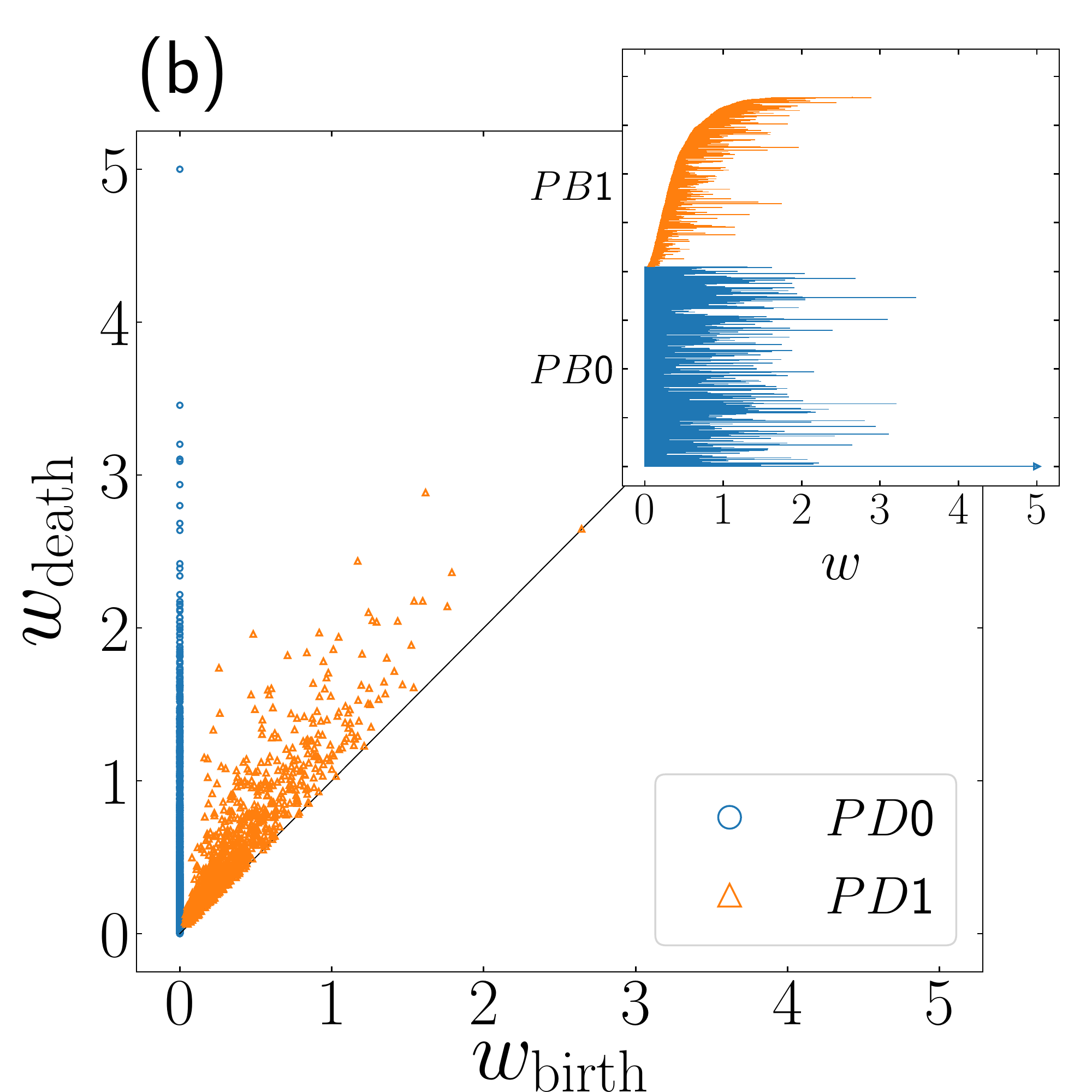}
            \includegraphics[width=0.3\textwidth]{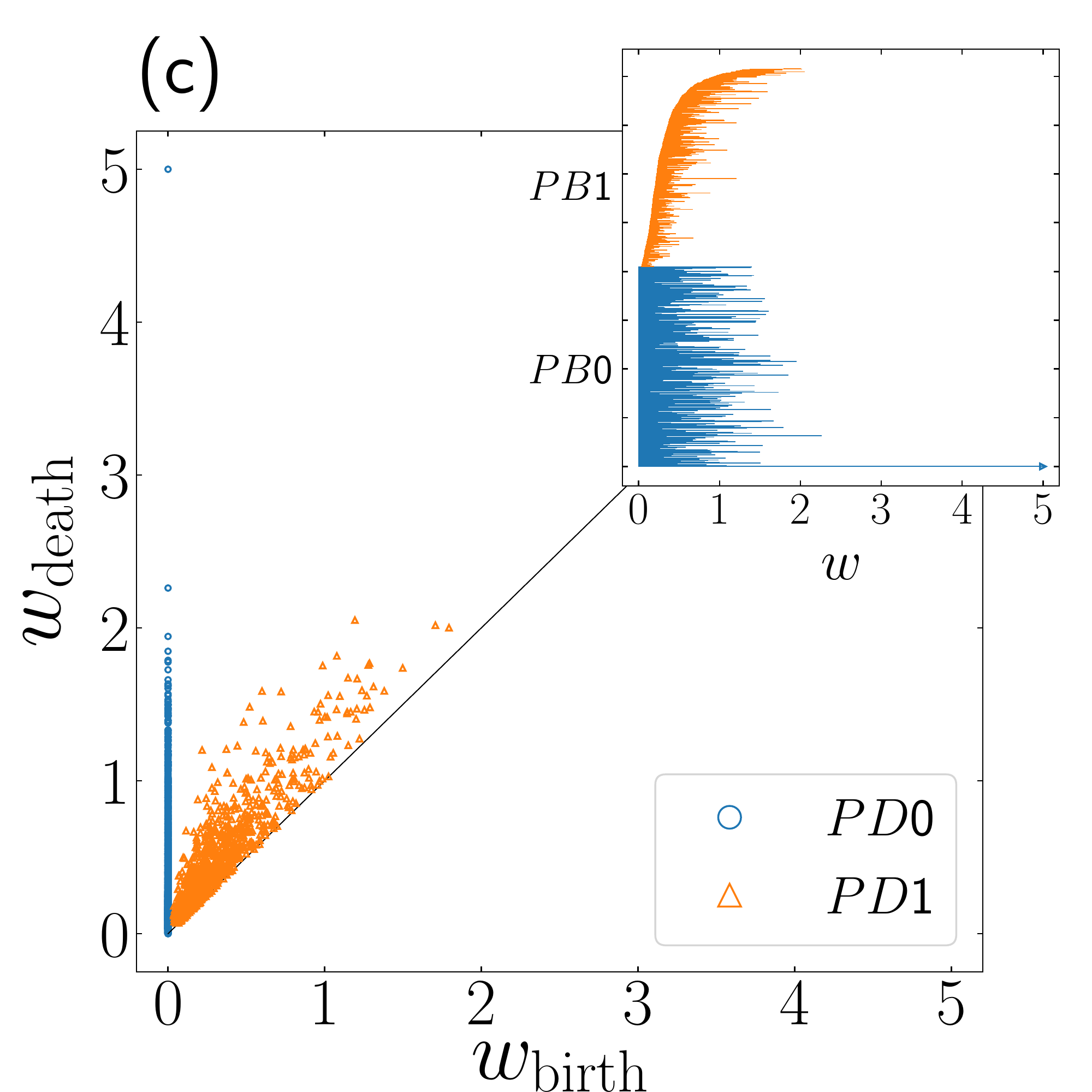}
            \caption{The persistence diagram and persistence barcode (inset) of weighted clique complex of WNVG corresponding to fGn with  (a) $H=0.2$ (anticorrelated noise), (b) $H=0.5$ (uncorrelated noise) and (c) $H=0.8$ (correlated noise).  The blue circle and orange triangle symbols are considered for  the  0-dimensional  and 1-dimensional  homology generators, respectively.}
            \label{fig:PB+PD}
        \end{center}
    \end{figure*}

    Figure~\ref{fig:PB+PD} indicates the persistence diagram (PD) and persistence barcode (PB) for three types of fGn signals for (a) $H=0.2$ (anticorrelated), (b) $H=0.5$ (uncorrelated) and (c) $H=0.8$ (correlated). The open circle and triangle symbols correspond, respectively to the $0$- and $1$-homology groups in a persistence diagram for WNVGs of fGn. Each symbol depicts a pair $(w_{\rm birth},w_{\rm death})$ of a $k$-dimensional hole. As expected, all 0-holes are born in $w_{\rm birth}=0$, and also always $w_{\rm death}>w_{\rm birth}$. In the barcode representation (inset plot), the horizontal lines (blue lines for $k=0$ and orange lines for $k=1$) start and end on the threshold values on which $k$-dimensional holes are born and die, respectively.

\begin{figure}
        \begin{center}
            \includegraphics[width=0.48\textwidth]{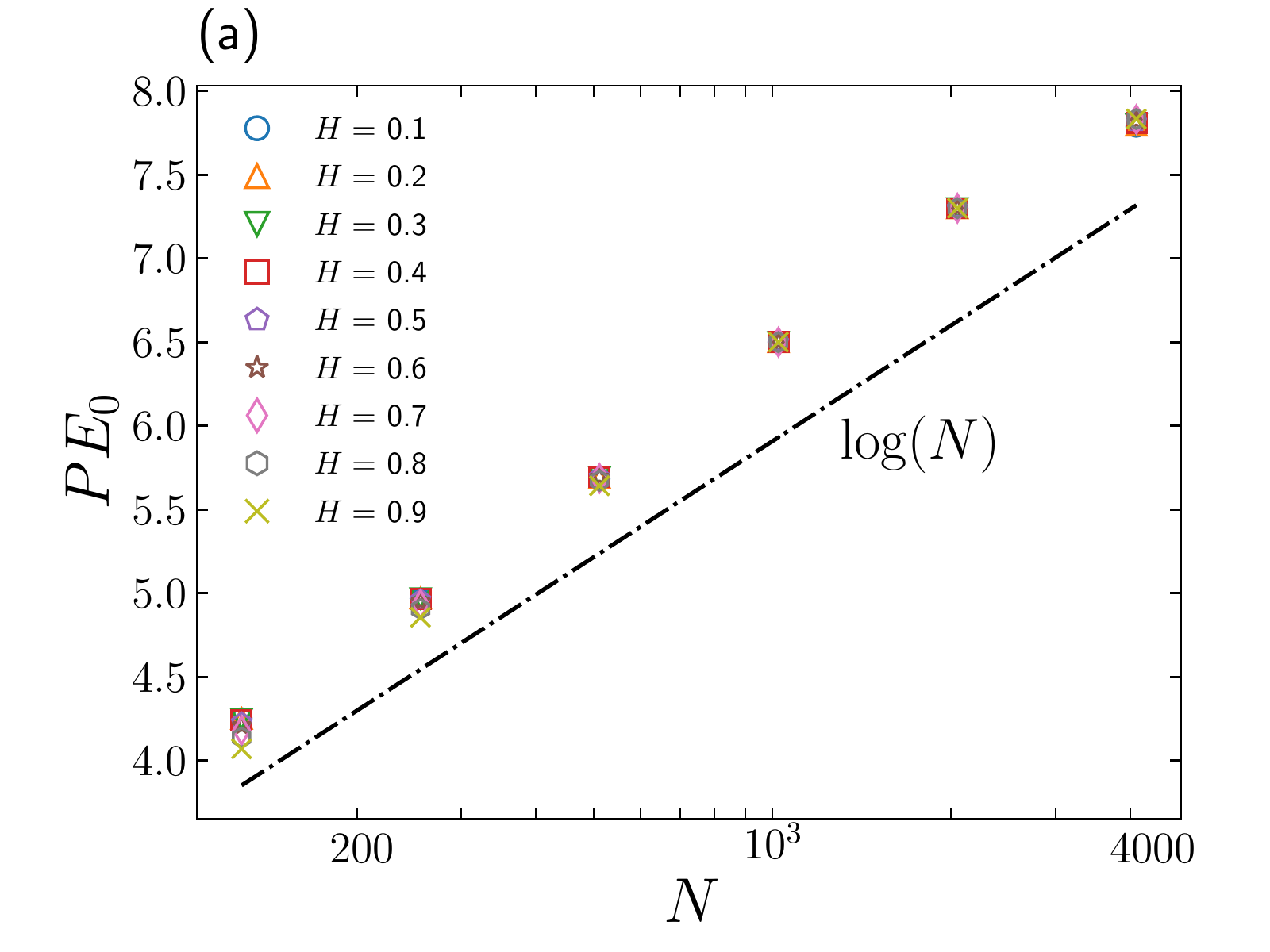}
            \includegraphics[width=0.48\textwidth]{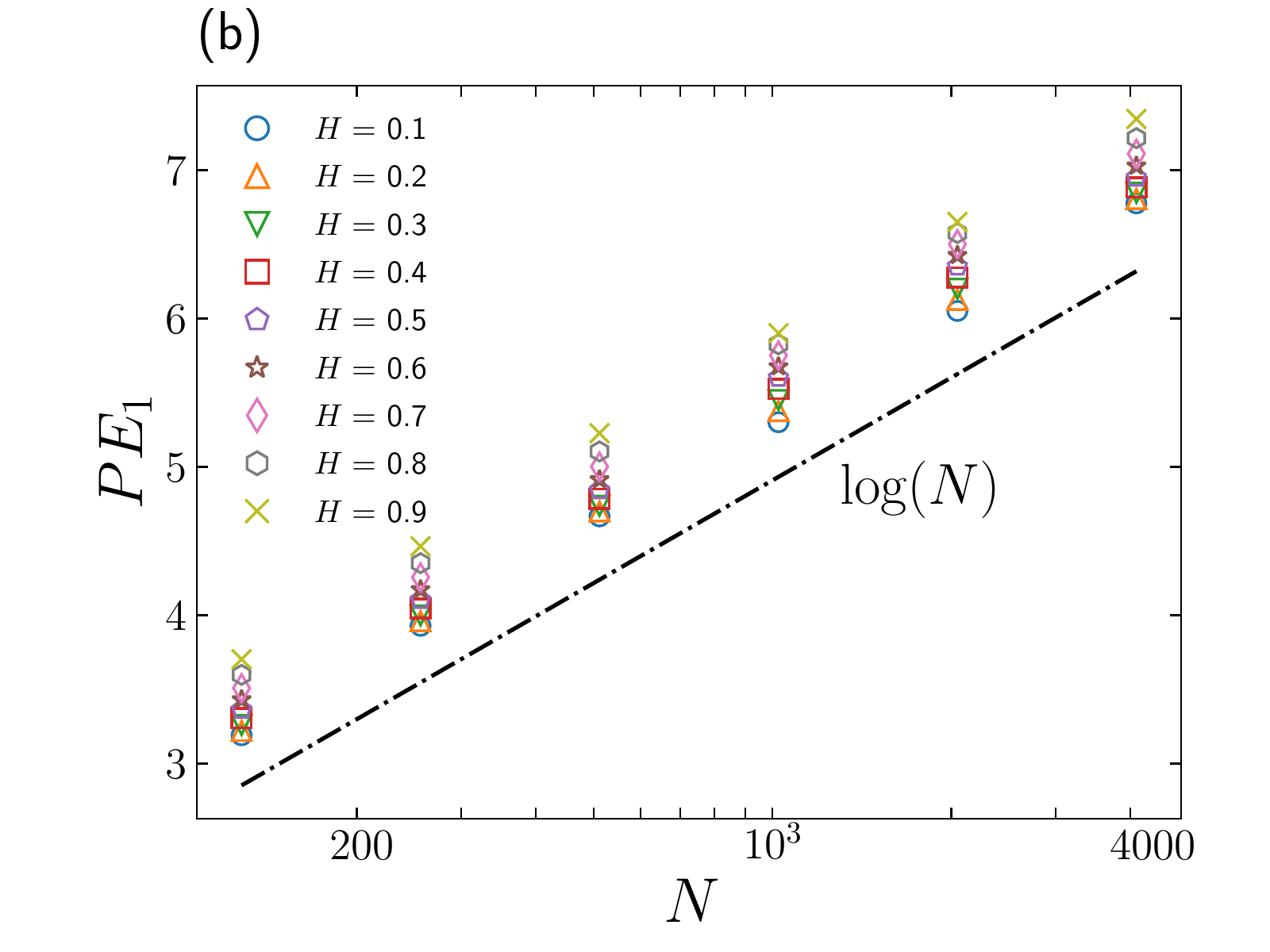}
            \includegraphics[width=0.48\textwidth]{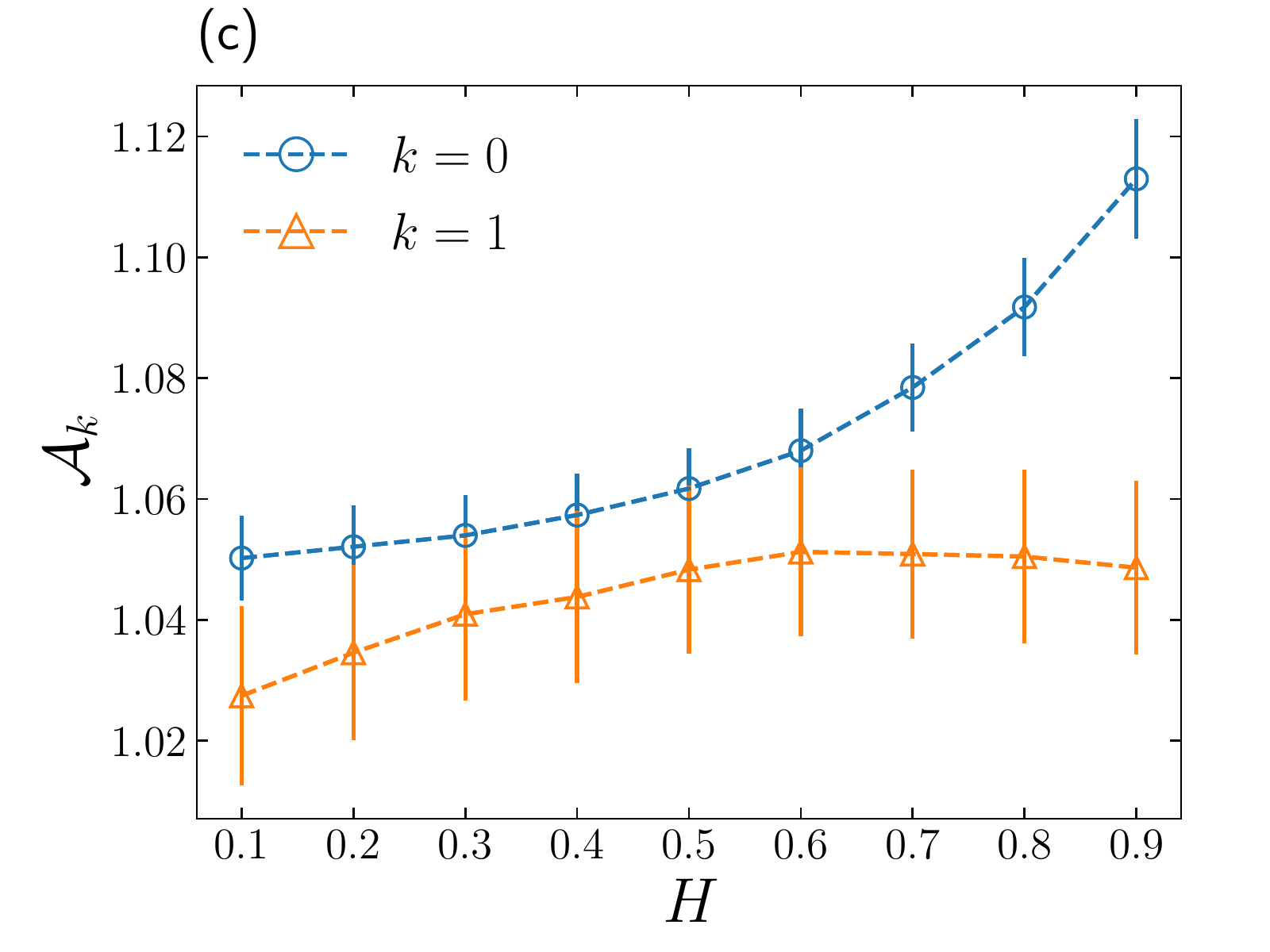}
            \caption{(a) The persistence entropy for the 0-homology group, $ PE_{0}$. (b) The $PE_1$ for 1-hole as a function of sample size in the log-scale. Different symbols correspond to various values of the Hurst exponent. (c) The prefactor of persistence entropy as a function of $H$ computed by the ensemble average.}
            \label{fig:entropy}
        \end{center}
    \end{figure}

    The associated persistence entropies (PEs) defined by Eq. (\ref{eq:PE}) are obtained using the persistence diagram. Panels (a) and (b) of Fig.~\ref{fig:entropy} depict the ${PE}_0$ and ${PE}_1$, as a function of sample size, respectively. We compute persistence entropy for all available Hurst exponent values represented by different symbols. Our results demonstrate that ${PE}_k=\mathcal{A}_k(H)\log_{10} N$ for $k=0,1$. The behavior of prefactor $\mathcal{A}_k$ versus $H$ is represented in panel (c) of Fig.~\ref{fig:entropy}. The $\mathcal{A}_0$ is almost an increasing function versus Hurst exponent, while the $\mathcal{A}_1$ becomes constant.

    %%%%%%%%%%%%%%%%%%%%%%%%%%%%%%%%%%%%%%%%%%%%%%%%%%%%%%%%%%%%%%%%%%
    %%%%%%%%%%%%%%%%%%%%%%%%%%%%%%%%%%%%%%%%%%%%%%%%%%%%%%%%%%%%%%%%%%

    %%%%%%%%%%%%%%%%%%%%%%%%%%%%%%%%%%%%%%%%%%%%%%%%%%%%%%%%%%%%%%%%%%%%%%%

    \section{Implementation on Realistic Data} \label{Sec:eeg}
\begin{figure*}
        \begin{center}
            \includegraphics[width=0.32\textwidth]{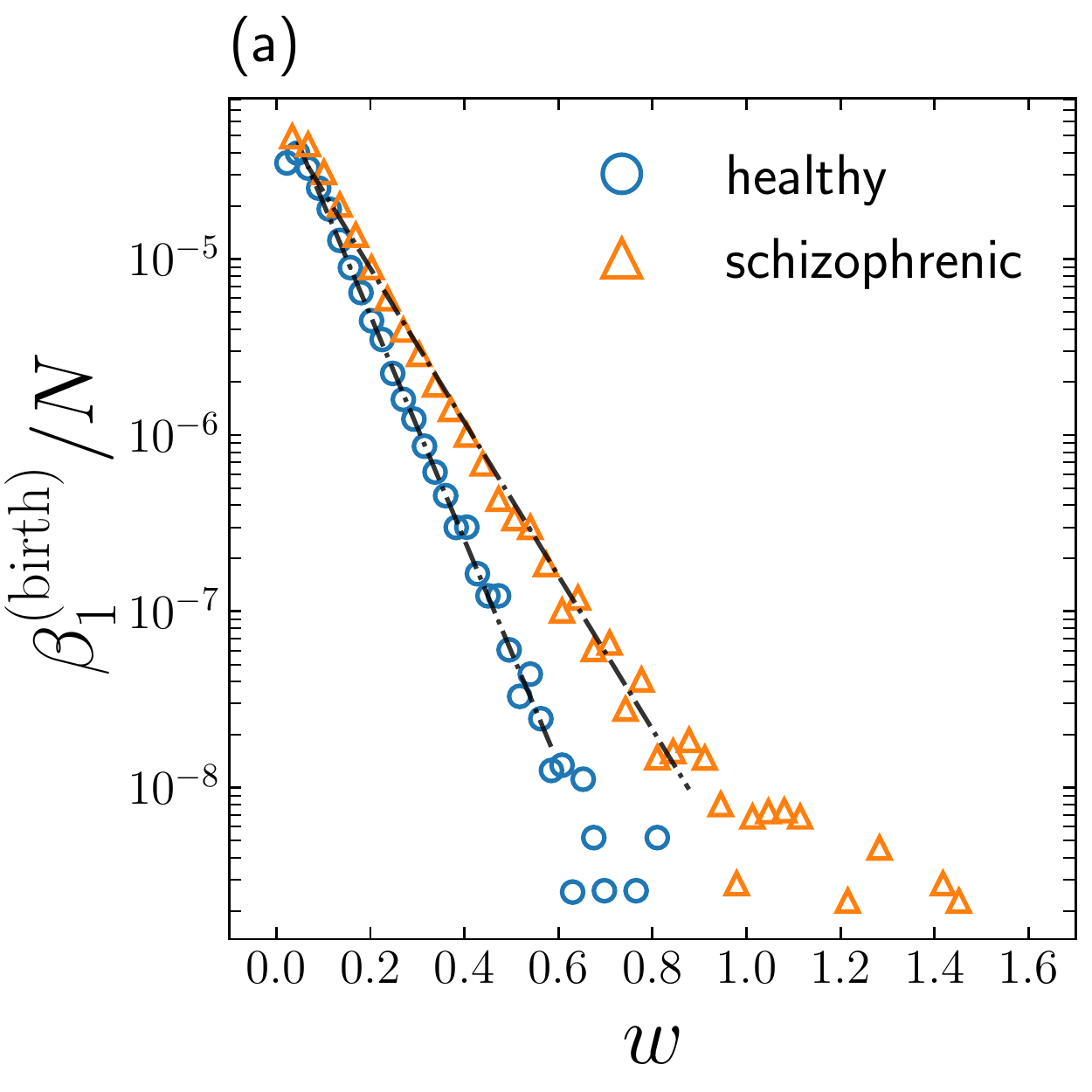}
            \includegraphics[width=0.32\textwidth]{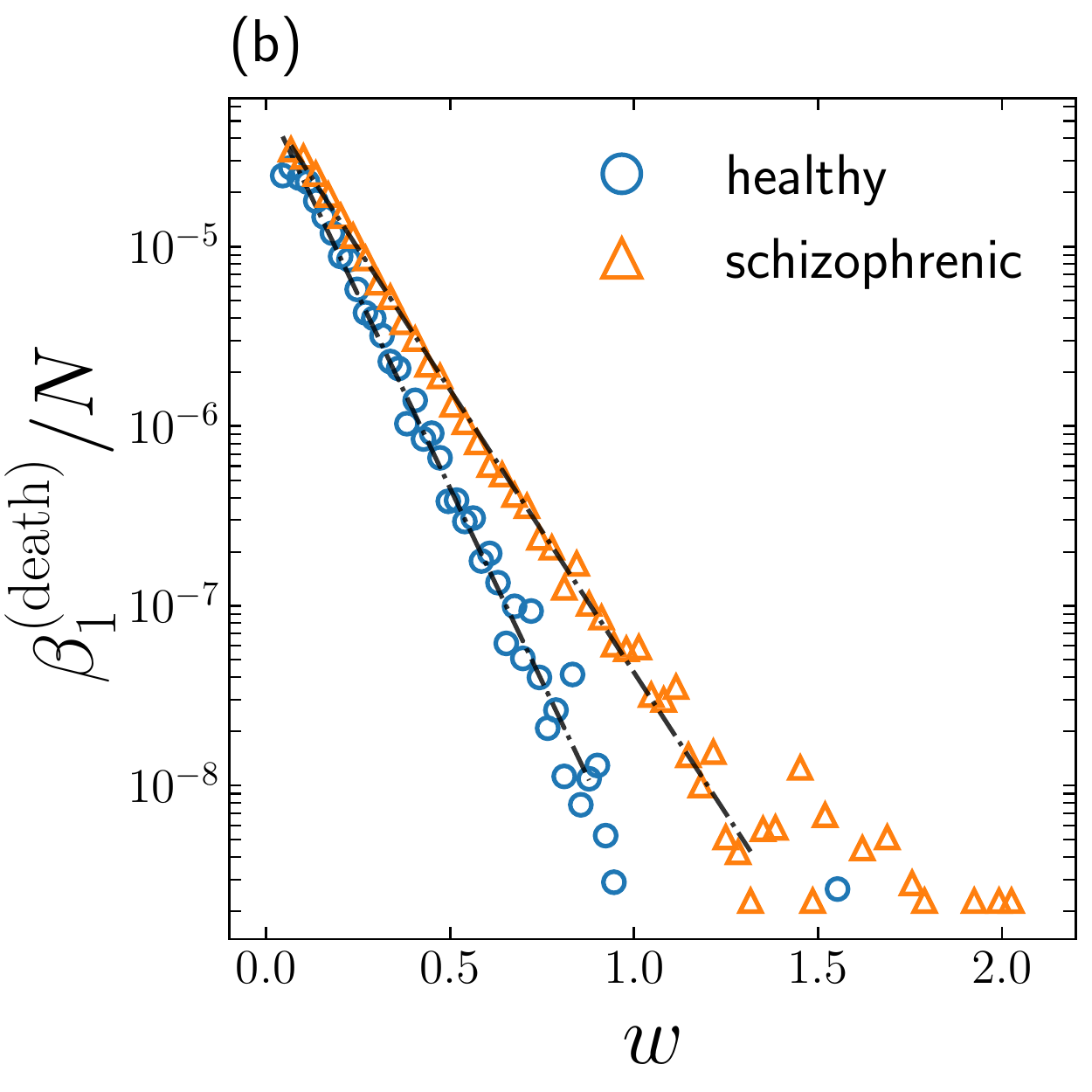}
            \includegraphics[width=0.32\textwidth]{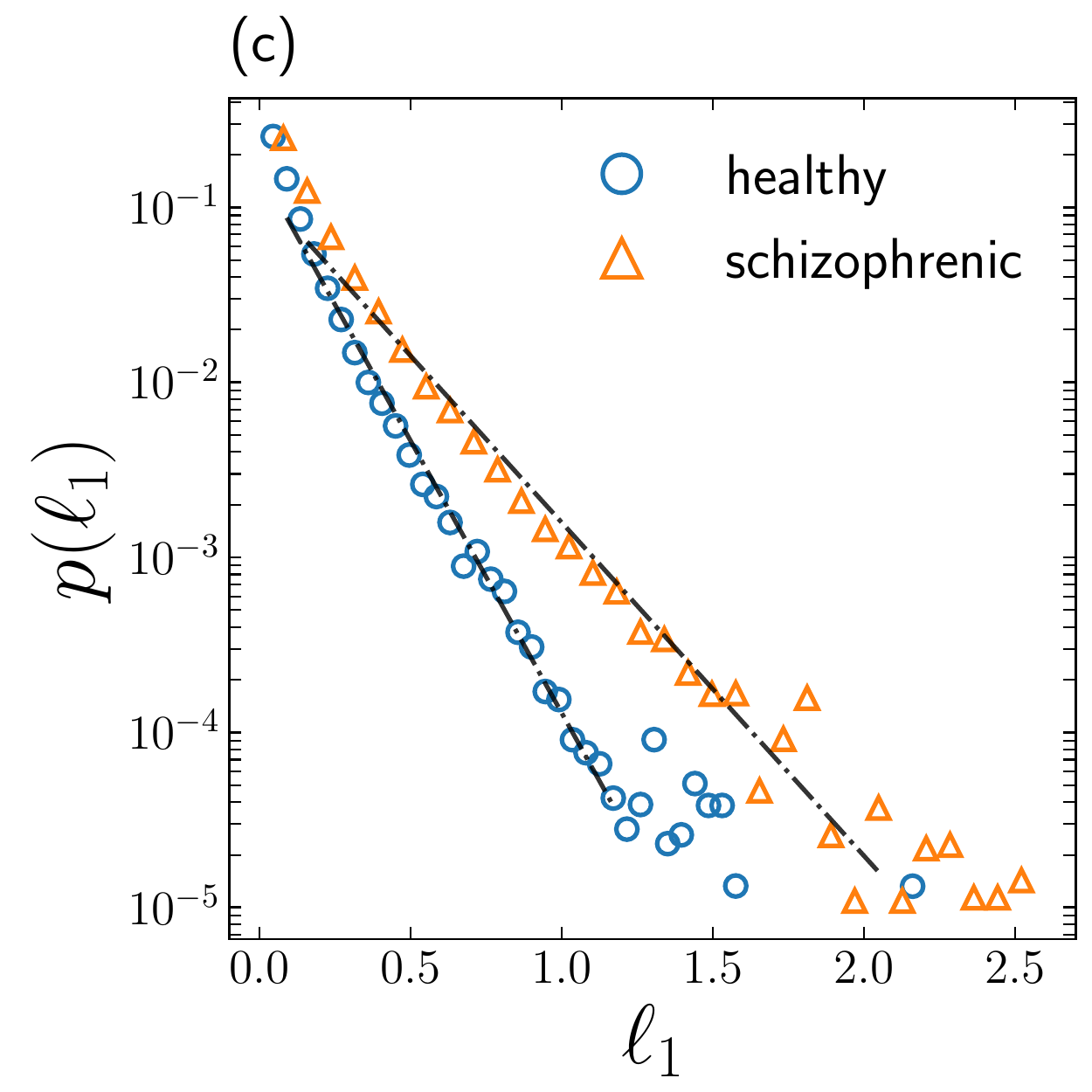}
            \caption{(a) The distribution of persistence diagram versus birth-axis ($\beta_1^{(\rm birth)}$) in log-scale for clique complexes of WNVGs associated with EEG signals for healthy (blue circle symbols) and schizophrenic cases (orange triangle symbols) as a function of threshold. (b) The $\beta_1^{(\rm death)}$ in log-scale as a function of threshold. (c) The lifetime distribution in log-scale for 1-homology generator statistics.} \label{fig:real_data}
        \end{center}
    \end{figure*}

    To illustrate the applicability of PH to characterize the self-similar nature of real data, in this section, we use the physiological data set studied in \cite{olejarczyk2017graph}. We implement the DFA algorithm to verify the scaling behavior of their correlation functions. Then, we apply our approach to the channel F8 of the standard 10–20 electroencephalography (EEG) montage recorded from 14 patients (7 males: 27.9 ± 3.3 years, 7 females: 28.3 ± 4.1 years) with paranoid schizophrenia and 14 healthy controls (7 males: 26.8 ± 2.9, 7 females: 28.7 ± 3.4 years). The EEG data were recorded in all subjects during an eyes-closed resting-state condition for 15 minutes with the sampling frequency of 250 Hz. We construct the WNVG from the EEG signals and the evolution of topological motifs through the filtration process is examined as depicted in Fig. \ref{fig:real_data}. To infer the statistical significance of the results, we carry  out the KS test  and we obtain the exponential behavior of results indicated in Fig. \ref{fig:real_data}. Then, according to the likelihood approach, we estimate the  topological coefficients and corresponding errorbars. The value of coefficients $\alpha_{1}^{(\rm birth)}$, $\alpha_{1}^{(\rm death)}$, $\alpha_{1}^{(\rm lifetime)}$ and associated Hurst exponents for both healthy and schizophrenic cases are reported in Table \ref{table:ttt}. Comparing the topological coefficients with our results for estimation of the Hurst exponent (Figs. \ref{fig:Betti-curve} and \ref{fig:betti1}), enables us to compute the associated Hurst exponent. The value of the Hurst exponent for each coefficient is reported in Table \ref{table:ttt}. Averaging on three kinds of Hurst exponents for EEG signals leads to $H = 0.86 \pm 0.02$ and $H= 0.69 \pm 0.03$ at the $1\sigma$ confidence interval for healthy and schizophrenic cases, respectively. Our results also reveal a correlated behavior in both data sets. It is worth noting that, the PH can classify different types of data.

    \begin{table}
        \begin{center}
            \begin{tabular}{|c|c|c|}\hline
                Measure & Healthy&  Schizophrenic \\
                \hline
                $\alpha_{1}^{(\rm birth)}$ & $4.85 \pm 0.04$ & $3.52 \pm 0.05$ \\
                \hline
                $H_{(\rm birth)}$ & $0.84 \pm 0.02$ & $0.66 \pm 0.03$ \\
                \hline
                $\alpha_{1}^{(\rm death)}$ & $3.96 \pm 0.04$ & $2.83 \pm 0.04$ \\
                \hline
                $H_{(\rm death)}$ & $0.85 \pm 0.02$ & $0.68 \pm 0.03$ \\
                \hline
                $\alpha_{1}^{(\rm lifetime)}$ & $7.03 \pm 0.03$ & $5.00 \pm 0.03$ \\
                \hline
                $H_{(\rm lifetime)}$ & $0.89 \pm 0.02$ & $0.73 \pm 0.03$ \\
                \hline
            \end{tabular}
        \end{center}
        \caption{\label{table:ttt}The value of topological coefficients and corresponding Hurst exponents for healthy and schizophrenic cases at $1\sigma$ level of confidence.}
    \end{table}

    %%%%%%%%%%%%%%%%%%%%%%%%%%%%%%%%%%%%%%%%%%%%%%%%%%%%%%%%%%%%%%%%%%%%%%%
    \section{Summary and Concluding Remarks} \label{sec:summery}

    Applying the various tools developed in network science on a network constructed from a typical time series reveals non-trivial information. Particularly, some of the features of the network for a generic fractional Gaussian noise (fGn), characterized by Hurst exponent $H$, exhibit a promising relation to the corresponding Hurst exponent. In this paper, we developed a method to generate a WNVG from the fGn (an increment of fBm) series which is a more general network compared to a binary network class. According to the filtration process, the evolution of topological holes in the associated WNVG can be captured. To this end, we used the PH technique, to examine the topological motifs.

    The statistical properties of the WNVG were analyzed using the standard network measures. Our main results are as follows:

    (1) The probability distribution functions of eigenvector, betweenness and closeness centralities computed for WNVGs constructed from fGn series [panel (b) of Fig. \ref{fig:HVG_NVG}] do not depend on $H$ and even the overall shape of mentioned distributions is highly size-dependent. The main message is that such kinds of statistical measures are not proper measures for determining the Hurst exponent of the fGn series (Fig.~\ref{fig:localstat1}).

    (2) Increasing the amount of correlation results in obtaining more dense networks and shifts the weights of the links to the lower values and, consequently, the width of the distribution function of the eigenvalues to be tighter. The $n$th moment of this distribution behaves as an $H$-dependent quantity. The higher the order of moments, the stronger the dependency on sample size (Fig.~\ref{fig:localstat2}).

    In the second part, we considered the topological properties of
    synthetic fGn. Our main results are summarized below:

    (3) The corresponding BNVG of a typical fGn is a \textit{topological tree} which means that all topological motifs have identically vanished while taking into account a weight function to construct the so-called weighted NVG (WNVG) leads to the survival of topological motifs and therefore their evolution with respect to self-similar exponents can be examined. Such evolution reveals a robust feature for determining the Hurst exponent reliably.

    (4) The Betti-0 is a representative of the number of connected components of the weighted clique simplicial complex associated with the WNVG with respect to the threshold (weight). The decreasing rate of Betti-0 increases by increasing $H$. The threshold value for which the underlying WNVG reaches the connected regime (path-connected) indicates an $H$-dependency [panels (a) and (b) of Fig. \ref{fig:beta0}].

    (5) Our result also confirms that the coefficient corresponding to the exponentially decaying function of the number of connected components dying (merging), $\alpha_0^{(\rm death)}$, as a function of threshold, depends on the Hurst exponent and interestingly it is almost size-independent [panels (c) and (d) of Fig. \ref{fig:beta0}].

    (6) The statistics of 1-holes (topological loops) show that the vanishing threshold for the number of topological loops, $w_1$, is $H$-dependent and it contains the sample size effect. The coefficients corresponding to the exponentially decaying function of the number of topological loops appearing ($\alpha_1^{({\rm birth})}$) and disappearing ($\alpha_1^{({\rm death})}$) as a function of threshold reveal the proper criteria for measuring the Hurst exponent (Fig.~\ref{fig:Betti-curve}). These coefficients also indicate that the topological motifs (upward concavity behavior data points of time-series satisfying loop condition) evolve (appear and disappear) in low weights for correlated signals.

    (7) The probability distribution of the lifetime of 1-holes ($\ell_1$) confirms that the corresponding coefficient ($\alpha_1^{({\rm lifetime})}$) is an increasing function versus $H$ emphasizing that the sample size effect is completely diminished in this quantity. Consequently, for a self-similar time-series in the absence of trends, this coefficient can be a reliable measure for estimating the Hurst exponent even for a small sample size irrespective of the value of the $H$ exponent (Fig.~\ref{fig:betti1}). This coefficient also indicates that the fGn signals living in negatively correlated regimes incorporate more robust topological motifs (1-holes) which shows a significant difference between the quality of visibility of inner and outer nodes corresponding to the local upward concavity behaving data points in these signals.

   (8) The persistence pairs (PPs) in the persistence diagram (PD) and the persistence barcode (PB) for the weighted clique complex of WNVGs which are indicators of persistent homology have been computed and with increasing the value of the Hurst exponent shrink to the origin of the coordinate (Fig.~\ref{fig:PB+PD}). We also computed persistence entropies (PEs) for 0-homology and 1-homology groups. Both quantities depend on sample size as expected and the corresponding slopes in semi-log scale were almost $H$-dependent (Fig. \ref{fig:entropy}).
    Implementing the PH method on real data and computing the topological parameters, we could determine the corresponding Hurst exponent. Interestingly, PH admitted its capability for the classification of various data sets.

(9) Employing PH on the constructed weighted network from the realistic series confirmed the correlated behavior of electroencephalography for both healthy and schizophrenic samples.
  
    Finally, we emphasize that the behavior of the local (statistical) observables depends weakly on $H$, whereas the exponents of global (topological) observables are almost strongly $H$-dependent and even for the $\alpha_1^{({\rm lifetime})}$, the size effect is completely diminished. A take-home message is that the TDA provides a new type of measure to quantify the Hurst exponent and, therefore, the scaling exponents of the correlation function, power spectrum, and fractal dimension can be specified with reliable approaches.

    In this paper, we have not verified whether the persistent homology technique is capable of recognizing that the underlying time-series is a self-similar set or not. Indeed, this purpose is beyond the scope of this paper. To address the capability of our method to discriminate between stationary and non-stationary times series,  more simulations for different types of non-stationary cases  should be performed; this is left for our future study. Also, it would be interesting to examine the effect of trends and irregularity which may occur in a wide range of events in nature in the context of the TDA and more precisely via the PH approach, and we leave this too for future research. The above analysis can be done on different phenomena ranging from cosmology, astrophysics, and economics to biology
    ~\cite{akrami2020planck,eghdami2018multifractal,sadr2018cosmic} with different approaches to make graphs from time-series. %In the context of self-organized critical (SOC) systems, the generated time-series of dynamical quantities observed in e.g. the sandpile models based on the Bak-Tang-Weisenfeld (BTW) dynamics can be mapped to the WNVGs, and finally, we explore the PH of topological holes.

    \section*{Acknowledgment}
    The authors are very grateful to  Marco Piangerelli  for his notice on TDA as a starting point in this research. In addition, we thank to the anonymous referees, who helped us to improve the paper, considerably. The numerical simulations were carried out on the computing cluster of the University of Mohaghegh Ardabili.

    \section*{Appendices}

    \subsection*{Appendix A:  Synthetic Fractional Gaussian Noise}\label{method:fGn}

    To model the stochastic fractal processes, Mandelbrot and Van Ness introduced the fractional Brownian motion (fBm) and fractional Gaussian noise (fGn) \cite{mandelbrot1968fractional}. The theory of fBm is a mathematical generalization of the classical random walk and Brownian motion \cite{mandelbrot1968fractional,kahane1993some}. A 1-dimensional fBm is represented by $B\equiv\{B(t):t\ge 0\}$, with power-law variance, for which $Var (B(a t))\triangleq Var(a^HB(t))=a^{2H}Var(B(t))$, where $H\in (0,1)$ is called the Hurst exponent. For this random force, the Markov property and the stationarity are violated (note that when we have domain Markov property, stationarity, and continuity for a time series, then it should be proportional to a 1-dimensional Brownian motion). A model for generating fBm (denoted by $B_H$ to emphasize on its Hurst exponent $H$) is a generalization of the Brownian motion which is non-stationary and non-Markovian \cite{reed1995spectral}, and is given by the Holmgren-Riemann-Liouville fractional integral,
    \begin{equation}
    B_H(t) = \frac{1}{{\Gamma (H + \frac{1}{2})}}\int\limits_0^t {{{(t - s)}^{H - \frac{1}{2}}}} \text{d}B(s)
    \label{Eq:FBM}
    \end{equation}
    where $\Gamma$ is the Gamma function, $\text{d}B(s)\equiv B(s+\text{d}s)-B(s)$ is the increment of 1-dimensional Brownian motion, and it has the following covariance:
    \begin{equation}
    \langle  {{B_H}(t){B_H}(s)} \rangle = \frac{\sigma^2}{2}\left( {{{\left| t \right|}^{2H}} + {{\left| s \right|}^{2H}}-{{\left| {t - s} \right|}^{2H}}} \right)
    \end{equation}
    where $\sigma^2\equiv \langle B(0)\rangle$  and also $\langle B_H\rangle =0$. The increments, $ x_H(t) \equiv \delta_t\left( {B_H}(t + \delta t) - {B_H}(t)\right)$, are known as fractional Gaussian noise (fGn). The power spectrum of fBm and fGn behaves as $S(f)\sim f^{-\xi}$, where $\xi(H)=2H+1$ and $\xi(H)=2H-1$ for fBm and fGn, respectively. For $H>\frac{1}{2}$ ($H<\frac{1}{2} $) the corresponding fGn is positively (negatively) correlated. According to the results provided by detrended fluctuation analysis (DFA), the relation between the scaling exponent derived by the DFA method and associated Hurst exponent of fBm is $H+1$, while constructing the fGn series by making the increment of fBm confirmed that the scaling exponent of fluctuation functions computed by DFA is directly related to the Hurst exponent \cite{taqqu1995estimators,Kantelhardt}.

    \subsection*{Appendix B: Statistical Network Analysis}
    Suppose that a network (graph) is represented by $G=(V,E,w)$.  Here,  $V \equiv \{v_i\}_{i=1}^{N}$ is a node (vertex) set, $E = V \times V  \equiv \Bigr\{ e_{ij}=(v_i,v_j) \Bigr| v_i,v_j \in V \Bigr\}_{i,j=1}^{N} $ is a link (edge) set, and $w: E \rightarrow \mathbb{R}$ is a weight function (threshold). Subsequently, the degree of $i$th node ($v_i$) is the number of nodes straightly connected with the underlying node by non-zero weight, and it is denoted by $k_i\equiv\sum_{j}(1-\delta_{0,w_{ij}})$. The degree centrality ($c_i^{D}$) is defined by  $\frac{k_i}{N-1}$, which is apparently related directly to how important the underlying node is, since it is the number of agents that have a connection with it. An important function concerning this quantity is the degree distribution showing  the probability distribution function for degree of all nodes in the network:
    \begin{equation}
    p(k) = \displaystyle \frac{1}{N} \displaystyle \sum_{i=1}^{N} \delta_{k,k_i}
    \end{equation}
    The eigenvector centrality ($c_{i}^{E}$) is also defined via the eigenvalue equation
    \begin{equation}
    \lambda c_i^E=\sum_{j=1}^Nw_{ij}c_j^E
    \label{Eq:eigenvalueproblem}
    \end{equation}
    where $\lambda$ is the eigenvalue. The maximum value of the $\lambda$ spectrum, i.e., $\lambda_{\rm max}$ plays the dominant role in the network properties, the corresponding eigenvectors of which are denoted by $c_{i,{\rm max}}^E$ revealing the importance of the nodes.  Let us denote the shortest distance between nodes $v_i$ and $v_j$ by $d_{ij}$ which is assumed to be $N$ when there is no path connecting them (disconnected graphs). Then the closeness centrality is defined by
    \begin{equation}
    c_i^C=\frac{N-1}{\sum_{j=1}^Nd_{ij}},
    \end{equation}
    and the betweenness centrality is as follows:
    \begin{equation}
    c_i^B=\frac{2}{(N-1)(N-2)}{\sum_{j=1,j\ne i}^N}{\sum^N_{k=1,k\ne i,j}}\frac{n_{jk}(i)}{n_{jk}},
    \end{equation}
    where $n_{jk}$ is the number of geodesics from $v_j$ to $v_k$, and $n_{jk}(i)$ is the number of geodesics from $v_j$ to $v_k$ which pass through node $v_i$.

    \subsection*{Appendix C: Algebraic Topology}
    Topology generally refers to the global features in contrast to the geometrical invariants of underlying objects or sets. We have two spaces represented by $X$ and $Y$, and they have the same local (geometric) features if any relevant features are invariant under \textit{congruence}, while the mentioned spaces are topologically equivalent if the associated features are invariant under \textit{homeomorphisms}. In other words, they are homeomorphic. Homology theory plays a crucial role in the mathematical description of the relevant building block of a typical topological space and reveals the connectedness of underlying space~\cite{Edels2010Comp,rote2006computational,zomorodian2009computational}. Based on such properties, for the sake of clarity, we will give a brief review on the building blocks of algebraic topology which are useful to set up homology groups.

\begin{figure}
        \centerline{\includegraphics[width=0.48\textwidth]{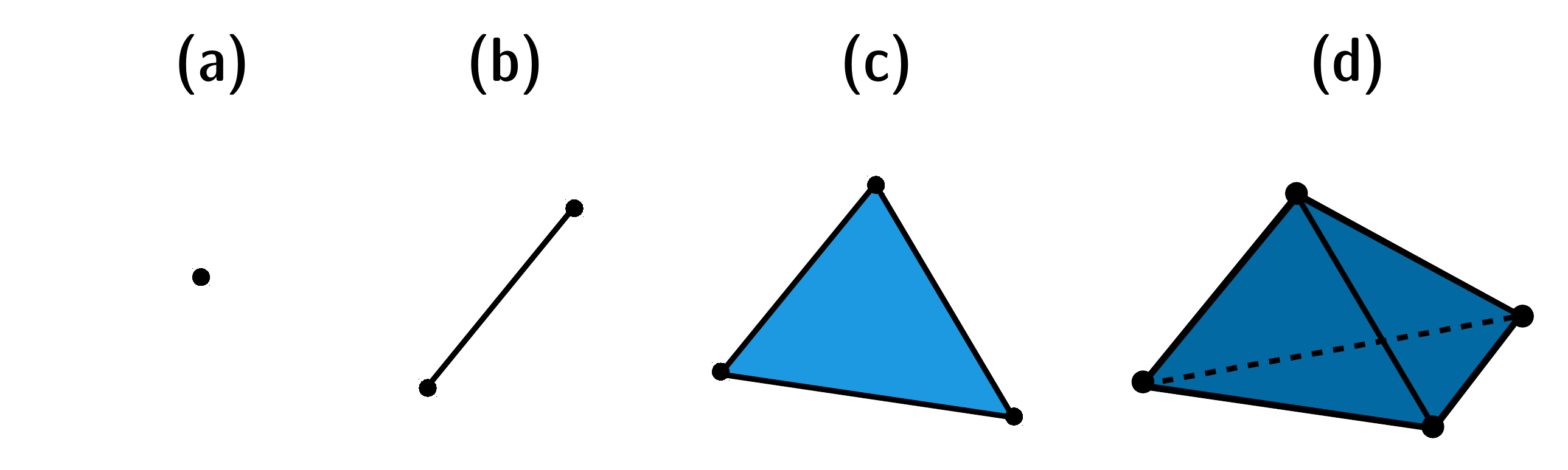}}
        \caption{Low dimensional simplices : (a) 0-simplex (point), (b) 1-simplex (line segment), (c) 2-simplex (filled triangle), and (d) 3-simplex (filled tetrahedron).}
        \label{fig:simplex1}
    \end{figure}

    \textbf{Simplex:} A $k$-simplex ($\sigma_k$) is a convex-hull of any geometrically independent subset, accordingly $\sigma_k\equiv[v_0,v_1,...,v_k]\subseteq\mathbb{R}^D$. By this definition, a 0-simplex is a point, a 1-simplex is a segment of a line, a 2-simplex is a filled triangle, a 3-simplex is a filled tetrahedron, and so on (Fig. \ref{fig:simplex1}).

    \textbf{Face:} An $l$-simplex which is denoted by $\sigma_l$ is a subset of the $k$-simplex ($\sigma_l \subseteq \sigma_k$) and  it is called the $l$-face of the $k$-simplex.

\begin{figure}
        \begin{center}
            \includegraphics[width=0.48\textwidth]{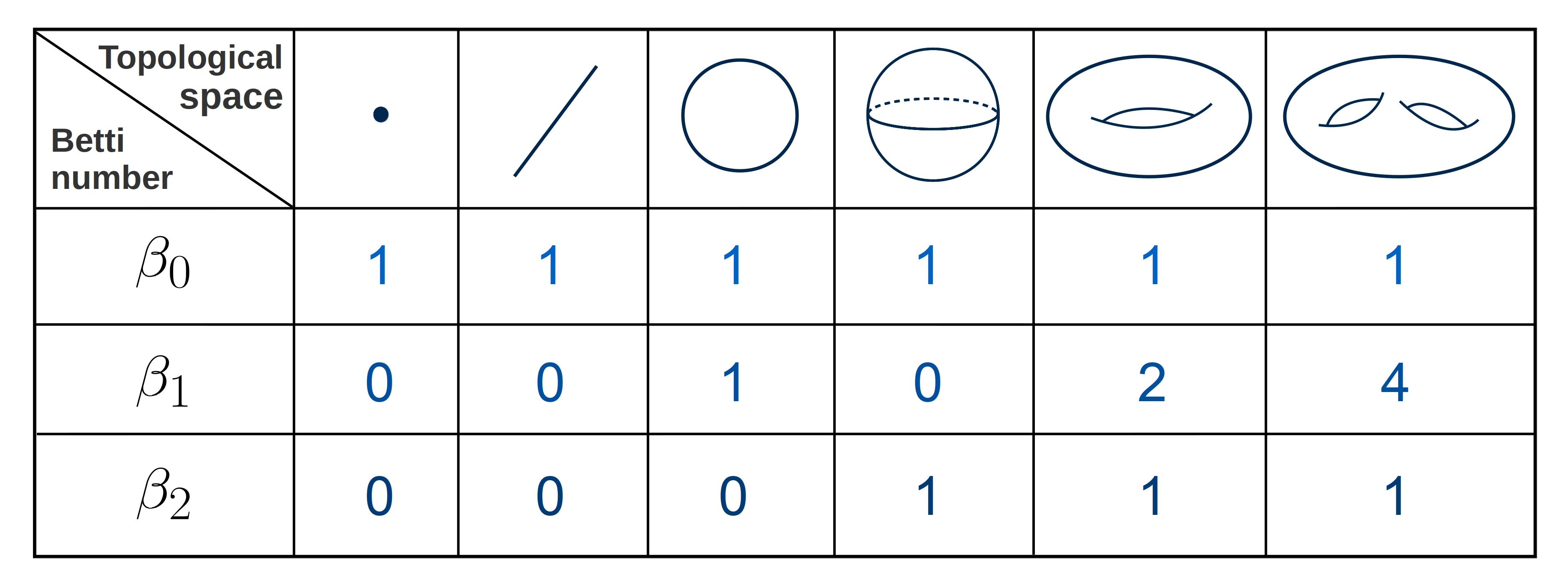}
            \caption{ Betti numbers of some topological spaces: point, line, circle, sphere, torus, 2-torus.}\label{fig:bettis}
        \end{center}
    \end{figure}

\begin{figure}
        \centerline{\includegraphics[width=0.48\textwidth]{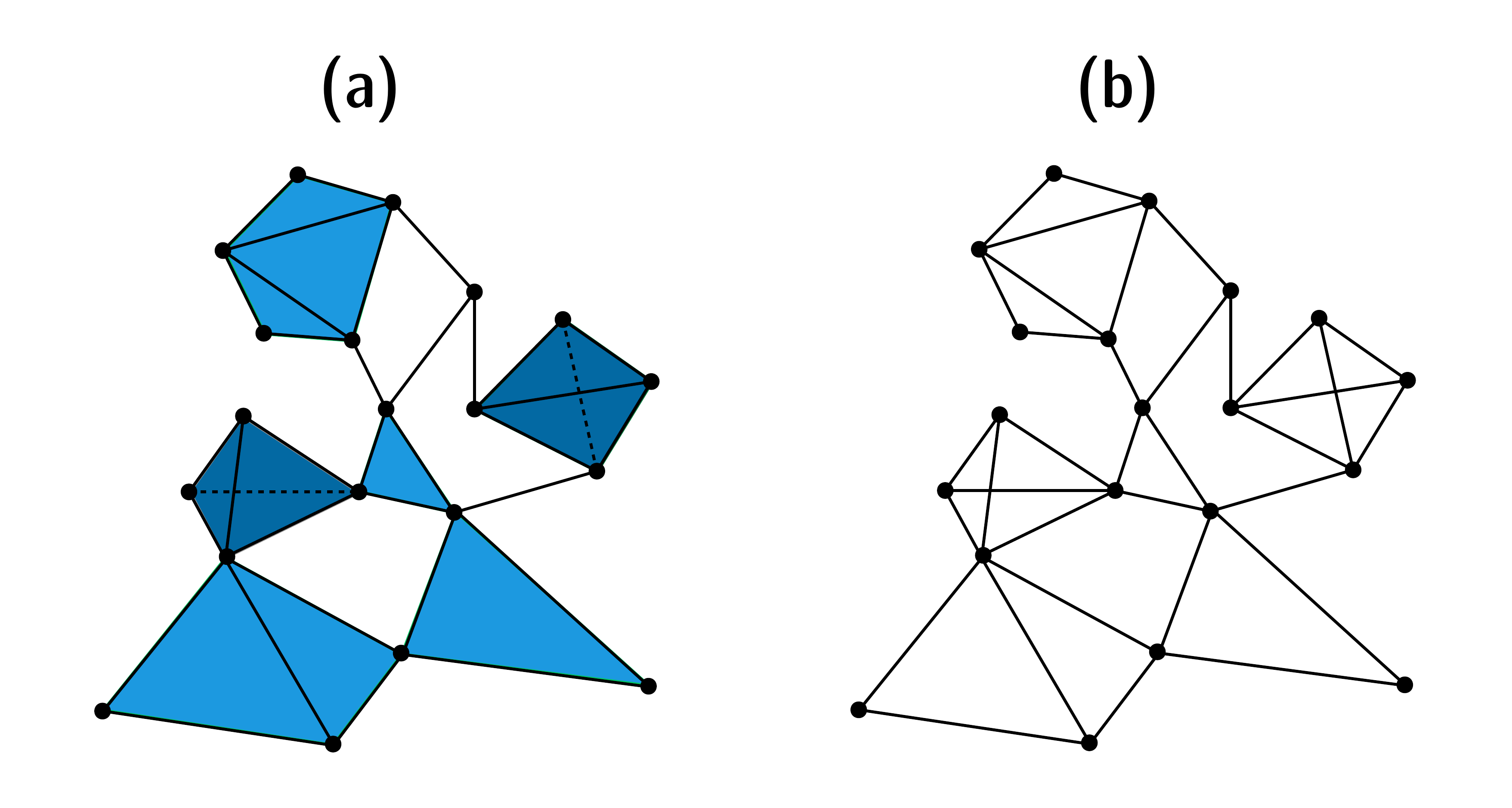}}
        \caption{(a) Clique complex of (b) a typical network.}
        \label{fig:cli}
    \end{figure}

    \textbf{Simplicial complex:} A simplicial complex ($\psi$) is a collection of simplices such that: any $l$-face of any $k$-simplex of a typical complex ($0<l<k$), is a member of the complex. In addition, the non-empty intersection of any two simplices, $\sigma_k$, and $\sigma_m$, from the complex is an $l$-face of both simplices. The dimension of a complex is the maximum dimension of all simplices of the complex.
    According to the definition of the complex, one can define its $k$-ordered subcollection of complexes $\psi$ as follows:
    \begin{equation}
    \Sigma_k(\psi) \equiv \Bigr \{\sigma \in \psi ~ \Bigr| ~ dim(\sigma)=k \Bigr \}
    \end{equation}
    \textbf{Chain:} For a given simplicial complex, a $k$-chain ($k$-dimensional chain) is a linear combination of $k$-simplices of $\psi$, defined by
    \begin{equation}
    c_k\equiv \displaystyle\sum_{i=1}^{|\Sigma_k (\psi)|} a_i ~ \sigma_{k}^{(i)} \quad ;\quad  \sigma_{k}^{(i)} \in \Sigma_k (\psi)
    \end{equation}
    where $|\Sigma_k (\psi)|$ corresponds to the cardinality of the $k$-ordered subcollection of complex and the coefficients, $a_i$s, belong to a field, which is usually considered as $\mathbb{F}=\mathbb{Z}_2 \equiv \{0,1\}$.
    The collection of all possible $k$-chains in the simplicial complex  is called the $k$-chain group as
    \begin{equation}
    C_k(\psi) \equiv \Bigr\{c_k ~\Bigr|~ c_k= \displaystyle\sum_{i=1}^{|\Sigma_k (\psi)|} a_i \sigma_{k}^{(i)} ; \sigma_{k}^{(i)} \in \Sigma_k (\psi), a_i \in \mathbb{F}\Bigr\}
    \end{equation}
    \textbf{Boundary operator:}
    For the simplices in any dimension, the boundary operator $\partial_k$ is an operator mapping $\sigma_k$ to its boundary according to
    \begin{equation}
    \partial_k(\sigma_{k}) \equiv \displaystyle\sum_{j=0}^{k}(-1)^j ~ [x_{0},x_{1}, ... ,x_{j-1},x_{j+1}, ... ,x_{k}] ~~ \subseteq ~ \sigma_{k}
    \end{equation}
    \textbf{Boundary:} A $k$-chain which is the boundary of a $(k+1)$-chain, is called the $k$-boundary, denoted by $b_k$.
    The $k$-boundary group is the collection of all $k$-boundaries in complex $\psi$,
    \begin{eqnarray}
    B_k (\psi) &\equiv& \Bigr\{c_k \in C_k(\psi) \Bigr|\exists c_{k+1} \in C_{k+1}(\psi) ; \partial_{k+1}(c_{k+1})= c_k \Bigr\} \nonumber\\
    &\equiv& \Bigr\{b_{k}^{(i)} \Bigr\}_{i}^{|B_k (\psi)|}\subseteq C_k(\psi)
    \end{eqnarray}
    \textbf{Cycle:} A $k$-chain that has no boundary, is called a $k$-cycle denoted by $z_k$ as
    \begin{equation}
    \partial_k(z_k) = \oslash
    \end{equation}
    The $k$-cycle group is defined as the collection of all $k$-cycles in complex $\psi$,
    \begin{eqnarray}
    Z_k(\psi) &\equiv& \Bigr\{c_k \in C_k(\psi) ~\Bigr|~ \partial_k(c_k) = \oslash \Bigr\} \nonumber\\
    &\equiv&\Bigr\{z_{k}^{(i)} \Bigr\}_{i}^{|Z_k(\psi)|}\subseteq C_k(\psi)
    \end{eqnarray}
    Since "boundaries have no boundary", therefore, we have
    \begin{equation}
    \partial_k(b_k) = \partial_k(\partial_{k+1}(c_{k+1})) = \oslash
    \end{equation}
    hence
    \begin{equation}
    B_k(\psi) \subseteq Z_k(\psi) \subseteq C_k(\psi)
    \end{equation}
    \textbf{Homology group:} The $k$-homology group is defined by the
    quotient group of the $k$-cycles group by the $k$-boundary group,
    \begin{equation}
    H_k(\psi) \equiv Z_k(\psi) / B_k(\psi).
    \end{equation}
    The $k$th \textit{Betti number} of a simplicial complex, denoted by $\beta_k (\psi)$ , is a topological invariant which counts the number of $k$-homology classes  corresponding to the number of $k$-dimensional holes of complex $\psi$ (Fig. \ref{fig:bettis}),
    \begin{equation}
    \beta_k(\psi) \equiv dim(H_k(\psi))
    \end{equation}
    \textbf{Clique (flag) simplicial complex} of an unweighted (binary) network, $G=(V,E, w\in\{0,1\}$), is a simplicial complex, denoted by $\psi^{(G)}$, such that any $k$-simplex of each dimension in the complex corresponds to a $(k+1)$-clique in the network and vice versa (Fig. \ref{fig:cli}).

    A binary network is a \textit{topological tree}, if and only if, it is topologically holeless in all dimensions. Namely,  the associated clique simplicial complex has the following property:
    \begin{equation}
    \beta_k(\psi^{(G)}) = min(\beta_k) = \left\{
    \begin{array}{l l}
    1 \qquad ; \qquad k=0 \\
    0  \qquad ; \qquad k>0 \\
    \end{array} \right.
    \end{equation}

    \subsection*{Appendix D: Persistent Homology}

    In the context of topological data analysis (TDA), we are interested in statistical analysis of the structure in data. Generally, in TDA-based analysis, data of any type (point cloud data, scalar field, time-series, network, \textit{etc.}) when mapped to a weighted simplicial complex are worked out topologically in terms of the parameters present inherently in the original data, e.g. the weight of links in a weighted network, or the pairwise distance between data points in point cloud data.  More precisely, TDA maps parameter-dependent data, $X(w)$, (where $w$ is a typical parameter, like the threshold value for any weighted network) to a weighted simplicial complex, $\psi^{X}(w)$. Such approach produces the chain group, the cycle group, the boundary group, the homology group  and particularly, the $k$th Betti number~\cite{nakahara2003geometry,Edels2010Comp}.

    Persistent homology (PH) as a powerful tool of TDA examines the creation (birth) and destruction (death) of topological invariants associated with homology classes during a mathematical process called filtration \cite{Edels2010Comp}. Filtration, $\phi$, is a nested sequence of weighted complex $\psi(w)$ in which any complex with a distinct weight is a subcollection of any complex with higher weight,
    \begin{equation}
    \phi(\psi(w)) \equiv \left( \psi(w) ~\Bigr|~ \forall w' < w'' : \psi(w') \subseteq \psi(w'') \right)_{w_{\rm min}}^{w_{\rm max}}
    \end{equation}
    More precisely, the PH technique enumerates the $k$th Betti number of any subcomplex in $\phi$ and assigns an ordered tuple $w^{(h_{k})}\equiv(w_{{\rm birth}}^{(h_{k})},w_{{\rm death}}^{(h_{k})})$ to the existing $k$-dimensional topological hole. Here  $w_{{\rm birth}}^{(h_{k})}$ and $w_{{\rm death}}^{(h_{k})}$ are the thresholds for which $h_{k}$ appears (birth) and disappears (death), respectively. Since $w_{{\rm birth}}^{(h_{k})} < w_{{\rm death}}^{(h_{k})}$, we can define the positive-value quantity $\ell^{(h_{k})} \equiv w_{{\rm death}}^{(h_{k})} - w_{{\rm birth}}^{(h_{k})}$ as persistency (lifetime) of a $k$-dimensional hole. The persistence barcode (PB) and equivalently the persistence diagram (PD) are  the famous representations of PH. As an illustration, the $k$-dimensional persistence diagram of weighted complex $\psi(w)$ is a multiset $ PD_k \Bigr( \phi(\psi(w)) \Bigr) \equiv (\mathcal{M} , \mathcal{N})$, where
    \begin{equation}
    \mathcal{M}~\equiv~\Bigr\{ w^{(h_{k})} ~\Bigr|~ h_k \in H_k(\psi(w)) ~,~ \psi(w) \in \phi  \Bigr\}
    \end{equation}
    and $\mathcal{N} : \mathcal{M} \rightarrow \mathbb{N}$ is the count function. Inspired by Shannon entropy for a typical state probability, one can define  the persistence entropy (PE) of the $k$th PD (PB). To this end, we construct the probability for lifetime of homology classes as
    \begin{align}
    &p(\ell^{(h_k)})\equiv \frac{\ell^{(h_k)}}{ \mathcal{L}} \quad ; \quad \mathcal{L}~ \equiv \displaystyle \sum_{w^{(h_{k})} \in \mathcal{M}({PD}_k)} \ell^{(h_k)}
    \end{align}
    Therefore, the PE for the $k$-dimensional topological hole is defined  by \cite{merelli2015topological,atienza2019persistent}
    \begin{align}\label{eq:PE}
    &PE_k = - \displaystyle \sum_{w^{(h_{k})} \in \mathcal{M}({PD}_k)} p(\ell^{(h_k)}) ~ \log_{10} p(\ell^{(h_k)})
    \end{align}

    %\newpage
%    \bibliography{refs}

\begin{thebibliography}{73}%
\makeatletter
\providecommand \@ifxundefined [1]{%
 \@ifx{#1\undefined}
}%
\providecommand \@ifnum [1]{%
 \ifnum #1\expandafter \@firstoftwo
 \else \expandafter \@secondoftwo
 \fi
}%
\providecommand \@ifx [1]{%
 \ifx #1\expandafter \@firstoftwo
 \else \expandafter \@secondoftwo
 \fi
}%
\providecommand \natexlab [1]{#1}%
\providecommand \enquote  [1]{``#1''}%
\providecommand \bibnamefont  [1]{#1}%
\providecommand \bibfnamefont [1]{#1}%
\providecommand \citenamefont [1]{#1}%
\providecommand \href@noop [0]{\@secondoftwo}%
\providecommand \href [0]{\begingroup \@sanitize@url \@href}%
\providecommand \@href[1]{\@@startlink{#1}\@@href}%
\providecommand \@@href[1]{\endgroup#1\@@endlink}%
\providecommand \@sanitize@url [0]{\catcode `\\12\catcode `\$12\catcode
  `\&12\catcode `\#12\catcode `\^12\catcode `\_12\catcode `\%12\relax}%
\providecommand \@@startlink[1]{}%
\providecommand \@@endlink[0]{}%
\providecommand \url  [0]{\begingroup\@sanitize@url \@url }%
\providecommand \@url [1]{\endgroup\@href {#1}{\urlprefix }}%
\providecommand \urlprefix  [0]{URL }%
\providecommand \Eprint [0]{\href }%
\providecommand \doibase [0]{http://dx.doi.org/}%
\providecommand \selectlanguage [0]{\@gobble}%
\providecommand \bibinfo  [0]{\@secondoftwo}%
\providecommand \bibfield  [0]{\@secondoftwo}%
\providecommand \translation [1]{[#1]}%
\providecommand \BibitemOpen [0]{}%
\providecommand \bibitemStop [0]{}%
\providecommand \bibitemNoStop [0]{.\EOS\space}%
\providecommand \EOS [0]{\spacefactor3000\relax}%
\providecommand \BibitemShut  [1]{\csname bibitem#1\endcsname}%
\let\auto@bib@innerbib\@empty
%</preamble>
\bibitem [{\citenamefont {Edelsbrunner}\ \emph {et~al.}(2001)\citenamefont
  {Edelsbrunner}, \citenamefont {Harer},\ and\ \citenamefont
  {Zomorodian}}]{edelsbrunner2000topological}%
  \BibitemOpen
  \bibfield  {author} {\bibinfo {author} {\bibfnamefont {H.}~\bibnamefont
  {Edelsbrunner}}, \bibinfo {author} {\bibfnamefont {J.}~\bibnamefont {Harer}},
  \ and\ \bibinfo {author} {\bibfnamefont {A.}~\bibnamefont {Zomorodian}},\
  }in\ \href {\doibase 10.1145/378583.378626} {\emph {\bibinfo {booktitle}
  {Proceedings of the Seventeenth Annual Symposium on Computational
  Geometry}}},\ \bibinfo {series and number} {SCG '01}\ (\bibinfo  {publisher}
  {Association for Computing Machinery},\ \bibinfo {address} {New York, NY,
  USA},\ \bibinfo {year} {2001})\ pp.\ \bibinfo {pages} {70--79}\BibitemShut
  {NoStop}%
\bibitem [{\citenamefont {Zomorodian}\ and\ \citenamefont
  {Carlsson}(2005)}]{zomorodian2005computing}%
  \BibitemOpen
  \bibfield  {author} {\bibinfo {author} {\bibfnamefont {A.}~\bibnamefont
  {Zomorodian}}\ and\ \bibinfo {author} {\bibfnamefont {G.}~\bibnamefont
  {Carlsson}},\ }\href {\doibase 10.1007/s00454-004-1146-y} {\bibfield
  {journal} {\bibinfo  {journal} {Discrete \& Computational Geometry}\ }\textbf
  {\bibinfo {volume} {33}},\ \bibinfo {pages} {249} (\bibinfo {year}
  {2005})}\BibitemShut {NoStop}%
\bibitem [{\citenamefont {Zomorodian}(2005)}]{zomorodian2005topology}%
  \BibitemOpen
  \bibfield  {author} {\bibinfo {author} {\bibfnamefont {A.~J.}\ \bibnamefont
  {Zomorodian}},\ }\href@noop {} {\emph {\bibinfo {title} {Topology for
  computing}}},\ Vol.~\bibinfo {volume} {16}\ (\bibinfo  {publisher} {Cambridge
  university press},\ \bibinfo {year} {2005})\BibitemShut {NoStop}%
\bibitem [{\citenamefont {Carlsson}(2009)}]{carlsson2009topology}%
  \BibitemOpen
  \bibfield  {author} {\bibinfo {author} {\bibfnamefont {G.}~\bibnamefont
  {Carlsson}},\ }\href@noop {} {\bibfield  {journal} {\bibinfo  {journal}
  {Bulletin of the American Mathematical Society}\ }\textbf {\bibinfo {volume}
  {46}},\ \bibinfo {pages} {255} (\bibinfo {year} {2009})}\BibitemShut
  {NoStop}%
\bibitem [{\citenamefont {Zomorodian}(2012)}]{zomorodian2012topological}%
  \BibitemOpen
  \bibfield  {author} {\bibinfo {author} {\bibfnamefont {A.}~\bibnamefont
  {Zomorodian}},\ }\href@noop {} {\bibfield  {journal} {\bibinfo  {journal}
  {Advances in applied and computational topology}\ }\textbf {\bibinfo {volume}
  {70}},\ \bibinfo {pages} {1} (\bibinfo {year} {2012})}\BibitemShut {NoStop}%
\bibitem [{\citenamefont {Wasserman}(2018)}]{wasserman2018topological}%
  \BibitemOpen
  \bibfield  {author} {\bibinfo {author} {\bibfnamefont {L.}~\bibnamefont
  {Wasserman}},\ }\href@noop {} {\bibfield  {journal} {\bibinfo  {journal}
  {Annual Review of Statistics and Its Application}\ }\textbf {\bibinfo
  {volume} {5}},\ \bibinfo {pages} {501} (\bibinfo {year} {2018})}\BibitemShut
  {NoStop}%
\bibitem [{\citenamefont {Nakahara}(2003)}]{nakahara2003geometry}%
  \BibitemOpen
  \bibfield  {author} {\bibinfo {author} {\bibfnamefont {M.}~\bibnamefont
  {Nakahara}},\ }\href@noop {} {\emph {\bibinfo {title} {Geometry, topology and
  physics}}}\ (\bibinfo  {publisher} {CRC Press},\ \bibinfo {year}
  {2003})\BibitemShut {NoStop}%
\bibitem [{\citenamefont {Munkres}(2018)}]{munkres2018elements}%
  \BibitemOpen
  \bibfield  {author} {\bibinfo {author} {\bibfnamefont {J.~R.}\ \bibnamefont
  {Munkres}},\ }\href@noop {} {\emph {\bibinfo {title} {Elements of algebraic
  topology}}}\ (\bibinfo  {publisher} {CRC Press},\ \bibinfo {year}
  {2018})\BibitemShut {NoStop}%
\bibitem [{\citenamefont {Hatcher}(2005)}]{hatcher2005algebraic}%
  \BibitemOpen
  \bibfield  {author} {\bibinfo {author} {\bibfnamefont {A.}~\bibnamefont
  {Hatcher}},\ }\href@noop {} {\emph {\bibinfo {title} {Algebraic topology}}}\
  (\bibinfo  {publisher} {Cambridge University Press},\ \bibinfo {year}
  {2005})\BibitemShut {NoStop}%
\bibitem [{\citenamefont {Edelsbrunner}\ and\ \citenamefont
  {Harer}(2010)}]{Edels2010Comp}%
  \BibitemOpen
  \bibfield  {author} {\bibinfo {author} {\bibfnamefont {H.}~\bibnamefont
  {Edelsbrunner}}\ and\ \bibinfo {author} {\bibfnamefont {J.}~\bibnamefont
  {Harer}},\ }\href@noop {} {\emph {\bibinfo {title} {Computational topology:
  an introduction}}}\ (\bibinfo  {publisher} {American Mathematical Soc.},\
  \bibinfo {year} {2010})\BibitemShut {NoStop}%
\bibitem [{\citenamefont {Ghrist}(2008)}]{ghrist2008barcodes}%
  \BibitemOpen
  \bibfield  {author} {\bibinfo {author} {\bibfnamefont {R.}~\bibnamefont
  {Ghrist}},\ }\href@noop {} {\bibfield  {journal} {\bibinfo  {journal}
  {Bulletin of the American Mathematical Society}\ }\textbf {\bibinfo {volume}
  {45}},\ \bibinfo {pages} {61} (\bibinfo {year} {2008})}\BibitemShut {NoStop}%
\bibitem [{\citenamefont {Carlsson}\ \emph {et~al.}(2005)\citenamefont
  {Carlsson}, \citenamefont {Zomorodian}, \citenamefont {Collins},\ and\
  \citenamefont {Guibas}}]{carlsson2005persistence}%
  \BibitemOpen
  \bibfield  {author} {\bibinfo {author} {\bibfnamefont {G.}~\bibnamefont
  {Carlsson}}, \bibinfo {author} {\bibfnamefont {A.}~\bibnamefont
  {Zomorodian}}, \bibinfo {author} {\bibfnamefont {A.}~\bibnamefont {Collins}},
  \ and\ \bibinfo {author} {\bibfnamefont {L.~J.}\ \bibnamefont {Guibas}},\
  }\href@noop {} {\bibfield  {journal} {\bibinfo  {journal} {International
  Journal of Shape Modeling}\ }\textbf {\bibinfo {volume} {11}},\ \bibinfo
  {pages} {149} (\bibinfo {year} {2005})}\BibitemShut {NoStop}%
\bibitem [{\citenamefont {Patel}(2018)}]{patel2018generalized}%
  \BibitemOpen
  \bibfield  {author} {\bibinfo {author} {\bibfnamefont {A.}~\bibnamefont
  {Patel}},\ }\href@noop {} {\bibfield  {journal} {\bibinfo  {journal} {Journal
  of Applied and Computational Topology}\ }\textbf {\bibinfo {volume} {1}},\
  \bibinfo {pages} {397} (\bibinfo {year} {2018})}\BibitemShut {NoStop}%
\bibitem [{\citenamefont {Bubenik}\ and\ \citenamefont
  {D{\l}otko}(2017)}]{bubenik2017persistence}%
  \BibitemOpen
  \bibfield  {author} {\bibinfo {author} {\bibfnamefont {P.}~\bibnamefont
  {Bubenik}}\ and\ \bibinfo {author} {\bibfnamefont {P.}~\bibnamefont
  {D{\l}otko}},\ }\href@noop {} {\bibfield  {journal} {\bibinfo  {journal}
  {Journal of Symbolic Computation}\ }\textbf {\bibinfo {volume} {78}},\
  \bibinfo {pages} {91} (\bibinfo {year} {2017})}\BibitemShut {NoStop}%
\bibitem [{\citenamefont {Adams}\ \emph {et~al.}(2017)\citenamefont {Adams},
  \citenamefont {Emerson}, \citenamefont {Kirby}, \citenamefont {Neville},
  \citenamefont {Peterson}, \citenamefont {Shipman}, \citenamefont
  {Chepushtanova}, \citenamefont {Hanson}, \citenamefont {Motta},\ and\
  \citenamefont {Ziegelmeier}}]{adams2017persistence}%
  \BibitemOpen
  \bibfield  {author} {\bibinfo {author} {\bibfnamefont {H.}~\bibnamefont
  {Adams}}, \bibinfo {author} {\bibfnamefont {T.}~\bibnamefont {Emerson}},
  \bibinfo {author} {\bibfnamefont {M.}~\bibnamefont {Kirby}}, \bibinfo
  {author} {\bibfnamefont {R.}~\bibnamefont {Neville}}, \bibinfo {author}
  {\bibfnamefont {C.}~\bibnamefont {Peterson}}, \bibinfo {author}
  {\bibfnamefont {P.}~\bibnamefont {Shipman}}, \bibinfo {author} {\bibfnamefont
  {S.}~\bibnamefont {Chepushtanova}}, \bibinfo {author} {\bibfnamefont
  {E.}~\bibnamefont {Hanson}}, \bibinfo {author} {\bibfnamefont
  {F.}~\bibnamefont {Motta}}, \ and\ \bibinfo {author} {\bibfnamefont
  {L.}~\bibnamefont {Ziegelmeier}},\ }\href@noop {} {\bibfield  {journal}
  {\bibinfo  {journal} {The Journal of Machine Learning Research}\ }\textbf
  {\bibinfo {volume} {18}},\ \bibinfo {pages} {218} (\bibinfo {year}
  {2017})}\BibitemShut {NoStop}%
\bibitem [{\citenamefont {Chazal}\ and\ \citenamefont
  {Michel}(2017)}]{chazal2017introduction}%
  \BibitemOpen
  \bibfield  {author} {\bibinfo {author} {\bibfnamefont {F.}~\bibnamefont
  {Chazal}}\ and\ \bibinfo {author} {\bibfnamefont {B.}~\bibnamefont
  {Michel}},\ }\href@noop {} {\bibfield  {journal} {\bibinfo  {journal} {arXiv
  preprint arXiv:1710.04019}\ } (\bibinfo {year} {2017})}\BibitemShut {NoStop}%
\bibitem [{\citenamefont {Munch}(2017)}]{munch2017user}%
  \BibitemOpen
  \bibfield  {author} {\bibinfo {author} {\bibfnamefont {E.}~\bibnamefont
  {Munch}},\ }\href@noop {} {\bibfield  {journal} {\bibinfo  {journal} {Journal
  of Learning Analytics}\ }\textbf {\bibinfo {volume} {4}},\ \bibinfo {pages}
  {47} (\bibinfo {year} {2017})}\BibitemShut {NoStop}%
\bibitem [{\citenamefont {Serrano}\ \emph {et~al.}(2020)\citenamefont
  {Serrano}, \citenamefont {Hern{\'a}ndez-Serrano},\ and\ \citenamefont
  {G{\'o}mez}}]{serrano2020simplicial}%
  \BibitemOpen
  \bibfield  {author} {\bibinfo {author} {\bibfnamefont {D.~H.}\ \bibnamefont
  {Serrano}}, \bibinfo {author} {\bibfnamefont {J.}~\bibnamefont
  {Hern{\'a}ndez-Serrano}}, \ and\ \bibinfo {author} {\bibfnamefont {D.~S.}\
  \bibnamefont {G{\'o}mez}},\ }\href@noop {} {\bibfield  {journal} {\bibinfo
  {journal} {Chaos, Solitons \& Fractals}\ }\textbf {\bibinfo {volume} {137}},\
  \bibinfo {pages} {109839} (\bibinfo {year} {2020})}\BibitemShut {NoStop}%
\bibitem [{\citenamefont {Saggar}\ \emph {et~al.}(2018)\citenamefont {Saggar},
  \citenamefont {Sporns}, \citenamefont {Gonzalez-Castillo}, \citenamefont
  {Bandettini}, \citenamefont {Carlsson}, \citenamefont {Glover},\ and\
  \citenamefont {Reiss}}]{saggar2018towards}%
  \BibitemOpen
  \bibfield  {author} {\bibinfo {author} {\bibfnamefont {M.}~\bibnamefont
  {Saggar}}, \bibinfo {author} {\bibfnamefont {O.}~\bibnamefont {Sporns}},
  \bibinfo {author} {\bibfnamefont {J.}~\bibnamefont {Gonzalez-Castillo}},
  \bibinfo {author} {\bibfnamefont {P.~A.}\ \bibnamefont {Bandettini}},
  \bibinfo {author} {\bibfnamefont {G.}~\bibnamefont {Carlsson}}, \bibinfo
  {author} {\bibfnamefont {G.}~\bibnamefont {Glover}}, \ and\ \bibinfo {author}
  {\bibfnamefont {A.~L.}\ \bibnamefont {Reiss}},\ }\href@noop {} {\bibfield
  {journal} {\bibinfo  {journal} {Nature communications}\ }\textbf {\bibinfo
  {volume} {9}},\ \bibinfo {pages} {1} (\bibinfo {year} {2018})}\BibitemShut
  {NoStop}%
\bibitem [{\citenamefont {Horak}\ \emph {et~al.}(2009)\citenamefont {Horak},
  \citenamefont {Maleti{\'c}},\ and\ \citenamefont
  {Rajkovi{\'c}}}]{horak2009persistent}%
  \BibitemOpen
  \bibfield  {author} {\bibinfo {author} {\bibfnamefont {D.}~\bibnamefont
  {Horak}}, \bibinfo {author} {\bibfnamefont {S.}~\bibnamefont {Maleti{\'c}}},
  \ and\ \bibinfo {author} {\bibfnamefont {M.}~\bibnamefont {Rajkovi{\'c}}},\
  }\href@noop {} {\bibfield  {journal} {\bibinfo  {journal} {Journal of
  Statistical Mechanics: Theory and Experiment}\ }\textbf {\bibinfo {volume}
  {2009}},\ \bibinfo {pages} {P03034} (\bibinfo {year} {2009})}\BibitemShut
  {NoStop}%
\bibitem [{\citenamefont {Topaz}\ \emph {et~al.}(2015)\citenamefont {Topaz},
  \citenamefont {Ziegelmeier},\ and\ \citenamefont
  {Halverson}}]{topaz2015topological}%
  \BibitemOpen
  \bibfield  {author} {\bibinfo {author} {\bibfnamefont {C.~M.}\ \bibnamefont
  {Topaz}}, \bibinfo {author} {\bibfnamefont {L.}~\bibnamefont {Ziegelmeier}},
  \ and\ \bibinfo {author} {\bibfnamefont {T.}~\bibnamefont {Halverson}},\
  }\href@noop {} {\bibfield  {journal} {\bibinfo  {journal} {PloS one}\
  }\textbf {\bibinfo {volume} {10}},\ \bibinfo {pages} {e0126383} (\bibinfo
  {year} {2015})}\BibitemShut {NoStop}%
\bibitem [{\citenamefont {Sizemore}\ \emph {et~al.}(2019)\citenamefont
  {Sizemore}, \citenamefont {Phillips-Cremins}, \citenamefont {Ghrist},\ and\
  \citenamefont {Bassett}}]{sizemore2019importance}%
  \BibitemOpen
  \bibfield  {author} {\bibinfo {author} {\bibfnamefont {A.~E.}\ \bibnamefont
  {Sizemore}}, \bibinfo {author} {\bibfnamefont {J.~E.}\ \bibnamefont
  {Phillips-Cremins}}, \bibinfo {author} {\bibfnamefont {R.}~\bibnamefont
  {Ghrist}}, \ and\ \bibinfo {author} {\bibfnamefont {D.~S.}\ \bibnamefont
  {Bassett}},\ }\href@noop {} {\bibfield  {journal} {\bibinfo  {journal}
  {Network Neuroscience}\ }\textbf {\bibinfo {volume} {3}},\ \bibinfo {pages}
  {656} (\bibinfo {year} {2019})}\BibitemShut {NoStop}%
\bibitem [{\citenamefont {Pranav}\ \emph {et~al.}(2017)\citenamefont {Pranav},
  \citenamefont {Edelsbrunner}, \citenamefont {Van~de Weygaert}, \citenamefont
  {Vegter}, \citenamefont {Kerber}, \citenamefont {Jones},\ and\ \citenamefont
  {Wintraecken}}]{pranav2017topology}%
  \BibitemOpen
  \bibfield  {author} {\bibinfo {author} {\bibfnamefont {P.}~\bibnamefont
  {Pranav}}, \bibinfo {author} {\bibfnamefont {H.}~\bibnamefont
  {Edelsbrunner}}, \bibinfo {author} {\bibfnamefont {R.}~\bibnamefont {Van~de
  Weygaert}}, \bibinfo {author} {\bibfnamefont {G.}~\bibnamefont {Vegter}},
  \bibinfo {author} {\bibfnamefont {M.}~\bibnamefont {Kerber}}, \bibinfo
  {author} {\bibfnamefont {B.~J.}\ \bibnamefont {Jones}}, \ and\ \bibinfo
  {author} {\bibfnamefont {M.}~\bibnamefont {Wintraecken}},\ }\href@noop {}
  {\bibfield  {journal} {\bibinfo  {journal} {Monthly Notices of the Royal
  Astronomical Society}\ }\textbf {\bibinfo {volume} {465}},\ \bibinfo {pages}
  {4281} (\bibinfo {year} {2017})}\BibitemShut {NoStop}%
\bibitem [{\citenamefont {Jaquette}\ and\ \citenamefont
  {Schweinhart}(2020)}]{jaquette2020fractal}%
  \BibitemOpen
  \bibfield  {author} {\bibinfo {author} {\bibfnamefont {J.}~\bibnamefont
  {Jaquette}}\ and\ \bibinfo {author} {\bibfnamefont {B.}~\bibnamefont
  {Schweinhart}},\ }\href@noop {} {\bibfield  {journal} {\bibinfo  {journal}
  {Communications in Nonlinear Science and Numerical Simulation}\ }\textbf
  {\bibinfo {volume} {84}},\ \bibinfo {pages} {105163} (\bibinfo {year}
  {2020})}\BibitemShut {NoStop}%
\bibitem [{\citenamefont {Speidel}\ \emph {et~al.}(2018)\citenamefont
  {Speidel}, \citenamefont {Harrington}, \citenamefont {Chapman},\ and\
  \citenamefont {Porter}}]{speidel2018topological}%
  \BibitemOpen
  \bibfield  {author} {\bibinfo {author} {\bibfnamefont {L.}~\bibnamefont
  {Speidel}}, \bibinfo {author} {\bibfnamefont {H.~A.}\ \bibnamefont
  {Harrington}}, \bibinfo {author} {\bibfnamefont {S.~J.}\ \bibnamefont
  {Chapman}}, \ and\ \bibinfo {author} {\bibfnamefont {M.~A.}\ \bibnamefont
  {Porter}},\ }\href@noop {} {\bibfield  {journal} {\bibinfo  {journal}
  {Physical Review E}\ }\textbf {\bibinfo {volume} {98}},\ \bibinfo {pages}
  {012318} (\bibinfo {year} {2018})}\BibitemShut {NoStop}%
\bibitem [{\citenamefont {Salnikov}\ \emph {et~al.}(2018)\citenamefont
  {Salnikov}, \citenamefont {Cassese},\ and\ \citenamefont
  {Lambiotte}}]{salnikov2018simplicial}%
  \BibitemOpen
  \bibfield  {author} {\bibinfo {author} {\bibfnamefont {V.}~\bibnamefont
  {Salnikov}}, \bibinfo {author} {\bibfnamefont {D.}~\bibnamefont {Cassese}}, \
  and\ \bibinfo {author} {\bibfnamefont {R.}~\bibnamefont {Lambiotte}},\
  }\href@noop {} {\bibfield  {journal} {\bibinfo  {journal} {European Journal
  of Physics}\ }\textbf {\bibinfo {volume} {40}},\ \bibinfo {pages} {014001}
  (\bibinfo {year} {2018})}\BibitemShut {NoStop}%
\bibitem [{\citenamefont {Bobrowski}\ and\ \citenamefont
  {Skraba}(2020)}]{bobrowski2020homological}%
  \BibitemOpen
  \bibfield  {author} {\bibinfo {author} {\bibfnamefont {O.}~\bibnamefont
  {Bobrowski}}\ and\ \bibinfo {author} {\bibfnamefont {P.}~\bibnamefont
  {Skraba}},\ }\href@noop {} {\bibfield  {journal} {\bibinfo  {journal}
  {Physical Review E}\ }\textbf {\bibinfo {volume} {101}},\ \bibinfo {pages}
  {032304} (\bibinfo {year} {2020})}\BibitemShut {NoStop}%
\bibitem [{\citenamefont {Tran}\ \emph {et~al.}(2021)\citenamefont {Tran},
  \citenamefont {Chen},\ and\ \citenamefont {Hasegawa}}]{tran2021topological}%
  \BibitemOpen
  \bibfield  {author} {\bibinfo {author} {\bibfnamefont {Q.~H.}\ \bibnamefont
  {Tran}}, \bibinfo {author} {\bibfnamefont {M.}~\bibnamefont {Chen}}, \ and\
  \bibinfo {author} {\bibfnamefont {Y.}~\bibnamefont {Hasegawa}},\ }\href@noop
  {} {\bibfield  {journal} {\bibinfo  {journal} {Physical Review E}\ }\textbf
  {\bibinfo {volume} {103}},\ \bibinfo {pages} {052127} (\bibinfo {year}
  {2021})}\BibitemShut {NoStop}%
\bibitem [{\citenamefont {Olsthoorn}\ \emph {et~al.}(2020)\citenamefont
  {Olsthoorn}, \citenamefont {Hellsvik},\ and\ \citenamefont
  {Balatsky}}]{olsthoorn2020finding}%
  \BibitemOpen
  \bibfield  {author} {\bibinfo {author} {\bibfnamefont {B.}~\bibnamefont
  {Olsthoorn}}, \bibinfo {author} {\bibfnamefont {J.}~\bibnamefont {Hellsvik}},
  \ and\ \bibinfo {author} {\bibfnamefont {A.~V.}\ \bibnamefont {Balatsky}},\
  }\href@noop {} {\bibfield  {journal} {\bibinfo  {journal} {Physical Review
  Research}\ }\textbf {\bibinfo {volume} {2}},\ \bibinfo {pages} {043308}
  (\bibinfo {year} {2020})}\BibitemShut {NoStop}%
\bibitem [{\citenamefont {Hurst}(1951)}]{hurst1951long}%
  \BibitemOpen
  \bibfield  {author} {\bibinfo {author} {\bibfnamefont {H.~E.}\ \bibnamefont
  {Hurst}},\ }\href@noop {} {\bibfield  {journal} {\bibinfo  {journal}
  {Transactions of the American society of civil engineers}\ }\textbf {\bibinfo
  {volume} {116}},\ \bibinfo {pages} {770} (\bibinfo {year}
  {1951})}\BibitemShut {NoStop}%
\bibitem [{\citenamefont {Mandelbrot}\ and\ \citenamefont
  {Van~Ness}(1968)}]{mandelbrot1968fractional}%
  \BibitemOpen
  \bibfield  {author} {\bibinfo {author} {\bibfnamefont {B.~B.}\ \bibnamefont
  {Mandelbrot}}\ and\ \bibinfo {author} {\bibfnamefont {J.~W.}\ \bibnamefont
  {Van~Ness}},\ }\href@noop {} {\bibfield  {journal} {\bibinfo  {journal} {SIAM
  review}\ }\textbf {\bibinfo {volume} {10}},\ \bibinfo {pages} {422} (\bibinfo
  {year} {1968})}\BibitemShut {NoStop}%
\bibitem [{\citenamefont {Eghdami}\ \emph {et~al.}(2018)\citenamefont
  {Eghdami}, \citenamefont {Panahi},\ and\ \citenamefont
  {Movahed}}]{eghdami2018multifractal}%
  \BibitemOpen
  \bibfield  {author} {\bibinfo {author} {\bibfnamefont {I.}~\bibnamefont
  {Eghdami}}, \bibinfo {author} {\bibfnamefont {H.}~\bibnamefont {Panahi}}, \
  and\ \bibinfo {author} {\bibfnamefont {S.}~\bibnamefont {Movahed}},\
  }\href@noop {} {\bibfield  {journal} {\bibinfo  {journal} {The Astrophysical
  Journal}\ }\textbf {\bibinfo {volume} {864}},\ \bibinfo {pages} {162}
  (\bibinfo {year} {2018})}\BibitemShut {NoStop}%
\bibitem [{\citenamefont {Jiang}\ \emph {et~al.}(2019)\citenamefont {Jiang},
  \citenamefont {Xie}, \citenamefont {Zhou},\ and\ \citenamefont
  {Sornette}}]{jiang2019multifractal}%
  \BibitemOpen
  \bibfield  {author} {\bibinfo {author} {\bibfnamefont {Z.-Q.}\ \bibnamefont
  {Jiang}}, \bibinfo {author} {\bibfnamefont {W.-J.}\ \bibnamefont {Xie}},
  \bibinfo {author} {\bibfnamefont {W.-X.}\ \bibnamefont {Zhou}}, \ and\
  \bibinfo {author} {\bibfnamefont {D.}~\bibnamefont {Sornette}},\ }\href@noop
  {} {\bibfield  {journal} {\bibinfo  {journal} {Reports on Progress in
  Physics}\ }\textbf {\bibinfo {volume} {82}},\ \bibinfo {pages} {125901}
  (\bibinfo {year} {2019})}\BibitemShut {NoStop}%
\bibitem [{\citenamefont {Peng}\ \emph {et~al.}(1994)\citenamefont {Peng},
  \citenamefont {Buldyrev}, \citenamefont {Havlin}, \citenamefont {Simons},
  \citenamefont {Stanley},\ and\ \citenamefont {Goldberger}}]{peng1994mosaic}%
  \BibitemOpen
  \bibfield  {author} {\bibinfo {author} {\bibfnamefont {C.-K.}\ \bibnamefont
  {Peng}}, \bibinfo {author} {\bibfnamefont {S.~V.}\ \bibnamefont {Buldyrev}},
  \bibinfo {author} {\bibfnamefont {S.}~\bibnamefont {Havlin}}, \bibinfo
  {author} {\bibfnamefont {M.}~\bibnamefont {Simons}}, \bibinfo {author}
  {\bibfnamefont {H.~E.}\ \bibnamefont {Stanley}}, \ and\ \bibinfo {author}
  {\bibfnamefont {A.~L.}\ \bibnamefont {Goldberger}},\ }\href@noop {}
  {\bibfield  {journal} {\bibinfo  {journal} {Physical review e}\ }\textbf
  {\bibinfo {volume} {49}},\ \bibinfo {pages} {1685} (\bibinfo {year}
  {1994})}\BibitemShut {NoStop}%
\bibitem [{\citenamefont {Peng}\ \emph {et~al.}(1995)\citenamefont {Peng},
  \citenamefont {Havlin}, \citenamefont {Stanley},\ and\ \citenamefont
  {Goldberger}}]{Peng95}%
  \BibitemOpen
  \bibfield  {author} {\bibinfo {author} {\bibfnamefont {C.-K.}\ \bibnamefont
  {Peng}}, \bibinfo {author} {\bibfnamefont {S.}~\bibnamefont {Havlin}},
  \bibinfo {author} {\bibfnamefont {H.~E.}\ \bibnamefont {Stanley}}, \ and\
  \bibinfo {author} {\bibfnamefont {A.~L.}\ \bibnamefont {Goldberger}},\
  }\href@noop {} {\bibfield  {journal} {\bibinfo  {journal} {Chaos: an
  interdisciplinary journal of nonlinear science}\ }\textbf {\bibinfo {volume}
  {5}},\ \bibinfo {pages} {82} (\bibinfo {year} {1995})}\BibitemShut {NoStop}%
\bibitem [{\citenamefont {Kantelhardt}\ \emph {et~al.}(2002)\citenamefont
  {Kantelhardt}, \citenamefont {Zschiegner}, \citenamefont {Koscielny-Bunde},
  \citenamefont {Havlin}, \citenamefont {Bunde},\ and\ \citenamefont
  {Stanley}}]{Kantelhardt}%
  \BibitemOpen
  \bibfield  {author} {\bibinfo {author} {\bibfnamefont {J.~W.}\ \bibnamefont
  {Kantelhardt}}, \bibinfo {author} {\bibfnamefont {S.~A.}\ \bibnamefont
  {Zschiegner}}, \bibinfo {author} {\bibfnamefont {E.}~\bibnamefont
  {Koscielny-Bunde}}, \bibinfo {author} {\bibfnamefont {S.}~\bibnamefont
  {Havlin}}, \bibinfo {author} {\bibfnamefont {A.}~\bibnamefont {Bunde}}, \
  and\ \bibinfo {author} {\bibfnamefont {H.~E.}\ \bibnamefont {Stanley}},\
  }\href@noop {} {\bibfield  {journal} {\bibinfo  {journal} {Physica A:
  Statistical Mechanics and its Applications}\ }\textbf {\bibinfo {volume}
  {316}},\ \bibinfo {pages} {87} (\bibinfo {year} {2002})}\BibitemShut
  {NoStop}%
\bibitem [{\citenamefont {Zhou}\ \emph {et~al.}(2008)\citenamefont {Zhou} \emph
  {et~al.}}]{mf-dxa}%
  \BibitemOpen
  \bibfield  {author} {\bibinfo {author} {\bibfnamefont {W.-X.}\ \bibnamefont
  {Zhou}} \emph {et~al.},\ }\href@noop {} {\bibfield  {journal} {\bibinfo
  {journal} {Physical Review E}\ }\textbf {\bibinfo {volume} {77}},\ \bibinfo
  {pages} {066211} (\bibinfo {year} {2008})}\BibitemShut {NoStop}%
\bibitem [{\citenamefont {Hu}\ \emph {et~al.}(2001)\citenamefont {Hu},
  \citenamefont {Ivanov}, \citenamefont {Chen}, \citenamefont {Carpena},\ and\
  \citenamefont {Stanley}}]{kunhu}%
  \BibitemOpen
  \bibfield  {author} {\bibinfo {author} {\bibfnamefont {K.}~\bibnamefont
  {Hu}}, \bibinfo {author} {\bibfnamefont {P.~C.}\ \bibnamefont {Ivanov}},
  \bibinfo {author} {\bibfnamefont {Z.}~\bibnamefont {Chen}}, \bibinfo {author}
  {\bibfnamefont {P.}~\bibnamefont {Carpena}}, \ and\ \bibinfo {author}
  {\bibfnamefont {H.~E.}\ \bibnamefont {Stanley}},\ }\href@noop {} {\bibfield
  {journal} {\bibinfo  {journal} {Physical Review E}\ }\textbf {\bibinfo
  {volume} {64}},\ \bibinfo {pages} {011114} (\bibinfo {year}
  {2001})}\BibitemShut {NoStop}%
\bibitem [{\citenamefont {Chen}\ \emph {et~al.}(2002)\citenamefont {Chen},
  \citenamefont {Ivanov}, \citenamefont {Hu},\ and\ \citenamefont
  {Stanley}}]{trend2}%
  \BibitemOpen
  \bibfield  {author} {\bibinfo {author} {\bibfnamefont {Z.}~\bibnamefont
  {Chen}}, \bibinfo {author} {\bibfnamefont {P.~C.}\ \bibnamefont {Ivanov}},
  \bibinfo {author} {\bibfnamefont {K.}~\bibnamefont {Hu}}, \ and\ \bibinfo
  {author} {\bibfnamefont {H.~E.}\ \bibnamefont {Stanley}},\ }\href@noop {}
  {\bibfield  {journal} {\bibinfo  {journal} {Physical review E}\ }\textbf
  {\bibinfo {volume} {65}},\ \bibinfo {pages} {041107} (\bibinfo {year}
  {2002})}\BibitemShut {NoStop}%
\bibitem [{\citenamefont {Kantelhardt}\ \emph {et~al.}(2001)\citenamefont
  {Kantelhardt}, \citenamefont {Koscielny-Bunde}, \citenamefont {Rego},
  \citenamefont {Havlin},\ and\ \citenamefont {Bunde}}]{physa}%
  \BibitemOpen
  \bibfield  {author} {\bibinfo {author} {\bibfnamefont {J.~W.}\ \bibnamefont
  {Kantelhardt}}, \bibinfo {author} {\bibfnamefont {E.}~\bibnamefont
  {Koscielny-Bunde}}, \bibinfo {author} {\bibfnamefont {H.~H.}\ \bibnamefont
  {Rego}}, \bibinfo {author} {\bibfnamefont {S.}~\bibnamefont {Havlin}}, \ and\
  \bibinfo {author} {\bibfnamefont {A.}~\bibnamefont {Bunde}},\ }\href@noop {}
  {\bibfield  {journal} {\bibinfo  {journal} {Physica A: Statistical Mechanics
  and its Applications}\ }\textbf {\bibinfo {volume} {295}},\ \bibinfo {pages}
  {441} (\bibinfo {year} {2001})}\BibitemShut {NoStop}%
\bibitem [{\citenamefont {Nagarajan}\ and\ \citenamefont
  {Kavasseri}(2005{\natexlab{a}})}]{trend3}%
  \BibitemOpen
  \bibfield  {author} {\bibinfo {author} {\bibfnamefont {R.}~\bibnamefont
  {Nagarajan}}\ and\ \bibinfo {author} {\bibfnamefont {R.~G.}\ \bibnamefont
  {Kavasseri}},\ }\href@noop {} {\bibfield  {journal} {\bibinfo  {journal}
  {Chaos, Solitons \& Fractals}\ }\textbf {\bibinfo {volume} {26}},\ \bibinfo
  {pages} {777} (\bibinfo {year} {2005}{\natexlab{a}})}\BibitemShut {NoStop}%
\bibitem [{\citenamefont {Xu}\ \emph {et~al.}(2005)\citenamefont {Xu},
  \citenamefont {Ivanov}, \citenamefont {Hu}, \citenamefont {Chen},
  \citenamefont {Carbone},\ and\ \citenamefont {Stanley}}]{xu2005quantifying}%
  \BibitemOpen
  \bibfield  {author} {\bibinfo {author} {\bibfnamefont {L.}~\bibnamefont
  {Xu}}, \bibinfo {author} {\bibfnamefont {P.~C.}\ \bibnamefont {Ivanov}},
  \bibinfo {author} {\bibfnamefont {K.}~\bibnamefont {Hu}}, \bibinfo {author}
  {\bibfnamefont {Z.}~\bibnamefont {Chen}}, \bibinfo {author} {\bibfnamefont
  {A.}~\bibnamefont {Carbone}}, \ and\ \bibinfo {author} {\bibfnamefont
  {H.~E.}\ \bibnamefont {Stanley}},\ }\href@noop {} {\bibfield  {journal}
  {\bibinfo  {journal} {Physical Review E}\ }\textbf {\bibinfo {volume} {71}},\
  \bibinfo {pages} {051101} (\bibinfo {year} {2005})}\BibitemShut {NoStop}%
\bibitem [{\citenamefont {Zou}\ \emph {et~al.}(2019)\citenamefont {Zou},
  \citenamefont {Donner}, \citenamefont {Marwan}, \citenamefont {Donges},\ and\
  \citenamefont {Kurths}}]{zou2019complex}%
  \BibitemOpen
  \bibfield  {author} {\bibinfo {author} {\bibfnamefont {Y.}~\bibnamefont
  {Zou}}, \bibinfo {author} {\bibfnamefont {R.~V.}\ \bibnamefont {Donner}},
  \bibinfo {author} {\bibfnamefont {N.}~\bibnamefont {Marwan}}, \bibinfo
  {author} {\bibfnamefont {J.~F.}\ \bibnamefont {Donges}}, \ and\ \bibinfo
  {author} {\bibfnamefont {J.}~\bibnamefont {Kurths}},\ }\href@noop {}
  {\bibfield  {journal} {\bibinfo  {journal} {Physics Reports}\ }\textbf
  {\bibinfo {volume} {787}},\ \bibinfo {pages} {1} (\bibinfo {year}
  {2019})}\BibitemShut {NoStop}%
\bibitem [{\citenamefont {Lacasa}\ \emph {et~al.}(2008)\citenamefont {Lacasa},
  \citenamefont {Luque}, \citenamefont {Ballesteros}, \citenamefont {Luque},\
  and\ \citenamefont {Nuno}}]{lacasa2008time}%
  \BibitemOpen
  \bibfield  {author} {\bibinfo {author} {\bibfnamefont {L.}~\bibnamefont
  {Lacasa}}, \bibinfo {author} {\bibfnamefont {B.}~\bibnamefont {Luque}},
  \bibinfo {author} {\bibfnamefont {F.}~\bibnamefont {Ballesteros}}, \bibinfo
  {author} {\bibfnamefont {J.}~\bibnamefont {Luque}}, \ and\ \bibinfo {author}
  {\bibfnamefont {J.~C.}\ \bibnamefont {Nuno}},\ }\href@noop {} {\bibfield
  {journal} {\bibinfo  {journal} {Proceedings of the National Academy of
  Sciences}\ }\textbf {\bibinfo {volume} {105}},\ \bibinfo {pages} {4972}
  (\bibinfo {year} {2008})}\BibitemShut {NoStop}%
\bibitem [{\citenamefont {Lacasa}\ \emph {et~al.}(2009)\citenamefont {Lacasa},
  \citenamefont {Luque}, \citenamefont {Luque},\ and\ \citenamefont
  {Nuno}}]{lacasa2009visibility}%
  \BibitemOpen
  \bibfield  {author} {\bibinfo {author} {\bibfnamefont {L.}~\bibnamefont
  {Lacasa}}, \bibinfo {author} {\bibfnamefont {B.}~\bibnamefont {Luque}},
  \bibinfo {author} {\bibfnamefont {J.}~\bibnamefont {Luque}}, \ and\ \bibinfo
  {author} {\bibfnamefont {J.~C.}\ \bibnamefont {Nuno}},\ }\href@noop {}
  {\bibfield  {journal} {\bibinfo  {journal} {EPL (Europhysics Letters)}\
  }\textbf {\bibinfo {volume} {86}},\ \bibinfo {pages} {30001} (\bibinfo {year}
  {2009})}\BibitemShut {NoStop}%
\bibitem [{\citenamefont {Nagarajan}\ and\ \citenamefont
  {Kavasseri}(2005{\natexlab{b}})}]{nagarajan2005minimizing}%
  \BibitemOpen
  \bibfield  {author} {\bibinfo {author} {\bibfnamefont {R.}~\bibnamefont
  {Nagarajan}}\ and\ \bibinfo {author} {\bibfnamefont {R.~G.}\ \bibnamefont
  {Kavasseri}},\ }\href@noop {} {\bibfield  {journal} {\bibinfo  {journal}
  {Physica A: Statistical Mechanics and its Applications}\ }\textbf {\bibinfo
  {volume} {354}},\ \bibinfo {pages} {182} (\bibinfo {year}
  {2005}{\natexlab{b}})}\BibitemShut {NoStop}%
\bibitem [{\citenamefont {Hu}\ \emph {et~al.}(2009)\citenamefont {Hu},
  \citenamefont {Gao},\ and\ \citenamefont {Wang}}]{hu2009multifractal}%
  \BibitemOpen
  \bibfield  {author} {\bibinfo {author} {\bibfnamefont {J.}~\bibnamefont
  {Hu}}, \bibinfo {author} {\bibfnamefont {J.}~\bibnamefont {Gao}}, \ and\
  \bibinfo {author} {\bibfnamefont {X.}~\bibnamefont {Wang}},\ }\href@noop {}
  {\bibfield  {journal} {\bibinfo  {journal} {Journal of Statistical Mechanics:
  Theory and Experiment}\ }\textbf {\bibinfo {volume} {2009}},\ \bibinfo
  {pages} {P02066} (\bibinfo {year} {2009})}\BibitemShut {NoStop}%
\bibitem [{\citenamefont {Wu}\ \emph {et~al.}(2007)\citenamefont {Wu},
  \citenamefont {Huang}, \citenamefont {Long},\ and\ \citenamefont
  {Peng}}]{wu2007trend}%
  \BibitemOpen
  \bibfield  {author} {\bibinfo {author} {\bibfnamefont {Z.}~\bibnamefont
  {Wu}}, \bibinfo {author} {\bibfnamefont {N.~E.}\ \bibnamefont {Huang}},
  \bibinfo {author} {\bibfnamefont {S.~R.}\ \bibnamefont {Long}}, \ and\
  \bibinfo {author} {\bibfnamefont {C.-K.}\ \bibnamefont {Peng}},\ }\href@noop
  {} {\bibfield  {journal} {\bibinfo  {journal} {Proceedings of the National
  Academy of Sciences}\ }\textbf {\bibinfo {volume} {104}},\ \bibinfo {pages}
  {14889} (\bibinfo {year} {2007})}\BibitemShut {NoStop}%
\bibitem [{\citenamefont {Costa}\ \emph {et~al.}(2011)\citenamefont {Costa},
  \citenamefont {Oliveira~Jr}, \citenamefont {Travieso}, \citenamefont
  {Rodrigues}, \citenamefont {Villas~Boas}, \citenamefont {Antiqueira},
  \citenamefont {Viana},\ and\ \citenamefont
  {Correa~Rocha}}]{costa2011analyzing}%
  \BibitemOpen
  \bibfield  {author} {\bibinfo {author} {\bibfnamefont {L.~d.~F.}\
  \bibnamefont {Costa}}, \bibinfo {author} {\bibfnamefont {O.~N.}\ \bibnamefont
  {Oliveira~Jr}}, \bibinfo {author} {\bibfnamefont {G.}~\bibnamefont
  {Travieso}}, \bibinfo {author} {\bibfnamefont {F.~A.}\ \bibnamefont
  {Rodrigues}}, \bibinfo {author} {\bibfnamefont {P.~R.}\ \bibnamefont
  {Villas~Boas}}, \bibinfo {author} {\bibfnamefont {L.}~\bibnamefont
  {Antiqueira}}, \bibinfo {author} {\bibfnamefont {M.~P.}\ \bibnamefont
  {Viana}}, \ and\ \bibinfo {author} {\bibfnamefont {L.~E.}\ \bibnamefont
  {Correa~Rocha}},\ }\href@noop {} {\bibfield  {journal} {\bibinfo  {journal}
  {Advances in Physics}\ }\textbf {\bibinfo {volume} {60}},\ \bibinfo {pages}
  {329} (\bibinfo {year} {2011})}\BibitemShut {NoStop}%
\bibitem [{\citenamefont {Gon{\c{c}}alves}\ \emph {et~al.}(2016)\citenamefont
  {Gon{\c{c}}alves}, \citenamefont {Carpi}, \citenamefont {Rosso},\ and\
  \citenamefont {Ravetti}}]{gonccalves2016time}%
  \BibitemOpen
  \bibfield  {author} {\bibinfo {author} {\bibfnamefont {B.~A.}\ \bibnamefont
  {Gon{\c{c}}alves}}, \bibinfo {author} {\bibfnamefont {L.}~\bibnamefont
  {Carpi}}, \bibinfo {author} {\bibfnamefont {O.~A.}\ \bibnamefont {Rosso}}, \
  and\ \bibinfo {author} {\bibfnamefont {M.~G.}\ \bibnamefont {Ravetti}},\
  }\href@noop {} {\bibfield  {journal} {\bibinfo  {journal} {Physica A:
  Statistical Mechanics and its Applications}\ }\textbf {\bibinfo {volume}
  {464}},\ \bibinfo {pages} {93} (\bibinfo {year} {2016})}\BibitemShut
  {NoStop}%
\bibitem [{\citenamefont {Gao}\ \emph {et~al.}(2016)\citenamefont {Gao},
  \citenamefont {Cai}, \citenamefont {Yang}, \citenamefont {Dang},\ and\
  \citenamefont {Zhang}}]{gao2016multiscale}%
  \BibitemOpen
  \bibfield  {author} {\bibinfo {author} {\bibfnamefont {Z.-K.}\ \bibnamefont
  {Gao}}, \bibinfo {author} {\bibfnamefont {Q.}~\bibnamefont {Cai}}, \bibinfo
  {author} {\bibfnamefont {Y.-X.}\ \bibnamefont {Yang}}, \bibinfo {author}
  {\bibfnamefont {W.-D.}\ \bibnamefont {Dang}}, \ and\ \bibinfo {author}
  {\bibfnamefont {S.-S.}\ \bibnamefont {Zhang}},\ }\href@noop {} {\bibfield
  {journal} {\bibinfo  {journal} {Scientific reports}\ }\textbf {\bibinfo
  {volume} {6}},\ \bibinfo {pages} {35622} (\bibinfo {year}
  {2016})}\BibitemShut {NoStop}%
\bibitem [{\citenamefont {Xie}\ and\ \citenamefont
  {Zhou}(2011)}]{xie2011horizontal}%
  \BibitemOpen
  \bibfield  {author} {\bibinfo {author} {\bibfnamefont {W.-J.}\ \bibnamefont
  {Xie}}\ and\ \bibinfo {author} {\bibfnamefont {W.-X.}\ \bibnamefont {Zhou}},\
  }\href@noop {} {\bibfield  {journal} {\bibinfo  {journal} {Physica A:
  Statistical Mechanics and its Applications}\ }\textbf {\bibinfo {volume}
  {390}},\ \bibinfo {pages} {3592} (\bibinfo {year} {2011})}\BibitemShut
  {NoStop}%
\bibitem [{\citenamefont {Zheng}\ \emph {et~al.}(2021)\citenamefont {Zheng},
  \citenamefont {Domanskyi}, \citenamefont {Piermarocchi},\ and\ \citenamefont
  {Mias}}]{zheng2020visibility}%
  \BibitemOpen
  \bibfield  {author} {\bibinfo {author} {\bibfnamefont {M.}~\bibnamefont
  {Zheng}}, \bibinfo {author} {\bibfnamefont {S.}~\bibnamefont {Domanskyi}},
  \bibinfo {author} {\bibfnamefont {C.}~\bibnamefont {Piermarocchi}}, \ and\
  \bibinfo {author} {\bibfnamefont {G.~I.}\ \bibnamefont {Mias}},\ }\href
  {\doibase 10.1038/s41598-021-84838-x} {\bibfield  {journal} {\bibinfo
  {journal} {Scientific Reports}\ }\textbf {\bibinfo {volume} {11}},\ \bibinfo
  {pages} {5623} (\bibinfo {year} {2021})}\BibitemShut {NoStop}%
\bibitem [{\citenamefont {Yang}\ \emph {et~al.}(2009)\citenamefont {Yang},
  \citenamefont {Wang}, \citenamefont {Yang},\ and\ \citenamefont
  {Mang}}]{yang2009visibility}%
  \BibitemOpen
  \bibfield  {author} {\bibinfo {author} {\bibfnamefont {Y.}~\bibnamefont
  {Yang}}, \bibinfo {author} {\bibfnamefont {J.}~\bibnamefont {Wang}}, \bibinfo
  {author} {\bibfnamefont {H.}~\bibnamefont {Yang}}, \ and\ \bibinfo {author}
  {\bibfnamefont {J.}~\bibnamefont {Mang}},\ }\href@noop {} {\bibfield
  {journal} {\bibinfo  {journal} {Physica A: Statistical Mechanics and its
  Applications}\ }\textbf {\bibinfo {volume} {388}},\ \bibinfo {pages} {4431}
  (\bibinfo {year} {2009})}\BibitemShut {NoStop}%
\bibitem [{\citenamefont {Ahmadlou}\ \emph {et~al.}(2010)\citenamefont
  {Ahmadlou}, \citenamefont {Adeli},\ and\ \citenamefont
  {Adeli}}]{ahmadlou2010new}%
  \BibitemOpen
  \bibfield  {author} {\bibinfo {author} {\bibfnamefont {M.}~\bibnamefont
  {Ahmadlou}}, \bibinfo {author} {\bibfnamefont {H.}~\bibnamefont {Adeli}}, \
  and\ \bibinfo {author} {\bibfnamefont {A.}~\bibnamefont {Adeli}},\
  }\href@noop {} {\bibfield  {journal} {\bibinfo  {journal} {Journal of neural
  transmission}\ }\textbf {\bibinfo {volume} {117}},\ \bibinfo {pages} {1099}
  (\bibinfo {year} {2010})}\BibitemShut {NoStop}%
\bibitem [{\citenamefont {Bianconi}\ and\ \citenamefont
  {Rahmede}(2016)}]{bianconi2016network}%
  \BibitemOpen
  \bibfield  {author} {\bibinfo {author} {\bibfnamefont {G.}~\bibnamefont
  {Bianconi}}\ and\ \bibinfo {author} {\bibfnamefont {C.}~\bibnamefont
  {Rahmede}},\ }\href@noop {} {\bibfield  {journal} {\bibinfo  {journal}
  {Physical Review E}\ }\textbf {\bibinfo {volume} {93}},\ \bibinfo {pages}
  {032315} (\bibinfo {year} {2016})}\BibitemShut {NoStop}%
\bibitem [{\citenamefont {Kovalenko}\ \emph {et~al.}(2021)\citenamefont
  {Kovalenko}, \citenamefont {Sendi{\~n}a-Nadal}, \citenamefont {Khalil},
  \citenamefont {Dainiak}, \citenamefont {Musatov}, \citenamefont
  {Raigorodskii}, \citenamefont {Alfaro-Bittner}, \citenamefont {Barzel},\ and\
  \citenamefont {Boccaletti}}]{kovalenko2021growing}%
  \BibitemOpen
  \bibfield  {author} {\bibinfo {author} {\bibfnamefont {K.}~\bibnamefont
  {Kovalenko}}, \bibinfo {author} {\bibfnamefont {I.}~\bibnamefont
  {Sendi{\~n}a-Nadal}}, \bibinfo {author} {\bibfnamefont {N.}~\bibnamefont
  {Khalil}}, \bibinfo {author} {\bibfnamefont {A.}~\bibnamefont {Dainiak}},
  \bibinfo {author} {\bibfnamefont {D.}~\bibnamefont {Musatov}}, \bibinfo
  {author} {\bibfnamefont {A.~M.}\ \bibnamefont {Raigorodskii}}, \bibinfo
  {author} {\bibfnamefont {K.}~\bibnamefont {Alfaro-Bittner}}, \bibinfo
  {author} {\bibfnamefont {B.}~\bibnamefont {Barzel}}, \ and\ \bibinfo {author}
  {\bibfnamefont {S.}~\bibnamefont {Boccaletti}},\ }\href@noop {} {\bibfield
  {journal} {\bibinfo  {journal} {Communications Physics}\ }\textbf {\bibinfo
  {volume} {4}},\ \bibinfo {pages} {1} (\bibinfo {year} {2021})}\BibitemShut
  {NoStop}%
\bibitem [{\citenamefont {Young}\ \emph {et~al.}(2020)\citenamefont {Young},
  \citenamefont {Petri},\ and\ \citenamefont {Peixoto}}]{young2020hypergraph}%
  \BibitemOpen
  \bibfield  {author} {\bibinfo {author} {\bibfnamefont {J.-G.}\ \bibnamefont
  {Young}}, \bibinfo {author} {\bibfnamefont {G.}~\bibnamefont {Petri}}, \ and\
  \bibinfo {author} {\bibfnamefont {T.~P.}\ \bibnamefont {Peixoto}},\
  }\href@noop {} {\bibfield  {journal} {\bibinfo  {journal} {arXiv preprint
  arXiv:2008.04948}\ } (\bibinfo {year} {2020})}\BibitemShut {NoStop}%
\bibitem [{\citenamefont {Rote}\ and\ \citenamefont
  {Vegter}(2006)}]{rote2006computational}%
  \BibitemOpen
  \bibfield  {author} {\bibinfo {author} {\bibfnamefont {G.}~\bibnamefont
  {Rote}}\ and\ \bibinfo {author} {\bibfnamefont {G.}~\bibnamefont {Vegter}},\
  }in\ \href@noop {} {\emph {\bibinfo {booktitle} {Effective Computational
  Geometry for curves and surfaces}}}\ (\bibinfo  {publisher} {Springer},\
  \bibinfo {year} {2006})\ pp.\ \bibinfo {pages} {277--312}\BibitemShut
  {NoStop}%
\bibitem [{\citenamefont {Zomorodian}(2009)}]{zomorodian2009computational}%
  \BibitemOpen
  \bibfield  {author} {\bibinfo {author} {\bibfnamefont {A.}~\bibnamefont
  {Zomorodian}},\ }\href@noop {} {\bibfield  {journal} {\bibinfo  {journal}
  {Algorithms and theory of computation handbook}\ }\textbf {\bibinfo {volume}
  {2}} (\bibinfo {year} {2009})}\BibitemShut {NoStop}%
\bibitem [{\citenamefont {Hosking}(1984)}]{hosking1984modeling}%
  \BibitemOpen
  \bibfield  {author} {\bibinfo {author} {\bibfnamefont {J.~R.}\ \bibnamefont
  {Hosking}},\ }\href@noop {} {\bibfield  {journal} {\bibinfo  {journal} {Water
  resources research}\ }\textbf {\bibinfo {volume} {20}},\ \bibinfo {pages}
  {1898} (\bibinfo {year} {1984})}\BibitemShut {NoStop}%
\bibitem [{\citenamefont {Dieker}\ and\ \citenamefont
  {Mandjes}(2003)}]{dieker2003spectral}%
  \BibitemOpen
  \bibfield  {author} {\bibinfo {author} {\bibfnamefont {A.~B.}\ \bibnamefont
  {Dieker}}\ and\ \bibinfo {author} {\bibfnamefont {M.}~\bibnamefont
  {Mandjes}},\ }\href@noop {} {\emph {\bibinfo {title} {On spectral simulation
  of fractional Brownian motion}}}\ (\bibinfo  {publisher} {Centrum voor
  Wiskunde en Informatica},\ \bibinfo {year} {2003})\BibitemShut {NoStop}%
\bibitem [{\citenamefont {Davies}\ and\ \citenamefont
  {Harte}(1987)}]{davies1987tests}%
  \BibitemOpen
  \bibfield  {author} {\bibinfo {author} {\bibfnamefont {R.~B.}\ \bibnamefont
  {Davies}}\ and\ \bibinfo {author} {\bibfnamefont {D.}~\bibnamefont {Harte}},\
  }\href@noop {} {\bibfield  {journal} {\bibinfo  {journal} {Biometrika}\
  }\textbf {\bibinfo {volume} {74}},\ \bibinfo {pages} {95} (\bibinfo {year}
  {1987})}\BibitemShut {NoStop}%
\bibitem [{Note1()}]{Note1}%
  \BibitemOpen
  \bibinfo {note} {\protect \texttt {https://networkx.org/}}\BibitemShut
  {NoStop}%
\bibitem [{\citenamefont {Dmitriy}(sus2)}]{dmitriydionysus}%
  \BibitemOpen
  \bibfield  {author} {\bibinfo {author} {\bibnamefont {Dmitriy}},\ }\href@noop
  {} {\  (\bibinfo {year} {https://mrzv.org/software/dionysus2/})}\BibitemShut
  {NoStop}%
\bibitem [{\citenamefont {Taqqu}\ \emph {et~al.}(1995)\citenamefont {Taqqu},
  \citenamefont {Teverovsky},\ and\ \citenamefont
  {Willinger}}]{taqqu1995estimators}%
  \BibitemOpen
  \bibfield  {author} {\bibinfo {author} {\bibfnamefont {M.~S.}\ \bibnamefont
  {Taqqu}}, \bibinfo {author} {\bibfnamefont {V.}~\bibnamefont {Teverovsky}}, \
  and\ \bibinfo {author} {\bibfnamefont {W.}~\bibnamefont {Willinger}},\
  }\href@noop {} {\bibfield  {journal} {\bibinfo  {journal} {Fractals}\
  }\textbf {\bibinfo {volume} {3}},\ \bibinfo {pages} {785} (\bibinfo {year}
  {1995})}\BibitemShut {NoStop}%
\bibitem [{\citenamefont {Olejarczyk}\ and\ \citenamefont
  {Jernajczyk}(2017)}]{olejarczyk2017graph}%
  \BibitemOpen
  \bibfield  {author} {\bibinfo {author} {\bibfnamefont {E.}~\bibnamefont
  {Olejarczyk}}\ and\ \bibinfo {author} {\bibfnamefont {W.}~\bibnamefont
  {Jernajczyk}},\ }\href@noop {} {\bibfield  {journal} {\bibinfo  {journal}
  {PLoS One}\ }\textbf {\bibinfo {volume} {12}},\ \bibinfo {pages} {e0188629}
  (\bibinfo {year} {2017})}\BibitemShut {NoStop}%
\bibitem [{\citenamefont {Akrami}\ \emph {et~al.}(2020)\citenamefont {Akrami},
  \citenamefont {Ashdown}, \citenamefont {Aumont}, \citenamefont {Baccigalupi},
  \citenamefont {Ballardini}, \citenamefont {Banday}, \citenamefont {Barreiro},
  \citenamefont {Bartolo}, \citenamefont {Basak}, \citenamefont {Benabed} \emph
  {et~al.}}]{akrami2020planck}%
  \BibitemOpen
  \bibfield  {author} {\bibinfo {author} {\bibfnamefont {Y.}~\bibnamefont
  {Akrami}}, \bibinfo {author} {\bibfnamefont {M.}~\bibnamefont {Ashdown}},
  \bibinfo {author} {\bibfnamefont {J.}~\bibnamefont {Aumont}}, \bibinfo
  {author} {\bibfnamefont {C.}~\bibnamefont {Baccigalupi}}, \bibinfo {author}
  {\bibfnamefont {M.}~\bibnamefont {Ballardini}}, \bibinfo {author}
  {\bibfnamefont {A.}~\bibnamefont {Banday}}, \bibinfo {author} {\bibfnamefont
  {R.}~\bibnamefont {Barreiro}}, \bibinfo {author} {\bibfnamefont
  {N.}~\bibnamefont {Bartolo}}, \bibinfo {author} {\bibfnamefont
  {S.}~\bibnamefont {Basak}}, \bibinfo {author} {\bibfnamefont
  {K.}~\bibnamefont {Benabed}},  \emph {et~al.},\ }\href@noop {} {\bibfield
  {journal} {\bibinfo  {journal} {Astronomy \& Astrophysics}\ }\textbf
  {\bibinfo {volume} {641}},\ \bibinfo {pages} {A4} (\bibinfo {year}
  {2020})}\BibitemShut {NoStop}%
\bibitem [{\citenamefont {Vafaei~Sadr}\ \emph {et~al.}(2018)\citenamefont
  {Vafaei~Sadr}, \citenamefont {Farhang}, \citenamefont {Movahed},
  \citenamefont {Bassett},\ and\ \citenamefont {Kunz}}]{sadr2018cosmic}%
  \BibitemOpen
  \bibfield  {author} {\bibinfo {author} {\bibfnamefont {A.}~\bibnamefont
  {Vafaei~Sadr}}, \bibinfo {author} {\bibfnamefont {M.}~\bibnamefont
  {Farhang}}, \bibinfo {author} {\bibfnamefont {S.}~\bibnamefont {Movahed}},
  \bibinfo {author} {\bibfnamefont {B.}~\bibnamefont {Bassett}}, \ and\
  \bibinfo {author} {\bibfnamefont {M.}~\bibnamefont {Kunz}},\ }\href@noop {}
  {\bibfield  {journal} {\bibinfo  {journal} {Monthly Notices of the Royal
  Astronomical Society}\ }\textbf {\bibinfo {volume} {478}},\ \bibinfo {pages}
  {1132} (\bibinfo {year} {2018})}\BibitemShut {NoStop}%
\bibitem [{\citenamefont {Kahane}\ and\ \citenamefont
  {Kahane}(1993)}]{kahane1993some}%
  \BibitemOpen
  \bibfield  {author} {\bibinfo {author} {\bibfnamefont {C.}~\bibnamefont
  {Kahane}}\ and\ \bibinfo {author} {\bibfnamefont {J.-P.}\ \bibnamefont
  {Kahane}},\ }\href@noop {} {\emph {\bibinfo {title} {Some random series of
  functions}}},\ Vol.~\bibinfo {volume} {5}\ (\bibinfo  {publisher} {Cambridge
  University Press},\ \bibinfo {year} {1993})\BibitemShut {NoStop}%
\bibitem [{\citenamefont {Reed}\ \emph {et~al.}(1995)\citenamefont {Reed},
  \citenamefont {Lee},\ and\ \citenamefont {Truong}}]{reed1995spectral}%
  \BibitemOpen
  \bibfield  {author} {\bibinfo {author} {\bibfnamefont {I.~S.}\ \bibnamefont
  {Reed}}, \bibinfo {author} {\bibfnamefont {P.}~\bibnamefont {Lee}}, \ and\
  \bibinfo {author} {\bibfnamefont {T.-K.}\ \bibnamefont {Truong}},\
  }\href@noop {} {\bibfield  {journal} {\bibinfo  {journal} {IEEE Transactions
  on Information Theory}\ }\textbf {\bibinfo {volume} {41}},\ \bibinfo {pages}
  {1439} (\bibinfo {year} {1995})}\BibitemShut {NoStop}%
\bibitem [{\citenamefont {Merelli}\ \emph {et~al.}(2015)\citenamefont
  {Merelli}, \citenamefont {Rucco}, \citenamefont {Sloot},\ and\ \citenamefont
  {Tesei}}]{merelli2015topological}%
  \BibitemOpen
  \bibfield  {author} {\bibinfo {author} {\bibfnamefont {E.}~\bibnamefont
  {Merelli}}, \bibinfo {author} {\bibfnamefont {M.}~\bibnamefont {Rucco}},
  \bibinfo {author} {\bibfnamefont {P.}~\bibnamefont {Sloot}}, \ and\ \bibinfo
  {author} {\bibfnamefont {L.}~\bibnamefont {Tesei}},\ }\href@noop {}
  {\bibfield  {journal} {\bibinfo  {journal} {Entropy}\ }\textbf {\bibinfo
  {volume} {17}},\ \bibinfo {pages} {6872} (\bibinfo {year}
  {2015})}\BibitemShut {NoStop}%
\bibitem [{\citenamefont {Atienza}\ \emph {et~al.}(2019)\citenamefont
  {Atienza}, \citenamefont {Gonzalez-Diaz},\ and\ \citenamefont
  {Rucco}}]{atienza2019persistent}%
  \BibitemOpen
  \bibfield  {author} {\bibinfo {author} {\bibfnamefont {N.}~\bibnamefont
  {Atienza}}, \bibinfo {author} {\bibfnamefont {R.}~\bibnamefont
  {Gonzalez-Diaz}}, \ and\ \bibinfo {author} {\bibfnamefont {M.}~\bibnamefont
  {Rucco}},\ }\href@noop {} {\bibfield  {journal} {\bibinfo  {journal} {Journal
  of Intelligent Information Systems}\ }\textbf {\bibinfo {volume} {52}},\
  \bibinfo {pages} {637} (\bibinfo {year} {2019})}\BibitemShut {NoStop}%
\end{thebibliography}
%merlin.mbs apsrev4-1.bst 2010-07-25 4.21a (PWD, AO, DPC) hacked
%Control: key (0)
%Control: author (72) initials jnrlst
%Control: editor formatted (1) identically to author
%Control: production of article title (-1) disabled
%Control: page (0) single
%Control: year (1) truncated
%Control: production of eprint (0) enabled
%

\end{document}